\documentstyle{article}

\def\noheaderplainsetup{%

      \topmargin=0pt \headheight=0pt \headsep=0pt  
      \oddsidemargin=0pt \evensidemargin=0pt       
      \textheight=8.9truein \textwidth=6.5truein}  

\noheaderplainsetup

\begin{document}

\newcommand{\lll}{\mbox{{\bf CL8}}}
\newcommand{\llls}{\mbox{{\bf CL8S}}}
\newcommand{\Bigmlc}{\mbox{{\Large $\wedge$}}}
\newcommand{\Bigmld}{\mbox{{\Large $\vee$}}}
\newcommand{\bigmlc}{\mbox{{\large $\wedge$}}}
\newcommand{\bigmld}{\mbox{{\large $\vee$}}}
\newcommand{\bigleftbrace}{\mbox{{\large $\{$}}}
\newcommand{\bigrightbrace}{\mbox{{\large $\}$}}}
\newcommand{\Bigleftbrace}{\mbox{{\Large $\{$}}}
\newcommand{\Bigrightbrace}{\mbox{{\Large $\}$}}}


\newcommand{\gneg}{\neg}                  
\newcommand{\mli}{\rightarrow}                     
\newcommand{\mld}{\vee}    
\newcommand{\mlc}{\wedge}  


\newtheorem{theoremm}{Theorem}[section]
\newtheorem{factt}[theoremm]{Fact}
\newtheorem{definitionn}[theoremm]{Definition}
\newtheorem{lemmaa}[theoremm]{Lemma}
\newtheorem{conventionn}[theoremm]{Convention}
\newtheorem{claimm}[theoremm]{Claim}

\newenvironment{definition}{\begin{definitionn} \em}{ \end{definitionn}}
\newenvironment{theorem}{\begin{theoremm}}{\end{theoremm}}
\newenvironment{lemma}{\begin{lemmaa}}{\end{lemmaa}}
\newenvironment{fact}{\begin{factt}}{\end{factt}}
\newenvironment{claim}{\begin{claimm}}{\end{claimm}}
\newenvironment{convention}{\begin{conventionn} \em}{\end{conventionn}}
\newenvironment{proof}{ {\bf Proof.} }{\  $\Box$ \vspace{.1in} }

\title{Cirquent calculus deepened}
\author{Giorgi Japaridze
}
\date{}
\maketitle

\begin{abstract}  {\em Cirquent calculus} is a new proof-theoretic and  semantic framework, whose main distinguishing feature is being based on circuit-style structures
(called {\em cirquents}), as opposed   to the more traditional approaches that deal with tree-like objects such as formulas, sequents or hypersequents. 
Among its advantages are greater efficiency, flexibility and expressiveness. 
This paper presents a detailed elaboration of a deep-inference cirquent logic, which is naturally and inherently resource conscious. It shows that classical logic, both syntactically and semantically, can be seen to be just a special, conservative fragment of this more general and, in a sense, more basic logic --- the logic of resources in the form of cirquent calculus. 
The reader will find various arguments in favor of switching to the new framework, such as arguments showing the insufficiency of the expressive power of linear logic or other formula-based approaches to developing resource logics, exponential improvements over the traditional approaches in both representational and proof complexities offered by cirquent calculus (including the existence of polynomial size cut-, substitution- and extension-free cirquent calculus proofs for the notoriously hard pigeonhole principle), and more. Among the main purposes of this paper is to provide an introductory-style  starting point for what, as the author wishes to hope, might have a chance to become a new line of research in proof theory --- a proof theory based on circuits instead of formulas.  

\end{abstract}


\  


\section{Introduction}\label{intr}
Among the main objectives of the introductory section of a well-written paper should be to help the reader determine whether he or she is willing to invest time into reading the rest of it.  The following rhetorical question\footnote{Asked by Alessio Guglielmi at http://news.gmane.org/gmane.science.mathematics.frogs on September 17, 2007 during a discussion of the preliminary version of the present paper.} may contribute to making such a determination in the present case:
\begin{center}{\em What is the most natural representation of 
Boolean functions, formulas or circuits?}\end{center}
Those who are not quite clear about the meaning of the word ``natural'', may try to replace it with ``direct'', ``reasonable''  or ``efficient'', and think about where the computer industry would be at present if, for some strange reason, computer engineers had insisted on tree- rather than graph-style cicuitries.
Or ask why one does not hear theoretical computer scientists speak about formula complexity nearly as often as about circuit complexity.

Should then proof-theoreticians continue sticking to formulas, especially now that logic is increasingly CS-oriented, and efficiency is of much greater  concern than it was in the days of Frege, Hilbert and Gentzen? 
The author believes that there are no good reasons for such conservatism other than habit and tradition, if not laziness.
And this paper is for those who might feel potentially ready to  accept the same view, or be  curious enough to be willing to take a look at what happens when one gives the idea a try. It is devoted to (re)introducing and advancing the foundations of {\em cirquent calculus}, the circuit-based proof theory.

Unlike the more traditional syntactic approaches that manipulate tree- or forest-like objects such as formulas or sequents and where proofs are often also trees, cirquent calculus deals with circuit-style constructs called {\em cirquents}, in which children may be shared between different parent nodes. Furthermore, being intrinsically a deep inference (see later) approach, it makes possible combining, within a single cirquent, what would otherwise be different parallel nodes (formulas, sequents) of a proof tree, meaning that eventually not only subcirquents, but also subtransformations are amenable to being shared. Sharing thus allows us to achieve higher efficiency, whether it be the compactness of representations of Boolean functions or other objects of study, or the numbers of steps in derivations and proofs. Indeed, in natural situations, specifically  ones arising in the world of computing,  
prohibitively long formulas typically owe their sizes to reoccurring subformulas, and explosively  
large proof trees often emerge as a result of the necessity to perform identical or similar steps over and over again.  

The possibility of compressing formulas or proofs is not the only --- in fact, not even the primary --- appeal of cirquent calculus. Generality, flexibility and expressiveness are other, more fundamental, advantages to point out. Cirquent calculus is more general than the calculus of structures (Guglielmi et al. \cite{Gug07,Brunnen,Gug01,Gug06}); the latter is more general than hypersequent calculus (Avron \cite{Avron}, Pottinger \cite{Pottinger}); and the latter, in turn, is more general than sequent calculus (Gentzen).  Each framework in this hierarchy permits to successfully axiomatize certain logics that the predecessor frameworks fail to tame. Cirquent calculus itself was originally introduced as a deductive system for the resource-conscious computability logic \cite{Jap03,Japdina,Japfin} after it had become evident that neither sequent calculus nor the more flexible and promising calculus of structures were sufficient to axiomatize it.  

While in classical logic circuits do not offer any additional expressive power, they --- more precisely, {\em cirquents} that are more general than circuits --- turn out to be properly more expressive than formulas when it comes to finer semantical approaches such as resource logics, with {\em computability logic} (\cite{Jap03,Japdina,Japfin,JapIC}) and {\em abstract resource semantics} (\cite{Jap06}) being two examples. Efficiency considerations totally aside, it was exactly this expressive power that in \cite{Jap06} made a difference between axiomatizability and unaxiomatizability for computability logic or abstract resource semantics: even if one is only trying to set up a deductive system that proves all (and only) valid {\em formulas}, intermediate steps in proofs of such formulas still inherently require using objects (cirquents) that cannot be written as formulas. 

Switching from formulas to cirquents indeed becomes imperative --- not only syntactically but also semantically --- if one wants to systematically develop resource logics. Fine-level resource-semantical approaches intrinsically require the ability to account for the possibility of {\em resource sharing}, the ability that linear logic or other formula- or sequent-based approaches do not and cannot possess. The following naive example may provide some insights. 

We are talking about a vending machine that has slots for 25-cent ($25c$) coins, with each slot taking a single coin. Coins can be authentic or counterfeited. Let us instead use the more generic terms {\em true} and {\em false} here, as there are various particular situations naturally and inevitably emerging in the world of resources corresponding to those two opposite values. Below are a few examples of real-world resources and the possible meanings of the two semantical values for them:   
\begin{itemize}
\item A financial debt, which may (true) or may not (false) be eventually paid; 
\item an electrical outlet or a battery, which may (true) or not (false) actually have sufficient power in it;
\item a standard task performed by a company's employee or an AI agent, which, eventually, may (true) or not (false) be successfully completed; 
\item a specified amount of computer memory required by a process, which may (true) or not (false) be available at a given time;
\item a promise, which may be kept (true) or broken (false).
\end{itemize}   
See Section 8 of \cite{Jap06} for detailed elaborations of these intuitions, as well as strict definitions of the concepts of the associated formal semantics,  which is the earlier-mentioned abstract resource semantics. 

Continuing the description of our vending machine, inserting a false coin into a slot fills the slot up (so that no other coins can be inserted into it until the operation is complete), but otherwise does not fool the machine into thinking that it has received 25 cents. A candy costs 50 cents, and the machine will dispense a candy if at least two of its slots receive true coins. Pressing the ``dispense'' button while having inserted anything less than 50 cents, such as a single coin, or one true and two false coins, results in a non-recoverable loss.   

Victor has three $25c$-coins, and he knows that two of them are true while one is perhaps false (but he has no way to tell which one is false). Could he get a candy? 

Well, expected or not, the answer depends on how many slots the machine has. Consider two cases: machine $M2$ with two slots, and machine $M3$ with three slots. Victor would have no problem with $M3$: he can insert his three coins into the three slots, and the machine, having received $\geq 50c$, will dispense a candy. With $M2$, however, Victor is in trouble. He can try inserting arbitrary two of his three coins into the two slots of the machine, but there is no guarantee that one of those two coins is not false, in which case Victor will end up with no candy and only $25$ cents remaining in his pocket. 

Both $M2$ and $M3$ can be understood as resources --- resources turning coins into a candy. And note that these two resources are not the same: $M3$ is obviously stronger (``better''), as it allows Victor to get a candy whereas $M2$ does not, while, at the same time, anyone rich enough to be able to make $M2$ dispense a candy would be able to do the same with $R3$ as well.  Yet, formulas fail to capture this important difference. 
With $\mli$, $\mlc$, $\mld$ here and later standing for multiplicative-style connectives (called {\em parallel connectives} in computability logic), $M2$ and $M3$ can be written as
\[\mbox{$R2\mli \mbox{\em Candy}$ \ and \ $R3\mli \mbox{\em Candy}$,}\] 
respectively: they consume a certain resource $R2$ or $R3$ and produce {\em Candy}. What makes $M3$ stronger than $M2$ is that the subresource $R3$ that it consumes is weaker (easier to supply) than the subresource $R2$ consumed by $M2$. Specifically, with one false and two true coins, Victor is able to satisfy $R3$ but not $R2$. 

The resource $R2$ can be represented as the following cirquent:
  
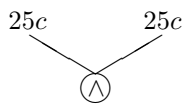
\begin{center} \begin{picture}(70,40)

\put(33,11){\line(5,3){25}}
\put(33,11){\line(-5,3){25}}

\put(0,28){$25c$}
\put(51,28){$25c$}

\put(30,3){$\mlc$}
\put(33,6){\circle{11}}
\end{picture}
\end{center}

\noindent which, due to being tree-like, can also be adequately written as the formula \[25c\mlc 25c.\] As for the resource $R3$, either one of the following two cirquents is an adequate representation of it, with one of them probably showing the relevant part of the actual physical circuitry used in $M3$: 

\begin{center} \begin{picture}(235,87)

\put(36,57){\line(-2,1){29}}
\put(36,57){\line(2,1){29}}
\put(7,57){\line(0,1){15}}
\put(7,57){\line(2,1){29}}

\put(36,31){\line(2,1){29}}
\put(36,31){\line(-2,1){29}}
\put(36,31){\line(0,1){14}}

\put(65,57){\line(-2,1){29}}
\put(65,57){\line(0,1){15}}

\put(0,75){$25c$}
\put(29,75){$25c$}
\put(58,75){$25c$}

\put(33,48){$\mlc$}
\put(36,51){\circle{11}}

\put(4,48){$\mlc$}
\put(7,51){\circle{11}}

\put(62,48){$\mlc$}
\put(65,51){\circle{11}}

\put(33,23){$\mld$}
\put(36,26){\circle{11}}
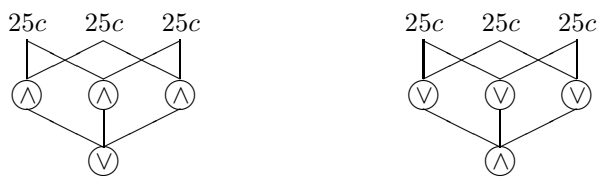
\put(-14,8){{\bf Figure 1:} Two equivalent cirquents for the resource $R3$}

\put(186,57){\line(-2,1){29}}
\put(186,57){\line(2,1){29}}
\put(157,57){\line(0,1){15}}
\put(157,57){\line(2,1){29}}

\put(186,31){\line(2,1){29}}
\put(186,31){\line(-2,1){29}}
\put(186,31){\line(0,1){14}}

\put(215,57){\line(-2,1){29}}
\put(215,57){\line(0,1){15}}

\put(150,75){$25c$}
\put(179,75){$25c$}
\put(208,75){$25c$}

\put(183,48){$\mld$}
\put(186,51){\circle{11}}

\put(154,48){$\mld$}
\put(157,51){\circle{11}}

\put(212,48){$\mld$}
\put(215,51){\circle{11}}

\put(183,23){$\mlc$}
\put(186,26){\circle{11}}
\end{picture}
\end{center}

\noindent Unlike $R2$, however, $R3$ cannot be represented through a formula. $25c\mlc 25c$ does not fit the bill, for it represents $R2$ which, as we already agreed, is not the same as $R3$. In another attempt to find a formula, we might try to rewrite one of the above two cirquents --- let it be the one on the right --- into an ``equivalent'' formula in the standard way, by duplicating and separating shared nodes. This results in  
\begin{equation}\label{f1}
(25c\mld 25c)\mlc(25c\mld 25c)\mlc (25c\mld 25c)\end{equation}
which, however, is not any more adequate than $25c\mlc 25c$. It expresses not $R3$ but the resource consumed by 
a    machine with six coin slots grouped into three pairs, where (at least) one slot in each of the three pairs needs to receive a true coin. Such a machine thus dispenses a candy for $\geq 75$ rather than $\geq 50$ cents, which makes Victor's resources insufficient. 

The trouble here is related to the inability of formulas to explicitly account for resource sharing or the absence thereof. The cirquent on the right of Figure 1 stands for a conjunction of three resources, each conjunct, in turn, being a disjunction of two subresources of type $25c$. However, altogether there are three rather than six $25c$-type subresources, each one being shared between two different conjuncts of the main resource. Formula (\ref{f1}) is inadequate because, for example, it fails to indicate that the first and the third occurrences of ``$25c$'' stand for the same resource while the second and the fifth (as well as the fourth and the sixth) occurrences stand for another resource, albeit a resource of the same $25c$-type. 

From the resource-philosophical point of view, classical logic and linear logic are two imperfect extremes.  In the former, all occurrences of a same subformula mean ``the same'' (represent the same resource), i.e., {\em everything is shared} that can be shared; and in the latter, each occurrence stands for a separate resource, i.e., {\em nothing is shared} at all. Neither approach does thus permit to account for mixed cases where certain occurrences are meant to represent the same resource while some other occurrences stand for different resources of the same type. And it is an absolute shame that linear logic or similar --- naive from the perspective of cirquent calculus and abstract resource semantics --- resource-oriented approaches fail  to express simple, natural and unavoidable things such as the ``two out of three'' combination expressed
by the cirquents of Figure 1.   

It was mentioned earlier that cirquents are more general than circuits --- otherwise there would be no need to invent a special name for them after all. In structures  commonly referred to as Boolean circuits, the label of each input is unique, while cirquents may have any number of ``inputs'' (called {\em ports} in abstract resource semantics) of any given {\em type} (label). So, strictly speaking, what we see in Figure 1, having three $25c$-ports, are cirquents but not circuits. Of course, in an attempt to make them meaningful as circuits in the traditional sense, one could think of renaming the three $25c$-ports 
 into, say, $P$, $Q$ and $R$. But then a crucial piece of information would be lost, specifically the information about all inputs being of the same type $25c$,  as opposed to, say, the three different types $25c$, $10c$, $5c$. This would make it impossible to match Victor's resources with those inputs.   

Among the main ideological merits of the present contribution is the unification and reconciliation of classical logic and the logic of resources  on the basis of one single semantics and one single syntax. The rather unsettling situation of conflict and disagreement between classical and resource logics, familiar from linear logic\footnote{What is meant by {\em linear logic} here and throughout this paper is the multiplicative fragment of linear logic, whose connectives are written using the classical symbols.} or even the predecessor \cite{Jap06} of the present paper, now gives way to a perfect peace an harmony. Specifically, we make the point that the language of classical logic can and should be seen as a proper fragment of the language of resource logics, obtained by considering only {\em circuits}, i.e., cirquents where multiple identical-label ports are not allowed. With this view, there is no need to have separate semantics and syntax for classical logic: they turn out to be the same as those of our cirquent calculus, only restricted to the cirquents that are circuits.  That is, our resource logic (cirquent calculus) is simply more expressive --- and thus more general --- than classical logic, but otherwise there are no semantic or syntactic differences or disagreements between the two: the former is a conservative extension of the latter. The following example may help get a feel of this. 

Let us consider the formula 
\begin{equation}\label{ff2}
\gneg P\mld (\gneg Q\mlc P)\mld (P\mlc \gneg R)\mld (Q\mlc R).
\end{equation}

\noindent This formula is valid in classical logic while linear logic, or the approach of \cite{Jap06}, would consider it invalid. But is it or is not it valid according to our present approach, which was promised to eliminate any disagreements between the classical and resource-conscious views? It should be remembered that we have dismissed formulas as imperfect means of expression. So, the intended meaning of (\ref{ff2}) must be first expressed through a cirquent before answering or even asking a question about its validity. And, as we have not (not yet, at least) agreed on any standard way of translating formulas into cirquents, only whoever wrote (\ref{ff2}) can tell what he or she wanted to express. Frege would probably explain his meaning through the left cirquent of the following Figure 2, while Girard through the right cirquent: 

\begin{center} \begin{picture}(271,87)
\put(0,75){$\gneg P$}
\put(19,75){$\gneg Q$}
\put(42,75){$P$}
\put(54,75){$\gneg R$}
\put(76,75){$Q$}
\put(92,75){$R$}

\put(28,57){\line(6,5){17}}
\put(28,57){\line(0,1){14}}
\put(25,48){$\mlc$}
\put(28,51){\circle{11}}

\put(62,57){\line(-6,5){17}}
\put(62,57){\line(0,1){14}}
\put(59,48){$\mlc$}
\put(62,51){\circle{11}}

\put(96,57){\line(-6,5){17}}
\put(96,57){\line(0,1){14}}
\put(93,48){$\mlc$}
\put(96,51){\circle{11}}

\put(9,41){\line(0,1){31}}
\put(45,32){\line(-4,1){36}}
\put(45,32){\line(4,1){52}}
\put(45,32){\line(-6,5){16}}
\put(45,32){\line(6,5){16}}
\put(42,23){$\mld$}
\put(45,26){\circle{11}}
\put(48,8){{\bf Figure 2:} Two possible meanings of (\ref{ff2})}

\put(158,75){$\gneg P$}
\put(177,75){$\gneg Q$}
\put(199,75){$P$}
\put(216,75){$P$}
\put(227,75){$\gneg R$}
\put(250,75){$Q$}
\put(265,75){$R$}

\put(194,57){\line(-1,2){7}}
\put(194,57){\line(1,2){7}}
\put(191,48){$\mlc$}
\put(194,51){\circle{11}}

\put(228,57){\line(-1,2){7}}
\put(228,57){\line(1,2){7}}
\put(225,48){$\mlc$}
\put(228,51){\circle{11}}

\put(262,57){\line(-1,2){7}}
\put(262,57){\line(1,2){7}}
\put(259,48){$\mlc$}
\put(262,51){\circle{11}}

\put(167,43){\line(0,1){29}}
\put(211,32){\line(-4,1){44}}
\put(211,32){\line(4,1){52}}
\put(211,32){\line(-6,5){16}}
\put(211,32){\line(6,5){16}}
\put(208,23){$\mld$}
\put(211,26){\circle{11}}
\end{picture}
\end{center}

\noindent Then, regarding the question on validity, as will be seen later, we would answer ``Yes'' to Frege and ``No'' to Girard. And our negative answer in the second case does not at all  conflict with the seemingly positive answer of classical logic. The right cirquent of Figure 2, for the reason of having two separate $P$-ports and thus not being a circuit, is simply not a meaningful or legal --- let alone valid --- expression for classical logic.  
 
The idea of cirquent calculus was born very recently in \cite{Jap06}.  That so far the only paper on the subject introduced cirquent calculus in a special, ``shallow'' form, where all cirquents were required to be of depth two, with a conjunctive gate at the root and disjunctive gates as second-level nodes. While the shallow version of cirquent calculus was sufficient to achieve the main goal of that paper --- axiomatizing the otherwise unaxiomatizable basic fragment of computability logic --- the paper also outlined the possibility and expediency of studying more general, deep versions of cirquent calculi. The present article contains a realization of that outline. It elaborates a deep cirquent calculus system {\bf CL8} for computability logic, which happens to coincide with the logic induced by  abstract resource semantics, and which is a conservative extension of classical logic, so that {\bf CL8} is also an alternative system for classical logic. {\bf CL8} permits cirquents of arbitrary depths and forms, which naturally invites inference rules that modify cirquents at any level rather than only around the root as is the case in sequent calculus. This is called {\em deep inference}, and is one of the central ideas in the earlier mentioned calculus of structures. The present paper also borrows many other useful ideas and techniques from the calculus of structures, which is the nearest precursor of cirquent calculus in its present, general form.

The rest of the paper is organized as follows. In Section 2 we (re)introduce the notion of cirquents which generalizes the cirquents from \cite{Jap06} by removing any restrictions on the depths and forms of cirquents.  Section 3 introduces and explains the rules of inference of the deep cirquent calculus system \lll. Section 4 defines the notions of derivation, proof, and admissibility  for \lll\ and similar systems. Section 5 generalizes both the semantics of classical logic and the abstract resource semantics of \cite{Jap06} to a common, unifying resource semantics (still called {\em abstract resource semantics}) for all cirquents, and proves the corresponding soundness and completeness result for \lll.  Section 6 discusses some possible variations of cirquent calculus systems, including a version of \lll\ where each rule of inference comes with its dual (symmetric) one, and systems that deal with cirquents with non-standard types of gates. Section 7 discusses the relation of \lll\ to classical sequents calculus (showing the p-simulation of the latter by the former) and the two shallow cirquent calculus systems constructed earlier in \cite{Jap06}. Finally, Section 8 presents polynomial size \lll-proofs of the notoriously hard-to-prove family of tautologies known as the {\em pigeonhole principle}. These are so far the only known efficient proofs of that family that employ neither cut nor extension or substitution --- the rules undesirable for their being ``highly non-analytic''.

\section{Cirquents, formulas and hyperformulas}\label{mul}

We fix some set of syntactic objects called {\bf atoms}, for which we will be using $P,Q,R,S,T$ as metavariables. An atom $P$ and its {\bf negation} $\gneg P$  are called {\bf literals}.
The two literals $P$ and $\gneg P$ are said to be {\bf opposite}.

Let us agree that in this paper a  {\bf graph} means a directed acyclic graph whose every node is labeled with either a literal or $\mlc$ or $\mld$. 
The $\mlc$- and $\mld$-labeled nodes of (such) a graph we call {\bf gates}, and the nodes labeled with literals we call {\bf ports}. Specifically, a node labeled with a literal $L$ is said to be a an {\bf $L$-port}; an $\mlc$-labeled node is said to be a {\bf conjunctive gate}; and an 
$\mld$-labeled node is said to be a {\bf disjunctive gate}. When there is an edge from a node $a$ to a node $b$, we say that $b$ is a {\bf child} of $a$ and $a$ is a {\bf parent} of $b$. 
The relations ``{\bf descendant}'' and ``{\bf ancestor}'' are the transitive closures of the relations ``child'' and ``parent'', respectively. The meanings of some other standard relations such as ``grandchild'', ``grandparent'', etc. should also be clear. 

A {\bf cirquent} is a graph (in the above sense) satisfying the following two conditions:
\begin{itemize}
\item Ports have no children.
\item There is a node, called the {\bf root}, which is an ancestor of all other nodes in the graph.
\end{itemize}

A cirquent is said to be a {\bf circuit} iff all of its ports have different labels (here the two opposite labels $P$ and $\gneg P$ count as different).

Graphically, we represent a port through the corresponding literal, a conjunctive gate through {\large $\circ$}, and a disjunctive gate through {\large $\bullet$}. We agree that the direction of an edge is always upward, which allows us to draw lines rather than arrows for edges. Below is an example of a cirquent with 4 ports  and 8 gates. Note that not only ports but also gates can be childless. A childless disjunctive gate semantically corresponds to $\bot$, and a childless conjunctive gate corresponds to $\top$.

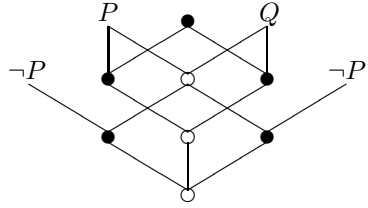
\begin{center} \begin{picture}(98,92)
\put(30,72){$P$}
\put(91,72){$Q$}
\put(-4,50){$\gneg P$}
\put(117,50){$\gneg P$}
\put(64,8){\line(0,3){18}}
\put(34,30){\line(-5,3){30}}
\put(34,30){\line(5,3){30}}
\put(94,30){\line(-5,3){30}}
\put(94,30){\line(5,3){30}}
\put(64,52){\line(-5,3){30}}
\put(64,52){\line(5,3){30}}
\put(34,52){\line(5,3){30}}
\put(94,52){\line(-5,3){30}}
\put(34,52){\line(0,3){18}}
\put(94,52){\line(0,3){18}}
\put(64,8){\line(5,3){30}}
\put(64,8){\line(-5,3){30}}
\put(64,30){\line(5,3){30}}
\put(64,30){\line(-5,3){30}}
\put(34,50){\circle*{5}}
\put(94,50){\circle*{5}}
\put(64,50){\circle{5}}
\put(34,28){\circle*{5}}
\put(64,28){\circle{5}}
\put(94,28){\circle*{5}}
\put(64,6){\circle{5}}
\put(64,72){\circle*{5}}
\end{picture}
\end{center}

In the introductory section, we emphasized the need for rejecting formulas in favor of cirquents: from the perspective of cirquent calculus, formulas are incomplete, inefficient  and not the most natural means of expression. In the process of formalizing a piece of the real world, a natural way to represent a Boolean function or whatever similar objects we study is to do so directly through a cirquent. It would be somewhat odd to first try to write it through a formula (if possible at all) and then translate that formula into a cirquent. So, in principle, we have no ``legal obligation'' to define the precise meanings of formulas
in terms of cirquents, as there is no need for formulas at all. But even if not ``legal'', we still do have a ``moral'' duty to agree on some standard way of translating formulas into cirquents, to pay tribute to the firmly established logical tradition of dealing with formulas.    

By a {\bf formula} in this paper we mean that of the language of classical propositional logic, built from literals and variable-arity operators $\mld$, $\mlc$ in the standard way. The disjunction of $F_1,\ldots,F_n$ can be written as either $F_1\mld\ldots\mld F_n$ or $\mld\{F_1,\ldots,F_n\}$. Similarly for conjunction. $\bot$ is considered an abbreviation of the empty disjunction $\mld\{\}$, and $\top$ an abbreviation of the empty conjunction $\mlc\{\}$. Further, we treat $E\mli F$ as an abbreviation of $\gneg E\mld F$, and $\gneg H$, when $H$ is not an atom, as an abbreviation defined by: $\gneg\gneg F=F$; \ $\gneg (F_1\mld\ldots\mld F_n)=\gneg F_1\mlc\ldots\mlc\gneg F_n$; \  $\gneg (F_1\mlc\ldots\mlc F_n)=\gneg F_1\mld\ldots\mld\gneg F_n$. 

We agree to understand (``translate'') each formula used in this paper as --- and identify with --- the cirquent which is nothing but the parse tree for that formula. More precisely, we have:

\begin{itemize}
\item A literal $L$ is understood as the cirquent whose only node (=root) is an $L$-port.
\item Let $F_1,\ldots,F_n$ be any formulas, and let graph $G$ be the disjoint union of those formulas understood as cirquents. Then:
\begin{itemize}
\item $F_1\mld\ldots\mld F_n$ is understood as the cirquent obtained by adding a new disjunctive gate (root) to $G$, and connecting it with an edge to each of the $n$ parentless nodes of $G$.
\item Similarly for $F_1\mlc\ldots\mlc F_n$, with the difference that the root gate here will be a conjunctive one.  
\end{itemize}      
\end{itemize}
Note that since we require the above $G$ to be a disjoint union, every formula is a tree-like cirquent,  with each non-root node having exactly one parent.  The above way of translating formulas into cirquents is thus in the spirit of formula-based resource logics (such as linear logic) rather than classical logic. For example, formula (\ref{ff2}) of Section 1 translates into the right rather than the left cirquent of Figure 2. So, we still need to separately clarify how to translate formulas when they appear in the context of classical logic (as they mostly do in the literature) rather than in the context of resource logics (as they always do in the present paper). For that purpose, we first generalize formulas to what we call ``hyperformulas''. 

A {\bf hyperformula} is the same as a formula, with the only difference that some subformulas in it may be {\em overlined} (double overlines are not allowed). Hyperformulas, just like formulas, are understood as cirquents. To translate a hyperformula $F$ into a corresponding cirquent, one should first ignore all overlines in $F$ and translate it into a tree-like cirquent according to the earlier prescriptions, and then merge all subcirquents that correspond to (originate from) identical {\em overlined} subformulas of $F$. Rather than trying to turn this semiformal explanation into a strict definition (which is certainly possible), below we just provide a few examples that should make the meaning of the above-said perfectly clear. 

The hyperformula 
\[(Q\mld\overline{R})\mlc(\overline{R}\mlc Q)\]
means (is translated into) the cirquent 
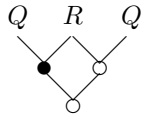
\begin{center} \begin{picture}(49,51)

\put(0,41){$Q$}
\put(21,41){$R$}
\put(43,41){$Q$}

\put(14,24){\circle*{5}}
\put(14,26){\line(-1,1){10}}
\put(14,26){\line(1,1){10}}

\put(35,24){\circle{5}}
\put(35,26){\line(-1,1){10}}
\put(35,26){\line(1,1){10}}

\put(25,12){\line(-1,1){10}}
\put(25,12){\line(1,1){10}}
\put(25,10){\circle{5}}
\end{picture}\end{center}
Here the two occurrences of $R$ in the hyperformula are considered ``the same'' as both are overlined; on the other hand, the two occurrences of $Q$ did not merge because they were not overlined. At the same time, each of the hyperformulas \((Q\mld R)\mlc(R\mlc Q)\), \ \((Q\mld \overline{R})\mlc(R\mlc Q)\), \  \((\overline{Q}\mld R)\mlc(\overline{R}\mlc Q)\), \ \(\overline{(Q\mld R)}\mlc(R\mlc Q)\) stands for the same tree-like cirquent 

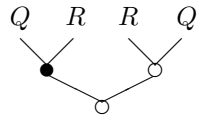
\begin{center} \begin{picture}(69,51)

\put(0,41){$Q$}
\put(21,41){$R$}

\put(41,41){$R$}
\put(63,41){$Q$}

\put(14,24){\circle*{5}}
\put(14,26){\line(-1,1){10}}
\put(14,26){\line(1,1){10}}

\put(55,24){\circle{5}}
\put(55,26){\line(-1,1){10}}
\put(55,26){\line(1,1){10}}

\put(35,12){\line(-2,1){20}}
\put(35,12){\line(2,1){19}}
\put(35,10){\circle{5}}
\end{picture}\end{center}
as there is nothing to merge within or across the overlined subexpressions. 

On the left of the following figure we see an overline-free hyperformula (i.e., a formula) and the corresponding tree-like cirquent; and, on the right, we see the same (hyper)formula fully overlined, resulting in a much more compressed cirquent. Note that in this case not only all identical-label ports have merged, but also the (two) identical-content gates, as both of the corresponding subformulas $\gneg P\mld P$ were found under an overline.

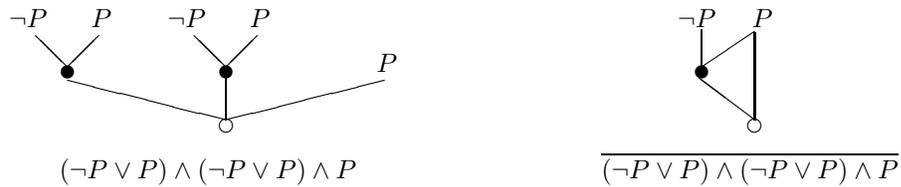
\begin{center} \begin{picture}(300,92)
\put(1,67){$\gneg P$}
\put(32,67){$P$}
\put(61,67){$\gneg P$}
\put(92,67){$P$}
\put(23,52){\line(-1,1){12}}
\put(23,52){\line(1,1){12}}
\put(83,52){\line(-1,1){12}}
\put(83,52){\line(1,1){12}}
\put(23,50){\circle*{5}}
\put(83,50){\circle*{5}}
\put(140,50){$P$}
\put(83,32){\line(-4,1){60}}
\put(83,32){\line(4,1){60}}
\put(83,32){\line(0,1){18}}
\put(83,30){\circle{5}}
\put(254,67){$\gneg P$}
\put(282,67){$P$}
\put(263,52){\line(0,1){14}}
\put(263,52){\line(3,2){20}}
\put(263,50){\circle*{5}}
\put(283,32){\line(-4,3){21}}
\put(283,32){\line(0,1){33}}
\put(283,30){\circle{5}}
\put(20,10){$(\gneg P\mld P)\mlc(\gneg P\mld P)\mlc P$}
\put(225,10){$\overline{(\gneg P\mld P)\mlc(\gneg P\mld P)\mlc P}$}
\end{picture}\end{center}

Anyway, now we are ready to explain how formulas should be translated into cirquents when they are used in the context of classical logic. We agree to see each formula $F$ of classical logic as the hyperformula --- and hence the corresponding cirquent --- obtained from $F$ by overlining all (and only)\footnote{Nothing would go wrong if we dropped this ``{\em only} literals'' condition, as long as the ``{\em all} literals'' condition is retained. For example, overlining the entire $F$ (which would automatically include all literals) would work for our purposes just as well. But overlining only literals is easier, so why bother.} literals. Such a hyperformula will be denoted by {\em underlining} (rather than overlining) $F$:
\[\underline{F}.\]
So, for example, if formula (\ref{ff2}) of Section 1 is found in a textbook on classical (as opposed to linear) logic, it should be understood as a ``lazy way'' to write the hyperformula 
\[\underline{\gneg P\mld (\gneg Q\mlc P)\mld (P\mlc \gneg R)\mld (Q\mlc R)},\]
i.e., the hyperformula 
\[\overline{\gneg P}\mld (\overline{\gneg Q}\mlc \overline{P})\mld (\overline{P}\mlc \overline{\gneg R})\mld (\overline{Q}\mlc \overline{R}),\]
and, correspondingly, should be translated as the left rather than the right cirquent of Figure 2. Generally, a notational synchronization of any traditional piece of writing on or in classical logic with our approach would take as little as just underlining --- explicitly or implicitly --- every formula appearing in it.

Before closing this section, we want to make the observation that, while hyperformulas are more expressive than formulas, they are still far from being expressive enough to be able to represent all cirquents. For example, the cirquents of Figure 1 cannot be written as hyperformulas. 

\section{The rules of \lll}\label{lll}

\begin{convention}\label{march15}
By a {\bf rule of inference} in this section we mean  a set $\cal R$ of $(2+m+n)$-tuples 
\[(A,B,a_1,\ldots,a_m,\Pi_1,\ldots,\Pi_n)\]
(fixed $m,n$ for a given rule), called {\bf instances} or {\bf applications}  of $\cal R$, where: 
\begin{enumerate}
\item $A$ and $B$ are cirquents, said to be the {\bf premise} and the {\bf conclusion} (of the given application of $\cal R$), respectively.
\item $a_1,\ldots,a_m$, said to be  {\bf central parameters}, are pairwise distinct nodes, each one being a node of either the premise or the conclusion or both.
\item Each $\Pi_i$, said to be a {\bf peripheral parameter}, is a set of nodes not containing any central parameters, such that every $b\in\Pi_i$ is a parent or a child of some central parameter in either the premise or the conclusion or both. 
\item All children and parents of each central parameter, whether it be in the premise or in the conclusion, are among the elements of  $\{a_1,\ldots,a_m\}\cup \Pi_1\cup\ldots\cup\Pi_n$. 
\end{enumerate} 
\end{convention}

When $(A,B,a_1,\ldots,a_m,\Pi_1,\ldots,\Pi_n)\in {\cal R}$, we say that $B$ {\bf follows from} $A$ by rule $\cal R$ with parameters $a_1,\ldots,a_m,\Pi_1,\ldots,\Pi_n$, or --- if lazy to specify the parameters --- simply that $B$ follows from $A$ by $\cal R$. 

Thus, each application of a rule has one premise and one conclusion. The conclusion is usually obtained from the premise (or vice versa) through modifying only a certain part, while leaving the rest of the cirquent unchanged. Specifically, depending on the particular stipulations of a given rule, some (and only) central-parameter nodes may appear or disappear when moving from the premise to the conclusion or vice versa, or may change their labels (say, turn from a conjunctive gate into a disjunctive one). Similarly, some (and only) arcs pointing to or from central parameters may appear or disappear. No other nodes or edges are affected. The only exception is when deleting  arcs from a central parameter to some of its children leaves those children parentless. As we do not allow non-root ``orphan'' nodes in cirquents, such nodes (together with the arcs incident with them, of course) should then also be deleted, along with their possibly further orphaned children, grandchildren, etc. Such a chain of deletions may delete nodes that are not among the central parameters and perhaps not even within the peripheral parameters. Other than this, we repeat, any node of the cirquent that does not happen to be a central parameter, and any arc of the cirquent that does not happen to be incident with a central parameter, remains unaffected when moving from premise to conclusion or vice versa. 

In view of conditions 3 and 4 of Convention \ref{march15}, the role of peripheral parameters is to list all parents and children of central parameters that do not themselves happen to be central parameters. The additional purpose that they sometimes serve is  dividing those parents or children into groups for reference purposes, as will be seen later.  

In general, the intuitive role played by parameters is telling us ``where the rule is exactly applied'' in the cirquent. Without this piece of information, determining whether one cirquent indeed follows from another by a given rule can be harder than it has to be.

When drawing cirquents, we typically do not bother to assign (unique) names to their nodes, as this is also commonly done in the literature when dealing with graphs in general: more often than not, one does not differentiate between {\em isomorphic} graphs --- graphs that only differ in the names of their nodes --- as such graphs behave in the same ways in all relevant aspects, and assigning names to their nodes can usually be done in an arbitrary fashion if and when necessary. However, when dealing with rules of inference, or any graph-transformation procedures, having names for nodes becomes necessary in order to be able to properly define and apply (machine-implement) those rules or procedures. Indeed, if we continue seeing cirquents not as particular graphs but rather as isomorphism classes of graphs  (as was implicitly done in the preceding section, whether the reader noticed it or not), then even deciding whether one cirquent is ``the same as'' another cirquent would take quite some work, let alone deciding whether one cirquent follows from another one by a given rule. 

So, officially we require that, when applying rules, all nodes of the involved cirquents had names, and that each transition from a premise to a conclusion be justified by not merely indicating the name of the corresponding rule, but also indicating the precise values of each of the parameters of the rule. With this requirement, the question on whether any given step (transition from a premise to a conclusion) is legal in a cirquent-calculus proof or derivation essentially reduces to nothing but checking whether the indicated parameters of the premise and the conclusion satisfy all conditions of the indicated rule, which, in the case of the rules of \lll\ or any other rules discussed in this paper, can be seen to be a rather easy (certainly polynomial time doable) task. 
   
We will be schematically representing rules of inference in the form 
\begin{center} \begin{picture}(23,21)
\put(5,16){$X$}
\put(0,11){\line(1,0){18}}
\put(5,0){$Y$}
\end{picture}\end{center}   
where $X$ stands for the relevant portion of the premise and $Y$ stands for the relevant portion of the conclusion. Here ``relevant portion'' is the fragment of the cirquent that contains all central parameters, all peripheral parameters, all edges incident with the central parameters, and no other edges or nodes. 
In such a representation, the letters $a,c,b,\Gamma,\Delta,\Pi,\Sigma,\Omega,\Theta$ will be used as variables for the parameters of the rule.  
 As noted, the conclusion is obtained from the premise (or vice versa) through replacing the $X$ part by $Y$ (or vice versa), leaving the rest of the cirquent unchanged. While $X$ and $Y$ represent not the premise and the conclusion but only the to-be-modified parts of those, by abuse of terminology, we may still sometimes refer to them as the ``premise'' and the ``conclusion''.  

Below is a full list of the rules of inference of \lll\ represented schematically for the convenience of quick future references. Certain necessary explanations of their meanings and examples of applications  follow. 
 
\begin{center} \begin{picture}(330,177)
\put(100,157){RESTRUCTURING RULES:}
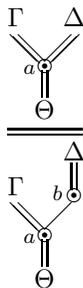
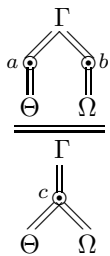
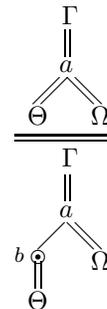
\put(-12,131){\bf Deepening}
\put(0,115){$\Gamma$}
\put(21,115){$\Delta$}
\put(14,102){\line(1,1){10}}
\put(16,101){\line(1,1){11}}
\put(14,102){\line(-1,1){10}}
\put(12,101){\line(-1,1){11}}
\put(14,100){\circle{5}}
\put(14,100){\circle*{2}}
\put(15,88){\line(0,1){10}}
\put(13,88){\line(0,1){10}}
\put(10,80){$\Theta$}
\put(17,49){\footnotesize $b$}
\put(6,33){\footnotesize $a$}
\put(6,97){\footnotesize $a$}

\put(0,76){\line(1,0){28}}
\put(0,74){\line(1,0){28}}
\put(25,51){\circle*{2}}
\put(25,51){\circle{5}}
\put(24,53){\line(0,1){10}}
\put(26,53){\line(0,1){10}}
\put(0,51){$\Gamma$}
\put(21,64){$\Delta$}
\put(14,38){\line(1,1){10}}
\put(14,38){\line(-1,1){10}}
\put(12,37){\line(-1,1){11}}
\put(14,36){\circle*{2}}
\put(14,36){\circle{5}}
\put(15,24){\line(0,1){10}}
\put(13,24){\line(0,1){10}}
\put(10,16){$\Theta$}
\put(-12,0){\bf Flattening}

\put(130,131){\bf Globalization}
\put(162,64){$\Gamma$}
\put(165,51){\line(0,1){10}}
\put(163,51){\line(0,1){10}}
\put(164,49){\circle{5}}
\put(164,49){\circle*{2}}
\put(152,37){\line(1,1){10}}
\put(155,37){\line(1,1){10}}
\put(149,28){$\Theta$}
\put(171,28){$\Omega$}
\put(173,37){\line(-1,1){10}}
\put(176,37){\line(-1,1){10}}

\put(147,76){\line(1,0){34}}
\put(147,74){\line(1,0){34}}
\put(175,100){\circle*{2}}
\put(175,100){\circle{5}}
\put(174,102){\line(-1,1){10}}
\put(177,102){\line(-1,1){10}}
\put(153,100){\circle*{2}}
\put(153,100){\circle{5}}
\put(151,102){\line(1,1){10}}
\put(154,102){\line(1,1){10}}
\put(162,114){$\Gamma$}
\put(152,88){\line(0,1){10}}
\put(154,88){\line(0,1){10}}
\put(174,88){\line(0,1){10}}
\put(176,88){\line(0,1){10}}
\put(149,80){$\Theta$}
\put(171,80){$\Omega$}

\put(156,49){\footnotesize $c$}
\put(179,98){\footnotesize $b$}
\put(144,98){\footnotesize $a$}

\put(133,0){\bf Localization}

\put(280,131){\bf Lengthening}
\put(310,100){$a$}
\put(300,88){\line(1,1){10}}
\put(303,88){\line(1,1){10}}
\put(323,88){\line(-1,1){10}}
\put(326,88){\line(-1,1){10}}
\put(298,80){$\Theta$}
\put(302,33){\line(1,1){10}}
\put(293,76){\line(1,0){37}}
\put(293,74){\line(1,0){37}}
\put(302,30){\circle*{2}}
\put(302,30){\circle{5}}
\put(310,45){$a$}
\put(301,18){\line(0,1){10}}
\put(303,18){\line(0,1){10}}
\put(298,10){$\Theta$}
\put(283,0){\bf Shortening}
\put(293,28){\footnotesize $b$}
\put(322,80){$\Omega$}
\put(322,25){$\Omega$}
\put(311,64){$\Gamma$}
\put(311,118){$\Gamma$}
\put(312,51){\line(0,1){10}}
\put(314,51){\line(0,1){10}}
\put(312,106){\line(0,1){10}}
\put(314,106){\line(0,1){10}}

\put(323,33){\line(-1,1){10}}
\put(326,33){\line(-1,1){10}}

\end{picture}\end{center}

\begin{center} \begin{picture}(345,240)
\put(137,201){MAIN RULES:}
\put(-4,175){\bf Coupling}
\put(20,114){\circle{5}}
\put(21,102){\line(0,1){10}}
\put(19,102){\line(0,1){10}}
\put(16,94){$\Theta$}
\put(4,90){\line(1,0){35}}
\put(24,72){$\gneg P$}
\put(6,72){$P$}
\put(20,59){\line(1,1){10}}
\put(20,59){\line(-1,1){10}}
\put(20,57){\circle*{5}}
\put(21,45){\line(0,1){10}}
\put(19,45){\line(0,1){10}}
\put(16,37){$\Theta$}
\put(12,55){\footnotesize $a$}
\put(12,112){\footnotesize $a$}
\put(7,82){\footnotesize $b$}
\put(31,82){\footnotesize $c$}

\put(137,175){\bf Weakening}
\put(161,129){$\Gamma$}
\put(165,116){\line(0,1){10}}
\put(163,116){\line(0,1){10}}
\put(164,114){\circle*{5}}
\put(165,102){\line(0,1){10}}
\put(163,102){\line(0,1){10}}
\put(160,94){$\Theta$}

\put(150,90){\line(1,0){30}}

\put(171,80){$\Delta$}
\put(151,80){$\Gamma$}
\put(166,67){\line(1,1){10}}
\put(163,67){\line(1,1){10}}
\put(162,67){\line(-1,1){10}}
\put(165,67){\line(-1,1){10}}
\put(164,65){\circle*{5}}
\put(165,53){\line(0,1){10}}
\put(163,53){\line(0,1){10}}
\put(160,45){$\Theta$}
\put(156,62){\footnotesize $a$}
\put(156,112){\footnotesize $a$}

\put(302,175){\bf Pulldown}

\put(324,161){$\Gamma$}
\put(339,161){$\Pi$}
\put(325,146){\line(0,1){12}}
\put(327,146){\line(0,1){12}}
\put(325,146){\line(3,2){16}}
\put(328,146){\line(3,2){16}}

\put(326,144){\circle*{5}}
\put(326,131){\line(0,1){10}}
\put(324,131){\line(-3,2){16}}
\put(327,131){\line(-3,2){16}}
\put(305,127){$\Delta$}
\put(326,129){\circle{5}}
\put(326,117){\line(0,1){10}}
\put(324,115){\line(-3,2){16}}
\put(327,115){\line(-3,2){16}}
\put(305,144){$\Sigma$}
\put(326,115){\circle*{5}}
\put(325,103){\line(0,1){10}}
\put(327,103){\line(0,1){10}}
\put(322,94){$\Theta$}

\put(300,90){\line(1,0){50}}

\put(324,78){$\Gamma$}
\put(327,63){\line(0,1){10}}
\put(325,63){\line(0,1){10}}
\put(326,61){\circle*{5}}
\put(305,61){$\Sigma$}
\put(326,48){\line(0,1){10}}
\put(327,48){\line(-3,2){16}}
\put(324,48){\line(-3,2){16}}
\put(305,44){$\Delta$}
\put(339,44){$\Pi$}
\put(326,46){\circle{5}}
\put(326,34){\line(0,1){10}}
\put(324,32){\line(-3,2){16}}
\put(327,32){\line(-3,2){16}}
\put(324,32){\line(3,2){16}}
\put(327,32){\line(3,2){16}}
\put(326,32){\circle*{5}}
\put(325,20){\line(0,1){10}}
\put(327,20){\line(0,1){10}}
\put(322,11){$\Theta$}

\put(318,142){\footnotesize $a$}
\put(318,59){\footnotesize $a$}
\put(318,42){\footnotesize $b$}
\put(318,125){\footnotesize $b$}
\put(318,27){\footnotesize $c$}
\put(318,110){\footnotesize $c$}

\end{picture}\end{center}

The double names and double horizontal lines in the restructuring rules indicate that these rules work in both top-down and bottom-up directions. The name on the top is for the direction where the top part is the premise, and the name at the bottom is for the direction where the bottom part is the premise.  Furthermore, 
\raisebox{.4ex}{\tiny $\odot$} is a variable over $\{\bullet,\circ\}$. This means that each restructuring rule comes in two versions: one for $\bullet$ and one for $\circ$. So, altogether there are 12 restructuring rules.
 
The following convention provides additional explanations and conditions, some essentially just reiterating (for safety) certain earlier-made stipulations:

\begin{convention}\label{conv} \ 
\begin{enumerate}
\item Lowercase Latin letters  stand for the central parameters of the rule. 
\item Uppercase Greek letters  stand for the peripheral parameters of the rule. We do not require the peripheral parameters to be non-empty, or --- when there are several peripheral parameters in the rule --- disjoint or even non-identical. We use a double rather than a single arc to indicate the presence of an arc between a given central parameter and each node of a given peripheral parameter. 
\item $P$ stands for any atom.
\item We assume that central parameters  have no children and parents other than those explicitly indicated (through single or double arcs) in the schematic representation of the rule. Hence, for example, as we see $b$ and $c$ only in the conclusion of the rule of coupling, these two nodes are simply absent in the premise. 
\item On the other hand, the nodes of peripheral parameters may have additional 
parents and/or children, not shown in the schematic representation of the rule. According to the earlier explanations, it is understood that the connections between such nodes with their invisible parents and/or children, just as all other invisible (``contextual'') connections and nodes, will be preserved when moving from premise to conclusion or vice versa. So, for example, while we do not see $\Delta$ in the premise of weakening, this does not necessarily mean that the nodes of $\Delta$ disappear when moving from conclusion to premise: those nodes may remain present in the premise if (and only if) they had some additional, invisible parents.       
\end{enumerate}
\end{convention}

Below come explanations of all rules. Such explanations can be provided either by saying how to obtain a conclusion from the premise, or saying how to obtain a premise from the conclusion. We choose one or the other way depending on which one appears to be more intuitive and convenient. For the same reason, for each of the three pairs of restructuring rules, we explain only one, with the other rule of the pair being symmetric, obtained by interchanging premise with conclusion.

\subsection{Deepening} As can be seen from the schematic representation, this rule has two central parameters $a,b$ and three peripheral parameters $\Gamma,\Delta,\Theta$. Its  meaning is that if a cirquent has a gate $b$ with exactly one parent $a$ such that $b$ and $a$ are of the same type (both conjunctive, or both disjunctive), then a premise can be obtained by deleting $b$ and connecting its children $\Delta$ directly to $a$.

\begin{center} \begin{picture}(27,129)
\put(0,109){$\Gamma$}
\put(21,109){$\Delta$}
\put(14,96){\line(1,1){10}}
\put(16,95){\line(1,1){11}}
\put(14,96){\line(-1,1){10}}
\put(12,95){\line(-1,1){11}}
\put(14,94){\circle{5}}
\put(14,94){\circle*{2}}
\put(6,90){\footnotesize $a$}
\put(15,82){\line(0,1){10}}
\put(13,82){\line(0,1){10}}
\put(10,74){$\Theta$}

\put(0,69){\line(1,0){28}}

\put(25,45){\circle*{2}}
\put(25,45){\circle{5}}
\put(17,43){\footnotesize $b$}
\put(24,47){\line(0,1){10}}
\put(26,47){\line(0,1){10}}
\put(0,45){$\Gamma$}
\put(21,58){$\Delta$}
\put(14,32){\line(1,1){10}}
\put(14,32){\line(-1,1){10}}
\put(12,31){\line(-1,1){11}}
\put(14,30){\circle*{2}}
\put(6,26){\footnotesize $a$}
\put(14,30){\circle{5}}
\put(15,18){\line(0,1){10}}
\put(13,18){\line(0,1){10}}
\put(10,10){$\Theta$}
\end{picture}\end{center}

Below are several examples of applications of this rule. 

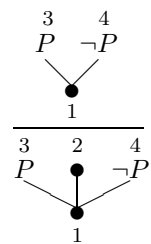
\begin{center} \begin{picture}(450,154)
\put(0,135){\bf Example 1}
\put(10,106){$P$}
\put(27,106){$\gneg P$}
\put(24,93){\line(1,1){10}}
\put(24,93){\line(-1,1){10}}
\put(24,91){\circle*{5}}
\put(22,81){\footnotesize $1$}
\put(12,116){\footnotesize $3$}
\put(33,116){\footnotesize $4$}

\put(10,77){\line(1,0){34}}

\put(35,45){\circle*{5}}
\put(39,43){\footnotesize $2$}
\put(35,47){\line(0,1){9}}
\put(10,45){$P$}
\put(27,58){$\gneg P$}
\put(24,32){\line(1,1){10}}
\put(24,32){\line(-1,1){10}}
\put(22,20){\footnotesize $1$}
\put(24,30){\circle*{5}}
\put(12,55){\footnotesize $3$}
\put(33,68){\footnotesize $4$}

\put(90,135){\bf Example 2}
\put(100,106){$P$}
\put(117,106){$\gneg P$}
\put(114,92){\line(1,1){10}}
\put(114,93){\line(-1,1){10}}
\put(114,91){\circle*{5}}
\put(112,81){\footnotesize $1$}

\put(102,68){\footnotesize $3$}
\put(123,68){\footnotesize $4$}

\put(102,116){\footnotesize $3$}
\put(123,116){\footnotesize $4$}

\put(99,77){\line(1,0){36}}

\put(125,45){\circle*{5}}
\put(129,43){\footnotesize $2$}
\put(125,47){\line(0,1){9}}
\put(125,47){\line(-5,2){23}}
\put(100,58){$P$}
\put(117,58){$\gneg P$}
\put(114,32){\line(1,1){10}}
\put(114,32){\line(-1,2){12}}
\put(111,20){\footnotesize $1$}
\put(114,30){\circle*{5}}

\put(190,135){\bf Example 3}
\put(200,106){$P$}
\put(217,106){$\gneg P$}
\put(214,93){\line(1,1){10}}
\put(214,93){\line(-1,1){10}}
\put(214,91){\circle*{5}}
\put(212,81){\footnotesize $1$}

\put(200,77){\line(1,0){34}}

\put(225,45){\circle*{5}}
\put(229,43){\footnotesize $2$}
\put(225,47){\line(0,1){9}}
\put(225,47){\line(-5,2){21}}
\put(200,58){$P$}
\put(217,58){$\gneg P$}
\put(204,32){\line(5,6){20}}
\put(204,32){\line(0,1){24}}
\put(204,32){\line(2,1){22}}
\put(202,20){\footnotesize $1$}
\put(204,30){\circle*{5}}
\put(201,116){\footnotesize $3$}
\put(223,116){\footnotesize $4$}
\put(201,68){\footnotesize $3$}
\put(223,68){\footnotesize $4$}

\put(290,135){\bf Example 4}
\put(300,106){$P$}
\put(317,106){$\gneg P$}
\put(314,93){\line(1,1){10}}
\put(314,93){\line(-1,1){10}}
\put(314,91){\circle*{5}}
\put(312,81){\footnotesize $1$}

\put(298,77){\line(1,0){34}}

\put(314,44){\circle*{5}}
\put(312,49){\footnotesize $2$}
\put(314,46){\line(1,1){10}}
\put(314,46){\line(-1,1){10}}
\put(300,58){$P$}
\put(317,58){$\gneg P$}
\put(314,32){\line(0,1){12}}
\put(312,20){\footnotesize $1$}
\put(314,30){\circle*{5}}
\put(301,116){\footnotesize $3$}
\put(323,116){\footnotesize $4$}
\put(301,68){\footnotesize $3$}
\put(323,68){\footnotesize $4$}

\put(398,135){\bf Example 5}
\put(408,106){$P$}
\put(425,106){$\gneg P$}
\put(422,93){\line(1,1){10}}
\put(422,93){\line(-1,1){10}}
\put(422,91){\circle*{5}}
\put(420,81){\footnotesize $1$}

\put(400,77){\line(1,0){49}}

\put(424,45){\circle*{5}}
\put(422,34){\footnotesize $1$}
\put(444,68){\footnotesize $4$}
\put(424,47){\line(2,1){20}}
\put(424,47){\line(-2,1){20}}
\put(424,47){\line(0,1){13}}
\put(400,58){$P$}
\put(437,58){$\gneg P$}
\put(424,61){\circle*{5}}
\put(411,116){\footnotesize $3$}
\put(432,116){\footnotesize $4$}
\put(402,68){\footnotesize $3$}
\put(422,68){\footnotesize $2$}

\end{picture}\end{center}

As we remember, a justification of an application of a rule should include a specification of the values of its parameters. Here are such specifications: 
\begin{itemize}
\item In Example 1: $a=1$, $b=2$, $\Gamma=\{3\}$, $\Delta=\{4\}$, $\Theta=\{\}$. 
\item In Example 2: $a=1$, $b=2$, $\Gamma=\{3\}$, $\Delta=\{3,4\}$, $\Theta=\{\}$.  
\item In Example 3: $a=1$, $b=2$, $\Gamma=\{3,4\}$, $\Delta=\{3,4\}$, $\Theta=\{\}$.  
\item In Example 4: $a=1$, $b=2$, $\Gamma=\{\}$, $\Delta=\{3,4\}$, $\Theta=\{\}$.
\item In Example 5:  $a=1$, $b=2$, $\Gamma=\{3,4\}$, $\Delta=\emptyset$,  $\Theta=\{\}$.
\end{itemize}

The following is an example of deepening applied to a bigger cirquent. Here we have $a=1$, $b=2$,  $\Gamma=\{5\}$, $\Delta=\{6,7\}$ and $\Theta=\{3,4\}$:

\begin{center} \begin{picture}(66,184)

\put(6,165){\bf Example 6}
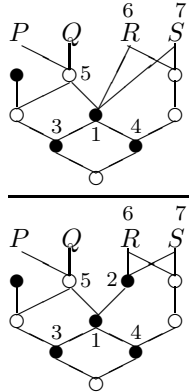
\put(23,123){\circle{5}}
\put(23,125){\line(0,1){10}}
\put(23,125){\line(-2,1){18}}
\put(0,136){$P$}
\put(20,136){$Q$}
\put(42,136){$R$}
\put(60,136){$S$}
\put(34,110){\line(6,5){29}}
\put(34,110){\line(1,2){12}}
\put(34,110){\line(-1,1){10}}
\put(31,98){\footnotesize $1$}
\put(33,108){\circle*{5}}

\put(27,121){\footnotesize $5$}
\put(63,123){\circle{5}}
\put(63,125){\line(0,1){10}}
\put(63,125){\line(-2,1){18}}

\put(63,108){\circle{5}}
\put(63,111){\line(0,1){10}}

\put(3,108){\circle{5}}
\put(3,111){\line(0,1){10}}
\put(3,110){\line(2,1){20}}
\put(3,123){\circle*{5}}

\put(18,96){\circle*{5}}
\put(18,98){\line(2,1){15}}
\put(18,98){\line(-2,1){15}}
\put(16,101){\footnotesize $3$}

\put(48,96){\circle*{5}}
\put(48,98){\line(2,1){15}}
\put(48,98){\line(-2,1){15}}
\put(46,101){\footnotesize $4$}

\put(33,84){\circle{5}}
\put(33,86){\line(2,1){15}}
\put(33,86){\line(-2,1){15}}

\put(0,77){\line(1,0){68}}

\put(45,45){\circle*{5}}
\put(23,45){\circle{5}}
\put(37,43){\footnotesize $2$}
\put(45,47){\line(0,1){10}}
\put(23,47){\line(0,1){10}}
\put(23,47){\line(-2,1){18}}
\put(45,47){\line(2,1){18}}
\put(0,58){$P$}
\put(20,58){$Q$}
\put(42,58){$R$}
\put(60,58){$S$}
\put(34,32){\line(1,1){10}}
\put(34,32){\line(-1,1){10}}
\put(31,20){\footnotesize $1$}
\put(33,30){\circle*{5}}

\put(27,43){\footnotesize $5$}
\put(63,45){\circle{5}}
\put(63,47){\line(0,1){10}}
\put(63,47){\line(-2,1){18}}

\put(63,30){\circle{5}}
\put(63,33){\line(0,1){10}}

\put(3,30){\circle{5}}
\put(3,33){\line(0,1){10}}
\put(3,32){\line(2,1){20}}
\put(3,45){\circle*{5}}

\put(18,18){\circle*{5}}
\put(18,20){\line(2,1){15}}
\put(18,20){\line(-2,1){15}}
\put(16,23){\footnotesize $3$}

\put(48,18){\circle*{5}}
\put(48,20){\line(2,1){15}}
\put(48,20){\line(-2,1){15}}
\put(46,23){\footnotesize $4$}

\put(33,6){\circle{5}}
\put(33,8){\line(2,1){15}}
\put(33,8){\line(-2,1){15}}

\put(43,145){\footnotesize $6$}
\put(63,145){\footnotesize $7$}
\put(43,68){\footnotesize $6$}
\put(63,68){\footnotesize $7$}

\end{picture}\end{center}

Most readers, no doubt, would (still) feel more comfortable with formulas than with cirquents. Therefore it would not hurt to also see a couple of examples where both the premise and the conclusion are (tree-like and hence can be written as) formulas. We are not providing the values of the five parameters for these instances of the rule, which are easy to guess anyway. Furthermore, note that Example 8 is simply the same as example 5.

\begin{center} \begin{picture}(306,55)

\put(12,44){\bf Example 7}
\put(5,25){$P\mlc\bigl(Q\mld R\mld S\bigr)$}
\put(0,19){\line(1,0){78}}
\put(0,6){$P\mlc\bigl(Q\mld (R\mld S)\bigr)$}

\put(200,44){\bf Example 8}
\put(160,25){$P\mld\gneg P$}
\put(150,19){\line(1,0){51}}
\put(205,19){, \ \ \ i.e., }
\put(150,6){$P\mld\bot\mld\gneg P$}

\put(256,25){$\mld\{P,\gneg P\}$}
\put(246,19){\line(1,0){62}}
\put(246,6){$\mld\{P,\mld\{\},\gneg P\}$}

\end{picture}\end{center}

\subsection{Localization} According to this rule, if a cirquent has two conjunctive or two disjunctive gates $a,b$ with exactly the same children $\Gamma$ (but not necessarily the same parents), then a premise can be obtained by merging $a$ and $b$ and calling the resulting node $c$. Here ``merging'' means that $c$ has the same type and same children as $a$ and $b$ have, and the set of the parents of $c$ is the union of those of $a$ and $b$.

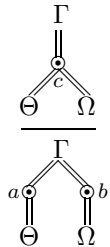
\begin{center} \begin{picture}(27,129)

\put(12,109){$\Gamma$}
\put(15,96){\line(0,1){10}}
\put(13,96){\line(0,1){10}}
\put(14,94){\circle{5}}
\put(14,94){\circle*{2}}
\put(12,85){\footnotesize $c$}
\put(3,82){\line(1,1){10}}
\put(5,82){\line(1,1){10}}
\put(-1,74){$\Theta$}
\put(21,74){$\Omega$}
\put(24,82){\line(-1,1){10}}
\put(26,82){\line(-1,1){10}}

\put(0,69){\line(1,0){28}}
\put(25,45){\circle*{2}}
\put(25,45){\circle{5}}
\put(29,43){\footnotesize $b$}
\put(24,47){\line(-1,1){10}}
\put(26,47){\line(-1,1){10}}
\put(3,45){\circle*{2}}
\put(3,45){\circle{5}}
\put(-5,43){\footnotesize $a$}
\put(2,47){\line(1,1){10}}
\put(4,47){\line(1,1){10}}
\put(12,58){$\Gamma$}
\put(2,33){\line(0,1){10}}
\put(4,33){\line(0,1){10}}
\put(24,33){\line(0,1){10}}
\put(26,33){\line(0,1){10}}
\put(-1,25){$\Theta$}
\put(21,25){$\Omega$}

\end{picture}\end{center}

Here are four examples of applications of this rule.

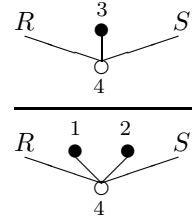
\begin{center} \begin{picture}(366,169)
\put(9,149){\bf Example 1}

\put(3,106){\circle{5}}
\put(33,106){\circle*{5}}
\put(31,111){\footnotesize $3$}
\put(63,106){\circle{5}}
\put(23,92){\circle{5}}
\put(43,92){\circle{5}}
\put(43,94){\line(2,1){19}}
\put(23,94){\line(1,1){10}}
\put(43,94){\line(-1,1){10}}
\put(3,108){\line(0,1){10}}
\put(63,108){\line(0,1){10}}
\put(0,120){$Q$}
\put(60,120){$Q$}
\put(20,120){$P$}
\put(35,120){$\gneg P$}
\put(33,80){\line(1,1){10}}
\put(33,80){\line(-1,1){10}}
\put(33,78){\circle*{5}}
\put(3,108){\line(2,1){19}}
\put(33,108){\line(1,1){10}}
\put(33,108){\line(-1,1){10}}
\put(63,108){\line(-2,1){19}}
\put(23,94){\line(-2,1){19}}
\put(15,86){\footnotesize $4$}
\put(46,86){\footnotesize $5$}

\put(21,129){\footnotesize $6$}
\put(42,129){\footnotesize $7$}
\put(21,62){\footnotesize $6$}
\put(42,62){\footnotesize $7$}

\put(0,71){\line(1,0){67}}

\put(3,38){\circle{5}}
\put(23,38){\circle*{5}}
\put(43,38){\circle*{5}}
\put(63,38){\circle{5}}
\put(23,24){\circle{5}}
\put(43,24){\circle{5}}
\put(43,26){\line(2,1){19}}
\put(23,26){\line(0,1){10}}
\put(43,26){\line(0,1){10}}
\put(15,36){\footnotesize $1$}
\put(43,40){\line(0,1){10}}
\put(23,40){\line(0,1){10}}
\put(3,40){\line(0,1){10}}
\put(63,40){\line(0,1){10}}
\put(0,52){$Q$}
\put(60,52){$Q$}
\put(20,52){$P$}
\put(35,52){$\gneg P$}
\put(33,12){\line(1,1){10}}
\put(33,12){\line(-1,1){10}}
\put(47,36){\footnotesize $2$}
\put(33,10){\circle*{5}}
\put(3,40){\line(2,1){19}}
\put(23,40){\line(2,1){19}}
\put(43,40){\line(-2,1){19}}
\put(63,40){\line(-2,1){19}}
\put(23,26){\line(-2,1){19}}
\put(15,18){\footnotesize $4$}
\put(46,18){\footnotesize $5$}

\put(109,149){\bf Example 2}

\put(103,106){\circle{5}}
\put(133,106){\circle*{5}}
\put(131,111){\footnotesize $3$}
\put(163,106){\circle{5}}
\put(123,92){\circle{5}}
\put(143,92){\circle{5}}
\put(143,94){\line(2,1){19}}
\put(123,94){\line(1,1){10}}
\put(143,94){\line(-1,1){10}}
\put(103,108){\line(0,1){10}}
\put(163,108){\line(0,1){10}}
\put(100,120){$Q$}
\put(160,120){$Q$}
\put(120,120){$P$}
\put(135,120){$\gneg P$}
\put(133,80){\line(1,1){10}}
\put(133,80){\line(-1,1){10}}
\put(133,78){\circle*{5}}
\put(103,108){\line(2,1){19}}
\put(133,108){\line(1,1){10}}
\put(133,108){\line(-1,1){10}}
\put(163,108){\line(-2,1){19}}
\put(123,94){\line(-2,1){19}}
\put(115,86){\footnotesize $4$}
\put(146,86){\footnotesize $5$}

\put(100,71){\line(1,0){67}}

\put(103,38){\circle{5}}
\put(123,38){\circle*{5}}
\put(143,38){\circle*{5}}
\put(163,38){\circle{5}}
\put(123,24){\circle{5}}
\put(143,24){\circle{5}}
\put(143,26){\line(2,1){19}}
\put(123,26){\line(0,1){10}}
\put(143,26){\line(0,1){10}}
\put(115,36){\footnotesize $1$}
\put(143,40){\line(0,1){10}}
\put(123,40){\line(0,1){10}}
\put(103,40){\line(0,1){10}}
\put(163,40){\line(0,1){10}}
\put(100,52){$Q$}
\put(160,52){$Q$}
\put(120,52){$P$}
\put(135,52){$\gneg P$}
\put(133,12){\line(1,1){10}}
\put(133,12){\line(-1,1){10}}
\put(147,36){\footnotesize $2$}
\put(133,10){\circle*{5}}
\put(103,40){\line(2,1){19}}
\put(123,40){\line(2,1){19}}
\put(143,40){\line(-2,1){19}}
\put(163,40){\line(-2,1){19}}
\put(123,26){\line(-2,1){19}}
\put(123,26){\line(2,1){19}}
\put(115,18){\footnotesize $4$}
\put(146,18){\footnotesize $5$}

\put(121,129){\footnotesize $6$}
\put(142,129){\footnotesize $7$}
\put(121,62){\footnotesize $6$}
\put(142,62){\footnotesize $7$}

\put(209,149){\bf Example 3}

\put(203,101){\circle{5}}
\put(233,101){\circle*{5}}
\put(231,106){\footnotesize $3$}
\put(263,101){\circle{5}}

\put(203,103){\line(0,1){10}}
\put(263,103){\line(0,1){10}}
\put(200,115){$Q$}
\put(260,115){$Q$}
\put(220,115){$P$}
\put(235,115){$\gneg P$}

\put(203,103){\line(2,1){19}}
\put(233,103){\line(1,1){10}}
\put(233,103){\line(-1,1){10}}
\put(263,103){\line(-2,1){19}}

\put(233,87){\circle{5}}
\put(233,89){\line(3,1){29}}
\put(233,89){\line(0,1){10}}
\put(233,89){\line(-3,1){29}}
\put(230,77){\footnotesize $4$}

\put(200,71){\line(1,0){67}}

\put(243,40){\line(0,1){10}}
\put(223,40){\line(0,1){10}}
\put(203,40){\line(0,1){10}}
\put(263,40){\line(0,1){10}}
\put(200,52){$Q$}
\put(260,52){$Q$}
\put(220,52){$P$}
\put(235,52){$\gneg P$}
\put(203,40){\line(2,1){19}}
\put(223,40){\line(2,1){19}}
\put(243,40){\line(-2,1){19}}
\put(263,40){\line(-2,1){19}}
\put(203,38){\circle{5}}
\put(223,38){\circle*{5}}
\put(243,38){\circle*{5}}
\put(263,38){\circle{5}}
\put(215,36){\footnotesize $1$}
\put(247,36){\footnotesize $2$}
\put(233,24){\circle{5}}
\put(233,26){\line(3,1){29}}
\put(233,26){\line(1,1){10}}
\put(233,26){\line(-1,1){10}}
\put(233,26){\line(-3,1){29}}
\put(230,14){\footnotesize $4$}

\put(221,125){\footnotesize $6$}
\put(242,125){\footnotesize $7$}
\put(221,62){\footnotesize $6$}
\put(242,62){\footnotesize $7$}

\put(309,149){\bf Example 4}

\put(333,101){\circle*{5}}
\put(331,106){\footnotesize $3$}

\put(300,101){$R$}
\put(360,101){$S$}

\put(333,87){\circle{5}}
\put(333,89){\line(3,1){29}}
\put(333,89){\line(0,1){10}}
\put(333,89){\line(-3,1){29}}
\put(330,77){\footnotesize $4$}

\put(300,71){\line(1,0){67}}

\put(300,55){$R$}
\put(360,55){$S$}

\put(323,55){\circle*{5}}
\put(343,55){\circle*{5}}

\put(321,61){\footnotesize $1$}
\put(340,61){\footnotesize $2$}
\put(333,41){\circle{5}}
\put(333,43){\line(3,1){29}}
\put(333,43){\line(1,1){10}}
\put(333,43){\line(-1,1){10}}
\put(333,43){\line(-3,1){29}}
\put(330,31){\footnotesize $4$}
\end{picture}\end{center}

\begin{itemize}
\item
In Example 1: $a=1$, $b=2$, $c=3$, $\Gamma=\{6,7\}$, $\Theta=\{4\}$,  $\Omega=\{5\}$.
\item
In Example 2: $a=1$, $b=2$, $c=3$,  $\Gamma=\{6,7\}$, $\Theta=\{4\}$, $\Omega=\{4,5\}$. 
\item
In Example 3: $a=1$, $b=2$, $c=3$,  $\Gamma=\{6,7\}$, $\Theta=\{4\}$, $\Omega=\{4\}$.  
\item
In Example 4: $a=1$, $b=2$, $c=3$, $\Gamma=\{\}$, $\Theta=\{4\}$,  $\Omega=\{4\}$.
\end{itemize}

The following is the same as Example 1, only with the premise and the conclusion written as hyperformulas (which, by good luck, is possible here, even though the same could not be done using just formulas):

\begin{center} \begin{picture}(210,40)
\put(0,29){$\bigl((Q\mlc \overline{P})\mlc\overline{(P\mld\gneg P)}\bigr)\mld \bigl(\overline{(P\mld\gneg P)}\mlc (\overline{\gneg P}\mlc Q)\bigr)$}
\put(0,21){\line(1,0){210}}
\put(0,6){$\bigl((Q\mlc \overline{P})\mlc(\overline{P}\mld\overline{\gneg P})\bigr)\mld \bigl((\overline{P}\mld\overline{\gneg P})\mlc (\overline{\gneg P}\mlc Q)\bigr)$}

\end{picture}\end{center}

\subsection{Lengthening} According to this rule, if a cirquent has a gate $b$ with exactly one child $a$, then a premise can be obtained by deleting $b$ and connecting $a$ directly to the parents $\Theta$ of $b$.

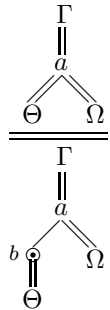
\begin{center} \begin{picture}(26,138)
\put(10,100){$a$}
\put(0,88){\line(1,1){10}}
\put(3,88){\line(1,1){10}}
\put(23,88){\line(-1,1){10}}
\put(26,88){\line(-1,1){10}}
\put(-2,80){$\Theta$}
\put(2,33){\line(1,1){10}}
\put(-7,76){\line(1,0){37}}
\put(-7,74){\line(1,0){37}}
\put(2,30){\circle*{2}}
\put(2,30){\circle{5}}
\put(10,45){$a$}
\put(1,18){\line(0,1){10}}
\put(3,18){\line(0,1){10}}
\put(-2,10){$\Theta$}
\put(-7,28){\footnotesize $b$}
\put(22,80){$\Omega$}
\put(22,25){$\Omega$}
\put(11,64){$\Gamma$}
\put(11,118){$\Gamma$}
\put(12,51){\line(0,1){10}}
\put(14,51){\line(0,1){10}}
\put(12,106){\line(0,1){10}}
\put(14,106){\line(0,1){10}}

\put(23,33){\line(-1,1){10}}
\put(26,33){\line(-1,1){10}}

\end{picture}\end{center}

Here are some illustrations:

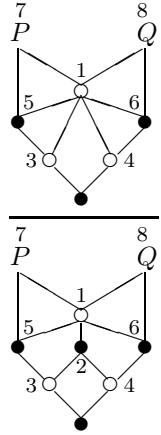
\begin{center} \begin{picture}(250,198)
\put(-12,178){\bf Example 1}
\put(10,93){$P$}
\put(14,59){\line(0,1){9}}
\put(0,88){\line(1,0){28}}
\put(14,56){\circle{5}}
\put(12,46){\footnotesize $2$}
\put(12,103){\footnotesize $1$}
\put(12,79){\footnotesize $1$}
\put(10,70){$P$}

\put(88,178){\bf Example 2}
\put(114,95){\circle{5}}
\put(114,63){\line(0,1){8}}
\put(100,88){\line(1,0){28}}
\put(114,61){\circle*{5}}
\put(112,51){\footnotesize $2$}
\put(114,73){\circle{5}}
\put(112,79){\footnotesize $1$}
\put(112,102){\footnotesize $1$}

\put(188,178){\bf Example 3}
\put(214,136){\circle{5}}
\put(214,138){\line(5,3){23}}
\put(214,138){\line(-5,3){23}}
\put(212,141){\footnotesize $1$}

\put(214,134){\line(-1,-2){11}}
\put(214,134){\line(1,-2){11}}

\put(202,110){\circle{5}}
\put(225,110){\circle{5}}

\put(214,97){\line(-1,1){10}}
\put(214,97){\line(1,1){10}}
\put(214,95){\circle*{5}}

\put(190,124){\circle*{5}}
\put(190,122){\line(1,-1){10}}
\put(190,126){\line(0,1){26}}
\put(190,126){\line(3,1){23}}
\put(238,126){\line(-3,1){23}}

\put(238,124){\circle*{5}}
\put(238,122){\line(-1,-1){10}}
\put(238,126){\line(0,1){26}}

\put(187,155){$P$}
\put(235,155){$Q$}

\put(193,22){\footnotesize $3$}
\put(230,22){\footnotesize $4$}

\put(193,107){\footnotesize $3$}
\put(230,107){\footnotesize $4$}

\put(192,44){\footnotesize $5$}
\put(232,44){\footnotesize $6$}
\put(189,79){\footnotesize $7$}
\put(235,79){\footnotesize $8$}

\put(192,129){\footnotesize $5$}
\put(232,129){\footnotesize $6$}
\put(189,164){\footnotesize $7$}
\put(235,164){\footnotesize $8$}

\put(214,40){\line(0,1){8}}
\put(187,88){\line(1,0){56}}
\put(214,39){\circle*{5}}
\put(212,29){\footnotesize $2$}
\put(214,51){\circle{5}}
\put(214,53){\line(5,3){23}}
\put(214,53){\line(-5,3){23}}
\put(212,56){\footnotesize $1$}

\put(214,37){\line(-1,-1){10}}
\put(214,37){\line(1,-1){10}}

\put(202,25){\circle{5}}
\put(225,25){\circle{5}}

\put(214,12){\line(-1,1){10}}
\put(214,12){\line(1,1){10}}
\put(214,10){\circle*{5}}

\put(190,39){\circle*{5}}
\put(190,37){\line(1,-1){10}}
\put(190,41){\line(0,1){26}}
\put(190,41){\line(3,1){23}}
\put(238,41){\line(-3,1){23}}

\put(238,39){\circle*{5}}
\put(238,37){\line(-1,-1){10}}
\put(238,41){\line(0,1){26}}

\put(187,70){$P$}
\put(235,70){$Q$}
\end{picture}\end{center}

\begin{itemize}
\item In Examples 1 and 2: $a=1$, $b=2$, $\Gamma=\{\}$, $\Theta=\{\}$ and $\Omega=\{\}$. 
\item
In Example 3: $a=1$, $b=2$, $\Gamma=\{7,8\}$, $\Theta=\{3,4\}$ and $\Omega=\{5,6\}$.
\end{itemize}

In terms of formulas, lengthening simply replaces a subformula $F$ by $\mld\{F\}$ or $\mlc\{F\}$. Shortening, of course, seems to be doing a more useful job than lengthening when it comes to formulas: it removes a ``dummy'' disjunction or conjunction that is applied to a single conjunct or disjunct. 

\subsection{Coupling} According to this rule, if a cirquent has a childless conjunctive gate $a$, then a conclusion can be obtained through making $a$ a disjunctive gate and adding to it two children $b$ and $c$ which are ports with opposite labels. 

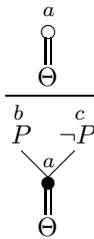
\begin{center} \begin{picture}(31,107)
\put(14,87){\circle{5}}
\put(15,75){\line(0,1){10}}
\put(13,75){\line(0,1){10}}
\put(10,67){$\Theta$}
\put(-2,63){\line(1,0){35}}
\put(18,45){$\gneg P$}
\put(0,45){$P$}
\put(14,32){\line(1,1){10}}
\put(14,32){\line(-1,1){10}}
\put(14,30){\circle*{5}}
\put(15,18){\line(0,1){10}}
\put(13,18){\line(0,1){10}}
\put(10,10){$\Theta$}
\put(12,36){\footnotesize $a$}
\put(12,93){\footnotesize $a$}
\put(1,54){\footnotesize $b$}
\put(24,54){\footnotesize $c$}
\end{picture}\end{center}

An important condition here is that the above $b$ and $c$ should be {\em new} nodes not present in the premise. That is, one cannot utilize some already existing node to make a child of $a$. Example 3 below violates this condition, and hence is an example of a wrong ``application'' of coupling.

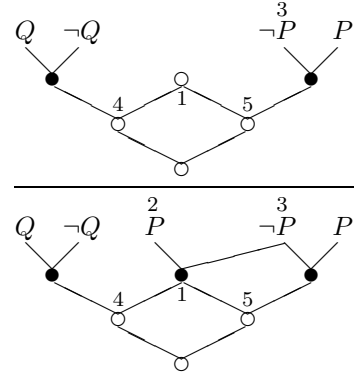
\begin{center} \begin{picture}(426,187)
\put(0,167){\bf Example 1}
\put(26,92){\circle{5}}
\put(24,98){\footnotesize $1$}

\put(7,85){\line(1,0){39}}

\put(26,54){\line(1,1){10}}
\put(26,54){\line(-1,1){10}}
\put(26,52){\circle*{5}}
\put(24,42){\footnotesize $1$}
\put(13,77){\footnotesize $2$}
\put(36,77){\footnotesize $3$}

\put(12,67){$P$}
\put(28,67){$\gneg P$}

\put(138,167){\bf Example 2}
\put(118,141){$\gneg Q$}
\put(100,141){$Q$}
\put(114,128){\line(1,1){10}}
\put(114,128){\line(-1,1){10}}
\put(114,126){\circle*{5}}
\put(139,111){\line(-2,1){24}}
\put(139,111){\line(2,1){24}}
\put(139,109){\circle{5}}

\put(163,126){\circle{5}}

\put(220,141){$P$}
\put(192,141){$\gneg P$}
\put(212,128){\line(1,1){10}}
\put(212,128){\line(-1,1){10}}
\put(212,126){\circle*{5}}

\put(188,111){\line(-2,1){24}}
\put(188,111){\line(2,1){24}}
\put(188,109){\circle{5}}

\put(163,94){\line(-2,1){24}}
\put(163,94){\line(2,1){24}}
\put(163,92){\circle{5}}

\put(161,116){\footnotesize $1$}
\put(137,114){\footnotesize $4$}
\put(186,114){\footnotesize $5$}
\put(150,77){\footnotesize $2$}
\put(173,77){\footnotesize $3$}

\put(100,85){\line(1,0){128}}
\put(118,67){$\gneg Q$}
\put(100,67){$Q$}
\put(114,54){\line(1,1){10}}
\put(114,54){\line(-1,1){10}}
\put(114,52){\circle*{5}}
\put(139,37){\line(-2,1){24}}
\put(139,37){\line(2,1){24}}
\put(139,35){\circle{5}}

\put(167,67){$\gneg P$}
\put(149,67){$P$}
\put(163,54){\line(1,1){10}}
\put(163,54){\line(-1,1){10}}
\put(163,52){\circle*{5}}

\put(220,67){$P$}
\put(192,67){$\gneg P$}
\put(212,54){\line(1,1){10}}
\put(212,54){\line(-1,1){10}}
\put(212,52){\circle*{5}}

\put(188,37){\line(-2,1){24}}
\put(188,37){\line(2,1){24}}
\put(188,35){\circle{5}}

\put(163,20){\line(-2,1){24}}
\put(163,20){\line(2,1){24}}
\put(163,18){\circle{5}}

\put(161,42){\footnotesize $1$}
\put(137,40){\footnotesize $4$}
\put(186,40){\footnotesize $5$}

\put(338,167){\bf Example 3}
\put(318,141){$\gneg Q$}
\put(300,141){$Q$}
\put(314,128){\line(1,1){10}}
\put(314,128){\line(-1,1){10}}
\put(314,126){\circle*{5}}
\put(339,111){\line(-2,1){24}}
\put(339,111){\line(2,1){24}}
\put(339,109){\circle{5}}

\put(363,126){\circle{5}}

\put(420,141){$P$}
\put(392,141){$\gneg P$}
\put(412,128){\line(1,1){10}}
\put(412,128){\line(-1,1){10}}
\put(412,126){\circle*{5}}

\put(388,111){\line(-2,1){24}}
\put(388,111){\line(2,1){24}}
\put(388,109){\circle{5}}

\put(363,94){\line(-2,1){24}}
\put(363,94){\line(2,1){24}}
\put(363,92){\circle{5}}

\put(361,116){\footnotesize $1$}
\put(337,114){\footnotesize $4$}
\put(386,114){\footnotesize $5$}

\put(300,85){\line(1,0){128}}
\put(318,67){$\gneg Q$}
\put(300,67){$Q$}
\put(314,54){\line(1,1){10}}
\put(314,54){\line(-1,1){10}}
\put(314,52){\circle*{5}}
\put(339,37){\line(-2,1){24}}
\put(339,37){\line(2,1){24}}
\put(339,35){\circle{5}}

\put(349,67){$P$}
\put(363,54){\line(4,1){39}}
\put(363,54){\line(-1,1){10}}
\put(363,52){\circle*{5}}

\put(420,67){$P$}
\put(392,67){$\gneg P$}
\put(412,54){\line(1,1){10}}
\put(412,54){\line(-1,1){10}}
\put(412,52){\circle*{5}}

\put(388,37){\line(-2,1){24}}
\put(388,37){\line(2,1){24}}
\put(388,35){\circle{5}}

\put(363,20){\line(-2,1){24}}
\put(363,20){\line(2,1){24}}
\put(363,18){\circle{5}}

\put(361,42){\footnotesize $1$}
\put(337,40){\footnotesize $4$}
\put(386,40){\footnotesize $5$}
\put(350,76){\footnotesize $2$}
\put(399,76){\footnotesize $3$}
\put(399,150){\footnotesize $3$}

\put(338,0){WRONG !!!}
\end{picture}\end{center}

\begin{itemize}
\item In Example 1: $a=1$, $b=2$, $c=3$ and $\Theta=\{\}$. 
\item In Example 2: $a=1$, $b=2$, $c=3$  and $\Theta=\{4,5\}$.
\item Example 3 (with the same parameters as Example 2) is wrong  because node $3$ was already in the premise.
\end{itemize}
Below we see Example 2 rewritten using hyperformulas (another ``lucky case'' where this is possible): 
 
\begin{center} \begin{picture}(210,40)
\put(30,29){$\bigl((Q\mld \gneg Q)\mlc\overline{\top}\bigr)\mlc\bigl(\overline{\top}\mlc(\gneg P\mld P)\bigr)$}
\put(0,21){\line(1,0){213}}
\put(0,6){$\bigl((Q\mld \gneg Q)\mlc\overline{(P\mld \gneg P)}\bigr)\mlc\bigl(\overline{( P\mld\gneg P)}\mlc(\gneg P\mld P)\bigr)$}

\end{picture}\end{center}

\subsection{Weakening} According to this rule, a premise can be obtained from the conclusion by deleting arcs from a disjunctive gate $a$ to some children $\Delta$ of it.

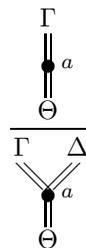
\begin{center} \begin{picture}(28,114)

\put(11,94){$\Gamma$}
\put(10,10){$\Theta$}
\put(15,18){\line(0,1){10}}
\put(13,18){\line(0,1){10}}
\put(15,81){\line(0,1){10}}
\put(13,81){\line(0,1){10}}
\put(0,55){\line(1,0){30}}
\put(14,79){\circle*{5}}
\put(15,67){\line(0,1){10}}
\put(13,67){\line(0,1){10}}
\put(10,59){$\Theta$}
\put(21,45){$\Delta$}
\put(1,45){$\Gamma$}
\put(16,32){\line(1,1){10}}
\put(13,32){\line(1,1){10}}
\put(12,32){\line(-1,1){10}}
\put(15,32){\line(-1,1){10}}
\put(14,30){\circle*{5}}
\put(19,29){\footnotesize $a$}
\put(19,78){\footnotesize $a$}

\end{picture}\end{center}

Deleting arcs from $a$ may make some children of $a$ parentless. As noted earlier, our present approach considers non-root parentless nodes (``orphan nodes'') meaningless and does not officially allow them in cirquents. So, deleting the arcs from $a$ to the nodes of $\Delta$ should be followed by (perhaps repeatedly) deleting all orphans as well, as done in Examples 1 and 3 below.

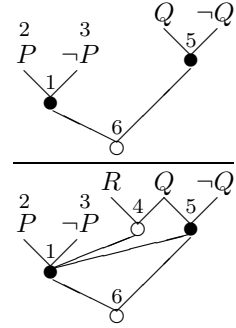
\begin{center} \begin{picture}(396,159)
\put(5,139){\bf Example 1}

\put(18,96){$\gneg P$}
\put(1,96){$P$}
\put(14,83){\line(1,1){10}}
\put(14,83){\line(-1,1){10}}
\put(14,81){\circle*{5}}
\put(12,72){\footnotesize $1$}
\put(1,105){\footnotesize $2$}
\put(26,105){\footnotesize $3$}

\put(-3,68){\line(1,0){63}}

\put(53,49){$Q$}
\put(18,49){$\gneg P$}
\put(1,49){$P$}
\put(14,36){\line(4,1){42}}
\put(14,36){\line(1,1){10}}
\put(14,36){\line(-1,1){10}}
\put(14,34){\circle*{5}}
\put(12,25){\footnotesize $1$}
\put(1,59){\footnotesize $2$}
\put(26,59){\footnotesize $3$}
\put(55,59){\footnotesize $4$}

\put(165,139){\bf Example 2}

\put(219,106){$\gneg Q$}
\put(203,106){$Q$}
\put(217,93){\line(1,1){10}}
\put(217,93){\line(-1,1){10}}
\put(217,91){\circle*{5}}

\put(168,106){$\gneg P$}
\put(151,106){$P$}
\put(164,93){\line(1,1){10}}
\put(164,93){\line(-1,1){10}}
\put(164,91){\circle*{5}}
\put(162,96){\footnotesize $1$}
\put(191,76){\line(-2,1){26}}
\put(191,76){\line(2,1){26}}
\put(189,79){\footnotesize $5$}
\put(191,74){\circle{5}}
\put(152,116){\footnotesize $2$}
\put(176,116){\footnotesize $3$}

\put(147,68){\line(1,0){83}}

\put(219,49){$\gneg Q$}
\put(203,49){$Q$}
\put(217,36){\line(1,1){10}}
\put(217,36){\line(-1,1){10}}
\put(217,34){\circle*{5}}

\put(168,49){$\gneg P$}
\put(151,49){$P$}
\put(164,36){\line(4,1){42}}
\put(164,36){\line(1,1){10}}
\put(164,36){\line(-1,1){10}}
\put(164,34){\circle*{5}}
\put(162,39){\footnotesize $1$}
\put(191,19){\line(-2,1){26}}
\put(191,19){\line(2,1){26}}
\put(189,22){\footnotesize $5$}
\put(191,16){\circle{5}}
\put(152,59){\footnotesize $2$}
\put(176,59){\footnotesize $3$}
\put(204,59){\footnotesize $4$}
\put(204,116){\footnotesize $4$}

\put(315,139){\bf Example 3}
\put(369,122){$\gneg Q$}
\put(353,122){$Q$}
\put(367,109){\line(1,1){10}}
\put(367,109){\line(-1,1){10}}
\put(367,107){\circle*{5}}
\put(318,106){$\gneg P$}
\put(301,106){$P$}
\put(314,93){\line(1,1){10}}
\put(314,93){\line(-1,1){10}}
\put(314,91){\circle*{5}}
\put(312,96){\footnotesize $1$}
\put(365,112){\footnotesize $5$}
\put(339,76){\line(-2,1){24}}
\put(339,76){\line(1,1){28}}
\put(337,79){\footnotesize $6$}
\put(339,74){\circle{5}}
\put(302,116){\footnotesize $2$}
\put(325,116){\footnotesize $3$}

\put(300,68){\line(1,0){83}}

\put(369,58){$\gneg Q$}
\put(353,58){$Q$}
\put(333,58){$R$}
\put(347,45){\line(1,1){10}}
\put(347,45){\line(-1,1){10}}
\put(347,43){\circle{5}}
\put(345,49){\footnotesize $4$}
\put(365,49){\footnotesize $5$}
\put(367,45){\line(1,1){10}}
\put(367,45){\line(-1,1){10}}
\put(367,43){\circle*{5}}

\put(318,42){$\gneg P$}
\put(301,42){$P$}
\put(314,28){\line(5,2){32}}
\put(314,28){\line(4,1){54}}
\put(314,29){\line(1,1){10}}
\put(314,29){\line(-1,1){10}}
\put(314,27){\circle*{5}}
\put(312,32){\footnotesize $1$}
\put(339,12){\line(-2,1){24}}
\put(339,12){\line(1,1){28}}
\put(337,15){\footnotesize $6$}
\put(339,10){\circle{5}}
\put(302,51){\footnotesize $2$}
\put(325,51){\footnotesize $3$}

\end{picture}\end{center}

\begin{itemize}
\item
In Example 1: $a=1$,   $\Gamma=\{2,3\}$, $\Delta=\{4\}$ and $\Theta=\{\}$. The arc from $1$ to $4$ was deleted (when moving from conclusion to premise), and so was node $4$ because it had no other parents.
\item
In Example 2: $a=1$,  $\Gamma=\{2,3\}$, $\Delta=\{4\}$ and $\Theta=\{5\}$. The arc from $1$ to $4$ was deleted but $4$ was preserved, as it had another parent in the cirquent. 
\item
In Example 3:  $a=1$,  $\Gamma=\{2,3\}$, $\Delta=\{4,5\}$ and $\Theta=\{6\}$.   The arcs from $1$ to $4$ and $5$ were deleted. This made $4$ an orphan, and $4$ was also deleted. But deleting $4$ made the $R$-labeled node an orphan, which resulted in further deleting that node as well. 
\end{itemize}

When applied to a (cirquent represented by a) formula, weakening can delete any number of disjuncts from a disjunctive subformula of the conclusion, as illustrated below:

\begin{center} \begin{picture}(135,36)
\put(0,6){$P\mld\bigl(Q\mlc(R_1\mld R_2\mld R_3\mld R_4)\bigr)$}
\put(0,19){\line(1,0){135}}
\put(21,25){$P\mld\bigl(Q\mlc(R_1\mld R_3)\bigr)$}

\end{picture}\end{center}
\subsection{Pulldown} This rule applies when the conclusion (as well as the premise) has a disjunctive gate $a$ with a single conjunctive parent $b$, which, in turn, has a single disjunctive parent $c$. Then a premise can be obtained by passing some (any) children $\Pi$ from $c$ to $a$. 

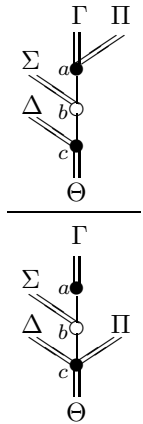
\begin{center} \begin{picture}(50,181)

\put(19,142){\footnotesize $a$}
\put(19,59){\footnotesize $a$}
\put(19,42){\footnotesize $b$}
\put(19,125){\footnotesize $b$}
\put(19,27){\footnotesize $c$}
\put(19,110){\footnotesize $c$}

\put(24,161){$\Gamma$}
\put(39,161){$\Pi$}
\put(25,146){\line(0,1){12}}
\put(27,146){\line(0,1){12}}
\put(25,146){\line(3,2){16}}
\put(28,146){\line(3,2){16}}
\put(26,144){\circle*{5}}
\put(26,131){\line(0,1){10}}
\put(24,131){\line(-3,2){16}}
\put(27,131){\line(-3,2){16}}
\put(5,127){$\Delta$}
\put(26,129){\circle{5}}
\put(26,117){\line(0,1){10}}
\put(24,115){\line(-3,2){16}}
\put(27,115){\line(-3,2){16}}
\put(5,144){$\Sigma$}
\put(26,115){\circle*{5}}
\put(25,103){\line(0,1){10}}
\put(27,103){\line(0,1){10}}
\put(22,94){$\Theta$}

\put(0,90){\line(1,0){50}}

\put(24,78){$\Gamma$}
\put(27,63){\line(0,1){10}}
\put(25,63){\line(0,1){10}}
\put(26,61){\circle*{5}}
\put(5,61){$\Sigma$}
\put(26,48){\line(0,1){10}}
\put(27,48){\line(-3,2){16}}
\put(24,48){\line(-3,2){16}}
\put(5,44){$\Delta$}
\put(39,44){$\Pi$}
\put(26,46){\circle{5}}
\put(26,34){\line(0,1){10}}
\put(24,32){\line(-3,2){16}}
\put(27,32){\line(-3,2){16}}
\put(24,32){\line(3,2){16}}
\put(27,32){\line(3,2){16}}
\put(26,32){\circle*{5}}
\put(25,20){\line(0,1){10}}
\put(27,20){\line(0,1){10}}
\put(22,11){$\Theta$}
\end{picture}\end{center}

When performing the above children-passing transformation, some or all nodes of $\Pi$ may still remain children of $c$. This is so because, according to  Convention \ref{conv}, $\Pi$ and $\Delta$ do not necessarily have to be disjoint. Similarly, $\Pi$ and $\Gamma$ do not have to be disjoint, meaning that some nodes of $\Pi$ may simply stop being children of $c$ without acquiring $a$ as a new parent, as $a$ already {\em was} a parent of them.

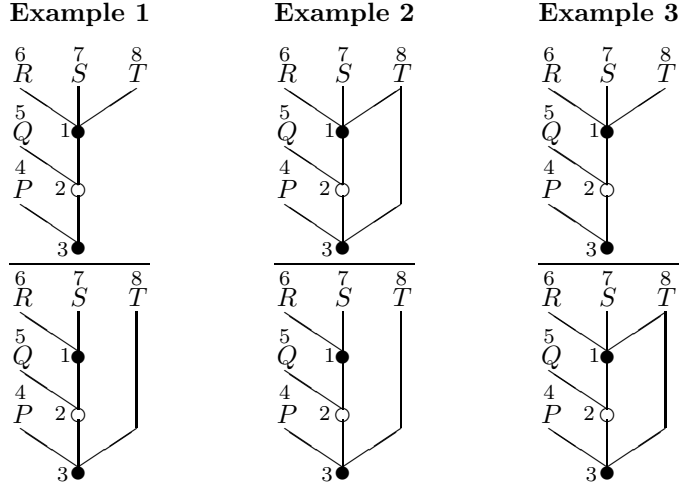
\begin{center} \begin{picture}(262,201)

\put(0,181){\bf Example 1}

\put(23,158){$S$}
\put(1,158){$R$}
\put(1,114){$P$}
\put(45,158){$T$}

\put(26,141){\line(0,1){15}}
\put(26,139){\circle*{5}}
\put(1,136){$Q$}
\put(26,119){\line(0,1){18}}
\put(26,141){\line(-3,2){22}}
\put(26,119){\line(-3,2){22}}

\put(26,117){\circle{5}}
\put(26,97){\line(0,1){18}}
\put(26,97){\line(-3,2){22}}
\put(26,141){\line(3,2){22}}
\put(26,95){\circle*{5}}

\put(2,123){\footnotesize $4$}
\put(2,144){\footnotesize $5$}
\put(2,166){\footnotesize $6$}
\put(24,166){\footnotesize $7$}
\put(46,166){\footnotesize $8$}

\put(19,137){\footnotesize $1$}
\put(17,115){\footnotesize $2$}
\put(18,92){\footnotesize $3$}

\put(0,89){\line(1,0){53}}

\put(23,73){$S$}
\put(1,73){$R$}
\put(1,29){$P$}
\put(45,73){$T$}
\put(1,51){$Q$}

\put(26,56){\line(0,1){15}}
\put(26,54){\circle*{5}}
\put(26,34){\line(0,1){18}}
\put(26,56){\line(-3,2){22}}
\put(26,34){\line(-3,2){22}}
\put(48,27){\line(0,1){44}}

\put(26,32){\circle{5}}
\put(26,12){\line(0,1){18}}
\put(26,12){\line(-3,2){22}}
\put(26,12){\line(3,2){22}}
\put(26,10){\circle*{5}}

\put(2,38){\footnotesize $4$}
\put(2,59){\footnotesize $5$}
\put(2,81){\footnotesize $6$}
\put(24,81){\footnotesize $7$}
\put(46,81){\footnotesize $8$}
\put(19,52){\footnotesize $1$}
\put(17,30){\footnotesize $2$}
\put(18,7){\footnotesize $3$}

\put(100,181){\bf Example 2}

\put(123,158){$S$}
\put(101,158){$R$}
\put(101,114){$P$}
\put(145,158){$T$}

\put(126,141){\line(0,1){15}}
\put(126,139){\circle*{5}}
\put(101,136){$Q$}
\put(126,119){\line(0,1){18}}
\put(126,141){\line(-3,2){22}}
\put(126,119){\line(-3,2){22}}

\put(126,117){\circle{5}}
\put(126,97){\line(0,1){18}}
\put(126,97){\line(-3,2){22}}
\put(126,97){\line(3,2){22}}
\put(126,141){\line(3,2){22}}
\put(148,112){\line(0,1){44}}
\put(126,95){\circle*{5}}

\put(102,123){\footnotesize $4$}
\put(102,144){\footnotesize $5$}
\put(102,166){\footnotesize $6$}
\put(124,166){\footnotesize $7$}
\put(146,166){\footnotesize $8$}

\put(119,137){\footnotesize $1$}
\put(117,115){\footnotesize $2$}
\put(118,92){\footnotesize $3$}

\put(100,89){\line(1,0){53}}

\put(123,73){$S$}
\put(101,73){$R$}
\put(101,29){$P$}
\put(145,73){$T$}
\put(101,51){$Q$}

\put(126,56){\line(0,1){15}}
\put(126,54){\circle*{5}}
\put(126,34){\line(0,1){18}}
\put(126,56){\line(-3,2){22}}
\put(126,34){\line(-3,2){22}}
\put(148,27){\line(0,1){44}}

\put(126,32){\circle{5}}
\put(126,12){\line(0,1){18}}
\put(126,12){\line(-3,2){22}}
\put(126,12){\line(3,2){22}}
\put(126,10){\circle*{5}}

\put(102,38){\footnotesize $4$}
\put(102,59){\footnotesize $5$}
\put(102,81){\footnotesize $6$}
\put(124,81){\footnotesize $7$}
\put(146,81){\footnotesize $8$}
\put(119,52){\footnotesize $1$}
\put(117,30){\footnotesize $2$}
\put(118,7){\footnotesize $3$}

\put(200,181){\bf Example 3}

\put(223,158){$S$}
\put(201,158){$R$}
\put(201,114){$P$}
\put(245,158){$T$}

\put(226,141){\line(0,1){15}}
\put(226,139){\circle*{5}}
\put(201,136){$Q$}
\put(226,119){\line(0,1){18}}
\put(226,141){\line(-3,2){22}}
\put(226,119){\line(-3,2){22}}

\put(226,117){\circle{5}}
\put(226,97){\line(0,1){18}}
\put(226,97){\line(-3,2){22}}
\put(226,141){\line(3,2){22}}
\put(226,95){\circle*{5}}

\put(202,123){\footnotesize $4$}
\put(202,144){\footnotesize $5$}
\put(202,166){\footnotesize $6$}
\put(224,166){\footnotesize $7$}
\put(246,166){\footnotesize $8$}

\put(219,137){\footnotesize $1$}
\put(217,115){\footnotesize $2$}
\put(218,92){\footnotesize $3$}

\put(200,89){\line(1,0){53}}

\put(223,73){$S$}
\put(201,73){$R$}
\put(201,29){$P$}
\put(245,73){$T$}
\put(201,51){$Q$}

\put(226,56){\line(0,1){15}}
\put(226,54){\circle*{5}}
\put(226,34){\line(0,1){18}}
\put(226,56){\line(-3,2){22}}
\put(226,34){\line(-3,2){22}}
\put(248,27){\line(0,1){44}}
\put(226,56){\line(3,2){22}}

\put(226,32){\circle{5}}
\put(226,12){\line(0,1){18}}
\put(226,12){\line(-3,2){22}}
\put(226,12){\line(3,2){22}}
\put(226,10){\circle*{5}}

\put(202,38){\footnotesize $4$}
\put(202,59){\footnotesize $5$}
\put(202,81){\footnotesize $6$}
\put(224,81){\footnotesize $7$}
\put(246,81){\footnotesize $8$}
\put(219,52){\footnotesize $1$}
\put(217,30){\footnotesize $2$}
\put(218,7){\footnotesize $3$}
\end{picture}\end{center}

\begin{itemize}
\item In Example 1: $a=1$, $b=2$, $c=3$, $\Gamma=\{6,7\}$, $\Delta=\{4\}$, $\Pi=\{8\}$, $\Sigma=\{5\}$, $\Theta=\{\}$.
\item In Example 2: $a=1$, $b=2$, $c=3$, $\Gamma=\{6,7\}$, $\Delta=\{4,8\}$, $\Pi=\{8\}$, $\Sigma=\{5\}$, $\Theta=\{\}$.
\item In Example 3: $a=1$, $b=2$, $c=3$, $\Gamma=\{6,7,8\}$, $\Delta=\{4\}$, $\Pi=\{8\}$, $\Sigma=\{5\}$, $\Theta=\{\}$.
\end{itemize}

Example 1 (and only this example) can also be written using formulas:

\begin{center} \begin{picture}(98,36)
\put(0,25){$P\mld\bigl(Q\mlc(R\mld S\mld T)\bigr)$}
\put(0,19){\line(1,0){98}}
\put(0,6){$P\mld\bigl(Q\mlc(R\mld S)\bigr)\mld T$}
\end{picture}\end{center}

\section{Derivability, provability and admissibility}

A {\bf \lll-derivation} of a cirquent $A$ from a cirquent $B$ is a sequence $C_1,\ldots,C_n$ of cirquents such that $C_1=B$, $C_n=A$ and each $C_{i+1}$ follows from $C_i$ by one of the rules of \lll. A derivation is usually required to come with an --- even if only implicit --- {\bf justification}, which is an indication of by which rule any given cirquent $C_{i+1}$ follows from $C_i$ and where that rule is applied, i.e., what the values of the parameters of the rule are.   

A {\bf \lll-proof} of a cirquent  $A$ is a \lll-derivation of $A$ from $\circ$. Thus, the single-node cirquent $\circ$, i.e., $\top$, is the (only) {\bf axiom} of \lll.

When a \lll-proof of a cirquent $A$ exists, we say that $A$ is {\bf provable} in \lll\ and write $\lll\vdash A$. Similar terminology applies to any other cirquent calculus system as well. When \lll\ is the only system we deal with in a given context (such as the present section), we usually omit ``\lll-'' and simply say ``derivation'', ``provable'' etc. 

Below is a \lll-proof of the left cirquent of Figure 2 in full detail, serving the purpose of giving the reader a better syntactic feel of cirquent calculus. All ports of each cirquent of the proof have unique labels, which allows us to unambiguously refer to those ports (in justifications) by their labels, without assigning names to them as we did in the examples of the previous section.

\begin{center} \begin{picture}(322,37)
\put(150,17){\footnotesize axiom}

\put(162,0){\circle{5}}
\put(160,5){\footnotesize $4$}
\end{picture}\end{center}

\begin{center} \begin{picture}(322,37)
\put(66,29){\footnotesize deepening: $a=4$, $b=2$, $\Gamma=\{\}$, $\Delta=\{\}$,  $\Theta=\{\}$}

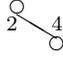
\put(162,0){\circle{5}}
\put(162,2){\line(-5,3){15}}

\put(147,14){\circle{5}}
\put(160,5){\footnotesize $4$}
\put(143,5){\footnotesize $2$}
\end{picture}\end{center}

\begin{center} \begin{picture}(322,37)
\put(66,29){\footnotesize deepening: $a=4$, $b=3$, $\Gamma=\{2\}$, $\Delta=\{\}$,  $\Theta=\{\}$}

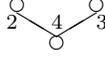
\put(162,0){\circle{5}}
\put(162,2){\line(-5,3){15}}
\put(162,2){\line(5,3){15}}

\put(177,14){\circle{5}}
\put(147,14){\circle{5}}
\put(160,5){\footnotesize $4$}
\put(177,5){\footnotesize $3$}
\put(143,5){\footnotesize $2$}
\end{picture}\end{center}

\begin{center} \begin{picture}(322,57)
\put(84,49){\footnotesize coupling: $a=2$, $b=Q$, $c=\gneg Q$, $\Theta=\{4\}$}
\put(115,29){$\gneg Q$}

\put(162,0){\circle{5}}
\put(162,2){\line(-5,3){15}}
\put(162,2){\line(5,3){15}}
\put(143,29){$Q$}

\put(147,16){\line(-5,2){22}}

\put(177,14){\circle{5}}
\put(147,14){\circle*{5}}
\put(147,16){\line(0,1){10}}
\put(160,5){\footnotesize $4$}
\put(177,5){\footnotesize $3$}
\put(143,5){\footnotesize $2$}
\end{picture}\end{center}

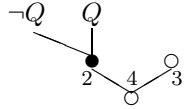
\begin{center} \begin{picture}(322,57)
\put(86,49){\footnotesize coupling: $a=3$, $b=R$, $c=\gneg R$, $\Theta=\{4\}$}
\put(115,29){$\gneg Q$}
\put(191,29){$\gneg R$}

\put(162,0){\circle{5}}
\put(162,2){\line(-5,3){15}}
\put(162,2){\line(5,3){15}}
\put(143,29){$Q$}
\put(174,29){$R$}

\put(177,16){\line(5,2){22}}
\put(147,16){\line(-5,2){22}}

\put(177,14){\circle*{5}}
\put(147,14){\circle*{5}}
\put(177,16){\line(0,1){10}}
\put(147,16){\line(0,1){10}}
\put(160,5){\footnotesize $4$}
\put(177,5){\footnotesize $3$}
\put(143,5){\footnotesize $2$}
\end{picture}\end{center}

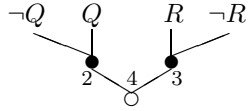
\begin{center} \begin{picture}(322,85)
\put(64,77){\footnotesize lengthening: $a=4$, $b=1$, $\Gamma=\{2,3\}$, $\Theta=\{\}$, $\Omega=\{\}$}
\put(115,57){$\gneg Q$}
\put(191,57){$\gneg R$}

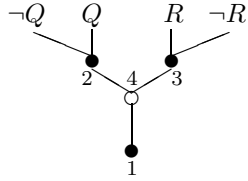
\put(162,28){\circle{5}}
\put(162,30){\line(-5,3){15}}
\put(162,30){\line(5,3){15}}
\put(162,10){\line(0,1){16}}
\put(143,57){$Q$}
\put(174,57){$R$}

\put(177,44){\line(5,2){22}}
\put(147,44){\line(-5,2){22}}

\put(162,8){\circle*{5}}
\put(177,42){\circle*{5}}
\put(147,42){\circle*{5}}
\put(177,44){\line(0,1){10}}
\put(147,44){\line(0,1){10}}
\put(160,-1){\footnotesize $1$}
\put(160,33){\footnotesize $4$}
\put(177,33){\footnotesize $3$}
\put(143,33){\footnotesize $2$}
\end{picture}\end{center}

\begin{center} \begin{picture}(322,85)
\put(16,77){\footnotesize pulldown: $a=2$, $b=4$, $c=1$, $\Gamma=\{Q\}$, $\Delta=\{\}$, $\Pi=\{\gneg Q\}$, $\Sigma=\{3\}$, $\Theta=\{\}$}
\put(92,28){$\gneg Q$}
\put(191,57){$\gneg R$}

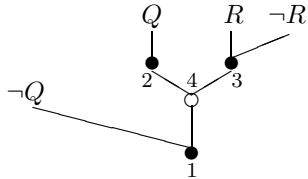
\put(162,28){\circle{5}}
\put(162,30){\line(-5,3){15}}
\put(162,30){\line(5,3){15}}
\put(162,10){\line(0,1){16}}
\put(162,10){\line(-4,1){60}}
\put(143,57){$Q$}
\put(174,57){$R$}

\put(177,44){\line(5,2){22}}

\put(162,8){\circle*{5}}
\put(177,42){\circle*{5}}
\put(147,42){\circle*{5}}
\put(177,44){\line(0,1){10}}
\put(147,44){\line(0,1){10}}
\put(160,-1){\footnotesize $1$}
\put(160,33){\footnotesize $4$}
\put(177,33){\footnotesize $3$}
\put(143,33){\footnotesize $2$}
\end{picture}\end{center}

\begin{center} \begin{picture}(322,85)
\put(10,77){\footnotesize pulldown: $a=3$, $b=4$, $c=1$, $\Gamma=\{R\}$, $\Delta=\{\gneg Q\}$, $\Pi=\{\gneg R\}$, $\Sigma=\{2\}$, $\Theta=\{\}$}
\put(92,28){$\gneg Q$}
\put(214,28){$\gneg R$}

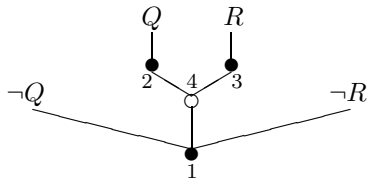
\put(162,28){\circle{5}}
\put(162,30){\line(-5,3){15}}
\put(162,30){\line(5,3){15}}
\put(162,10){\line(0,1){16}}
\put(162,10){\line(4,1){60}}
\put(162,10){\line(-4,1){60}}
\put(143,57){$Q$}
\put(174,57){$R$}

\put(162,8){\circle*{5}}
\put(177,42){\circle*{5}}
\put(147,42){\circle*{5}}
\put(177,44){\line(0,1){10}}
\put(147,44){\line(0,1){10}}
\put(160,-1){\footnotesize $1$}
\put(160,33){\footnotesize $4$}
\put(177,33){\footnotesize $3$}
\put(143,33){\footnotesize $2$}
\end{picture}\end{center}

\begin{center} \begin{picture}(322,85)
\put(70,77){\footnotesize shortening: $a=Q$, $b=2$, $\Gamma=\{\}$, $\Theta=\{4\}$, $\Omega=\{\}$}
\put(92,28){$\gneg Q$}
\put(214,28){$\gneg R$}

\put(162,28){\circle{5}}
\put(162,30){\line(-5,3){15}}
\put(162,30){\line(5,3){15}}
\put(162,10){\line(0,1){16}}
\put(162,10){\line(4,1){60}}
\put(162,10){\line(-4,1){60}}
\put(143,42){$Q$}
\put(174,57){$R$}

\put(162,8){\circle*{5}}
\put(177,42){\circle*{5}}
\put(177,44){\line(0,1){10}}
\put(160,-1){\footnotesize $1$}
\put(160,33){\footnotesize $4$}
\put(177,33){\footnotesize $3$}
\end{picture}\end{center}

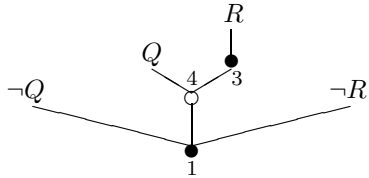
\begin{center} \begin{picture}(322,70)
\put(70,62){\footnotesize shortening: $a=R$, $b=3$, $\Gamma=\{\}$, $\Theta=\{4\}$, $\Omega=\{\}$}
\put(92,28){$\gneg Q$}
\put(214,28){$\gneg R$}

\put(162,28){\circle{5}}
\put(162,30){\line(-5,3){15}}
\put(162,30){\line(5,3){15}}
\put(162,10){\line(0,1){16}}
\put(162,10){\line(4,1){60}}
\put(162,10){\line(-4,1){60}}
\put(143,42){$Q$}
\put(174,42){$R$}

\put(162,8){\circle*{5}}
\put(160,-1){\footnotesize $1$}
\put(160,33){\footnotesize $4$}
\end{picture}\end{center}

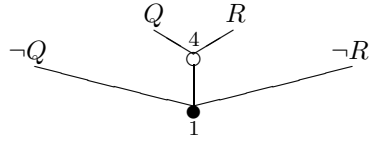
\begin{center} \begin{picture}(322,70)
\put(64,62){\footnotesize lengthening: $a=\gneg Q$, $b=2$, $\Gamma=\{\}$, $\Theta=\{1\}$, $\Omega=\{\}$}
\put(92,42){$\gneg Q$}
\put(214,28){$\gneg R$}

\put(101,28){\circle{5}}
\put(101,30){\line(0,1){10}}
\put(162,28){\circle{5}}
\put(162,30){\line(-5,3){15}}
\put(162,30){\line(5,3){15}}
\put(162,10){\line(0,1){16}}
\put(162,10){\line(4,1){60}}
\put(162,10){\line(-4,1){60}}
\put(143,42){$Q$}
\put(174,42){$R$}

\put(162,8){\circle*{5}}
\put(160,-1){\footnotesize $1$}
\put(160,33){\footnotesize $4$}
\put(93,26){\footnotesize $2$}
\end{picture}\end{center}

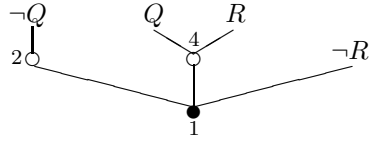
\begin{center} \begin{picture}(322,70)
\put(64,62){\footnotesize lengthening: $a=\gneg R$, $b=3$, $\Gamma=\{\}$, $\Theta=\{1\}$, $\Omega=\{\}$}
\put(92,42){$\gneg Q$}
\put(214,42){$\gneg R$}

\put(101,28){\circle{5}}
\put(101,30){\line(0,1){10}}
\put(162,28){\circle{5}}
\put(162,30){\line(-5,3){15}}
\put(162,30){\line(5,3){15}}
\put(223,28){\circle{5}}
\put(223,30){\line(0,1){10}}
\put(162,10){\line(0,1){16}}
\put(162,10){\line(4,1){60}}
\put(162,10){\line(-4,1){60}}
\put(143,42){$Q$}
\put(174,42){$R$}

\put(162,8){\circle*{5}}
\put(160,-1){\footnotesize $1$}
\put(160,33){\footnotesize $4$}
\put(92,26){\footnotesize $2$}
\put(227,26){\footnotesize $3$}
\end{picture}\end{center}

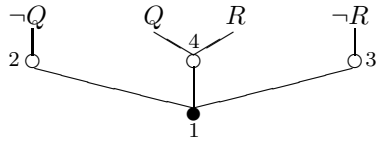
\begin{center} \begin{picture}(322,70)
\put(67,62){\footnotesize deepening: $a=2$, $b=5$, $\Gamma=\{\gneg Q\}$, $\Delta=\{\}$, $\Theta=\{1\}$}
\put(92,42){$\gneg Q$}
\put(214,42){$\gneg R$}

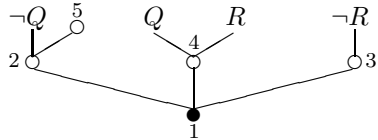
\put(101,28){\circle{5}}
\put(101,30){\line(0,1){10}}
\put(101,30){\line(5,3){15}}
\put(162,28){\circle{5}}
\put(162,30){\line(-5,3){15}}
\put(162,30){\line(5,3){15}}
\put(223,28){\circle{5}}
\put(223,30){\line(0,1){10}}
\put(162,10){\line(0,1){16}}
\put(162,10){\line(4,1){60}}
\put(162,10){\line(-4,1){60}}

\put(118,41){\circle{5}}
\put(143,42){$Q$}
\put(174,42){$R$}
\put(116,45){\footnotesize $5$}

\put(162,8){\circle*{5}}
\put(160,-1){\footnotesize $1$}
\put(160,33){\footnotesize $4$}
\put(92,26){\footnotesize $2$}
\put(227,26){\footnotesize $3$}
\end{picture}\end{center}

\begin{center} \begin{picture}(322,70)
\put(67,62){\footnotesize deepening: $a=3$, $b=6$, $\Gamma=\{\gneg R\}$, $\Delta=\{\}$, $\Theta=\{1\}$}
\put(92,42){$\gneg Q$}
\put(214,42){$\gneg R$}

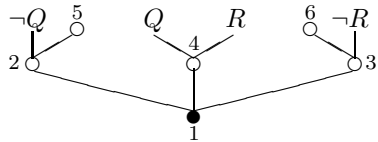
\put(101,28){\circle{5}}
\put(101,30){\line(0,1){10}}
\put(101,30){\line(5,3){15}}
\put(162,28){\circle{5}}
\put(162,30){\line(-5,3){15}}
\put(162,30){\line(5,3){15}}
\put(223,28){\circle{5}}
\put(223,30){\line(0,1){10}}
\put(223,30){\line(-5,3){15}}
\put(162,10){\line(0,1){16}}
\put(162,10){\line(4,1){60}}
\put(162,10){\line(-4,1){60}}

\put(118,41){\circle{5}}
\put(206,41){\circle{5}}
\put(143,42){$Q$}
\put(174,42){$R$}
\put(116,45){\footnotesize $5$}
\put(204,45){\footnotesize $6$}

\put(162,8){\circle*{5}}
\put(160,-1){\footnotesize $1$}
\put(160,33){\footnotesize $4$}
\put(92,26){\footnotesize $2$}
\put(227,26){\footnotesize $3$}
\end{picture}\end{center}

\begin{center} \begin{picture}(322,81)
\put(112,73){\footnotesize just redrawing the cirquent}
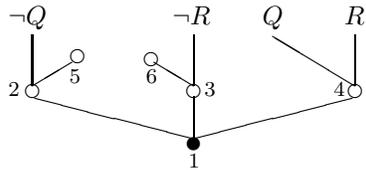
\put(92,53){$\gneg Q$}
\put(154,53){$\gneg R$}
\put(188,53){$Q$}
\put(219,53){$R$}
\put(101,28){\circle{5}}
\put(101,30){\line(0,1){19}}
\put(101,30){\line(5,3){15}}
\put(162,28){\circle{5}}
\put(162,30){\line(0,1){19}}
\put(162,30){\line(-5,3){15}}
\put(223,28){\circle{5}}
\put(223,30){\line(0,1){19}}
\put(223,30){\line(-5,3){31}}
\put(162,10){\line(0,1){16}}
\put(162,10){\line(4,1){60}}
\put(162,10){\line(-4,1){60}}

\put(118,41){\circle{5}}
\put(146,40){\circle{5}}
\put(115,31){\footnotesize $5$}
\put(144,31){\footnotesize $6$}

\put(162,8){\circle*{5}}
\put(160,-1){\footnotesize $1$}
\put(166,26){\footnotesize $3$}
\put(92,26){\footnotesize $2$}
\put(215,26){\footnotesize $4$}
\end{picture}\end{center}

\begin{center} \begin{picture}(322,81)
\put(50,73){\footnotesize globalization: $a=5$, $b=6$, $c=7$, $\Gamma=\{\}$, $\Theta=\{2\}$, $\Omega=\{3\}$}
\put(92,53){$\gneg Q$}
\put(154,53){$\gneg R$}
\put(188,53){$Q$}
\put(219,53){$R$}
\put(101,28){\circle{5}}
\put(101,30){\line(0,1){19}}
\put(101,30){\line(5,3){31}}
\put(162,28){\circle{5}}
\put(162,30){\line(0,1){19}}
\put(162,30){\line(-5,3){31}}
\put(223,28){\circle{5}}
\put(223,30){\line(0,1){19}}
\put(223,30){\line(-5,3){31}}
\put(162,10){\line(0,1){16}}
\put(162,10){\line(4,1){60}}
\put(162,10){\line(-4,1){60}}

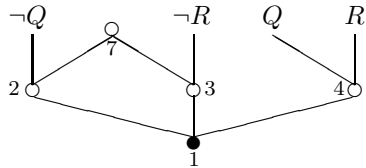
\put(162,8){\circle*{5}}
\put(131,51){\circle{5}}
\put(160,-1){\footnotesize $1$}
\put(166,26){\footnotesize $3$}
\put(92,26){\footnotesize $2$}
\put(215,26){\footnotesize $4$}
\put(129,41){\footnotesize $7$}
\end{picture}\end{center}

\begin{center} \begin{picture}(322,95)
\put(80,87){\footnotesize coupling: $a=7$, $b=P$, $c=\gneg P$, $\Theta=\{2,3\}$}
\put(113,67){$\gneg P$}
\put(92,53){$\gneg Q$}
\put(139,67){$P$}
\put(154,53){$\gneg R$}
\put(188,53){$Q$}
\put(219,53){$R$}
\put(101,28){\circle{5}}
\put(101,30){\line(0,1){19}}
\put(101,30){\line(5,3){31}}
\put(162,28){\circle{5}}
\put(162,30){\line(0,1){19}}
\put(162,30){\line(-5,3){31}}
\put(223,28){\circle{5}}
\put(223,30){\line(0,1){19}}
\put(223,30){\line(-5,3){31}}
\put(162,10){\line(0,1){16}}
\put(162,10){\line(4,1){60}}
\put(162,10){\line(-4,1){60}}

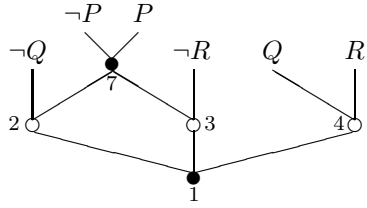
\put(131,53){\line(1,1){10}}
\put(131,53){\line(-1,1){10}}
\put(162,8){\circle*{5}}
\put(131,51){\circle*{5}}
\put(160,-1){\footnotesize $1$}
\put(166,26){\footnotesize $3$}
\put(92,26){\footnotesize $2$}
\put(215,26){\footnotesize $4$}
\put(129,41){\footnotesize $7$}
\end{picture}\end{center}

\begin{center} \begin{picture}(322,101)
\put(40,93){\footnotesize localization: $a=5$, $b=6$, $c=7$, $\Gamma=\{\gneg P,P\}$, $\Theta=\{2\}$, $\Omega=\{3\}$}
\put(123,73){$\gneg P$}
\put(92,53){$\gneg Q$}
\put(127,53){$P$}
\put(154,53){$\gneg R$}
\put(188,53){$Q$}
\put(219,53){$R$}
\put(101,28){\circle{5}}
\put(101,30){\line(0,1){19}}
\put(101,30){\line(5,3){31}}
\put(162,28){\circle{5}}
\put(162,30){\line(0,1){19}}
\put(162,30){\line(-5,3){31}}
\put(223,28){\circle{5}}
\put(223,30){\line(0,1){19}}
\put(223,30){\line(-5,3){31}}
\put(162,10){\line(0,1){16}}
\put(162,10){\line(4,1){60}}
\put(162,10){\line(-4,1){60}}

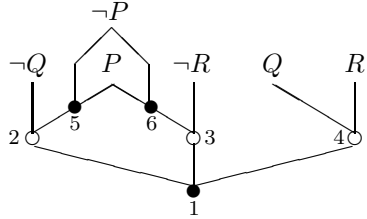
\put(117,56){\line(1,1){14}}
\put(117,40){\line(0,1){16}}
\put(145,56){\line(-1,1){14}}
\put(145,40){\line(0,1){16}}
\put(162,8){\circle*{5}}
\put(117,40){\circle*{5}}
\put(146,40){\circle*{5}}
\put(160,-1){\footnotesize $1$}
\put(166,26){\footnotesize $3$}
\put(92,26){\footnotesize $2$}
\put(215,26){\footnotesize $4$}
\put(115,31){\footnotesize $5$}
\put(144,31){\footnotesize $6$}
\end{picture}\end{center}

\begin{center} \begin{picture}(322,101)
\put(0,93){\footnotesize pulldown: $a=6$, $b=3$, $c=1$, $\Gamma=\{P\}$, $\Delta=\{2,4\}$, $\Pi=\{\gneg P\}$, $\Sigma=\{\gneg R\}$, $\Theta=\{\}$}
\put(123,73){$\gneg P$}
\put(92,53){$\gneg Q$}
\put(127,53){$P$}
\put(154,53){$\gneg R$}
\put(188,53){$Q$}
\put(219,53){$R$}
\put(101,28){\circle{5}}
\put(101,30){\line(0,1){19}}
\put(101,30){\line(5,3){31}}
\put(162,28){\circle{5}}
\put(162,30){\line(0,1){19}}
\put(162,30){\line(-5,3){31}}
\put(223,28){\circle{5}}
\put(223,30){\line(0,1){19}}
\put(223,30){\line(-5,3){31}}
\put(162,10){\line(0,1){16}}
\put(162,10){\line(4,1){60}}
\put(162,10){\line(-4,1){60}}
\put(162,10){\line(-6,1){89}}
\put(73,25){\line(0,1){33}}
\put(73,58){\line(5,1){58}}
\put(117,56){\line(1,1){13}}
\put(117,40){\line(0,1){16}}
\put(162,8){\circle*{5}}
\put(117,40){\circle*{5}}
\put(146,40){\circle*{5}}
\put(160,-1){\footnotesize $1$}
\put(166,26){\footnotesize $3$}
\put(92,26){\footnotesize $2$}
\put(215,26){\footnotesize $4$}
\put(115,31){\footnotesize $5$}
\put(144,31){\footnotesize $6$}
\end{picture}\end{center}

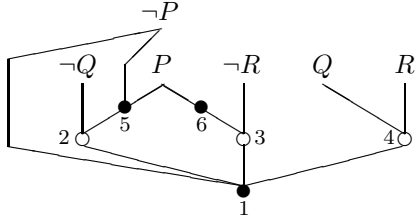
\begin{center} \begin{picture}(322,81)
\put(-8,73){\footnotesize pulldown: $a=5$, $b=2$, $c=1$, $\Gamma=\{P\}$, $\Delta=\{\gneg P,3,4\}$, $\Pi=\{\gneg P\}$, $\Sigma=\{\gneg Q\}$, $\Theta=\{\}$}
\put(63,53){$\gneg P$}
\put(92,53){$\gneg Q$}
\put(127,53){$P$}
\put(154,53){$\gneg R$}
\put(188,53){$Q$}
\put(219,53){$R$}
\put(101,28){\circle{5}}
\put(101,30){\line(0,1){19}}
\put(101,30){\line(5,3){31}}
\put(162,28){\circle{5}}
\put(162,30){\line(0,1){19}}
\put(162,30){\line(-5,3){31}}
\put(223,28){\circle{5}}
\put(223,30){\line(0,1){19}}
\put(223,30){\line(-5,3){31}}
\put(162,10){\line(0,1){16}}
\put(162,10){\line(4,1){60}}
\put(162,10){\line(-4,1){60}}
\put(162,10){\line(-6,1){89}}
\put(73,25){\line(0,1){24}}
\put(162,8){\circle*{5}}
\put(117,40){\circle*{5}}
\put(146,40){\circle*{5}}
\put(160,-1){\footnotesize $1$}
\put(166,26){\footnotesize $3$}
\put(92,26){\footnotesize $2$}
\put(215,26){\footnotesize $4$}
\put(115,31){\footnotesize $5$}
\put(144,31){\footnotesize $6$}

\end{picture}\end{center}

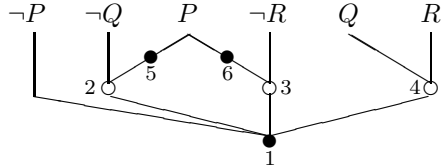
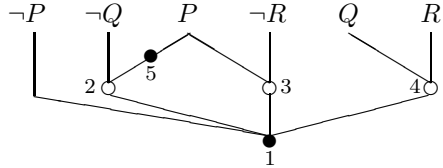
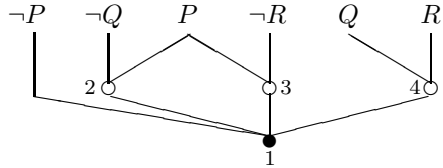
\begin{center} \begin{picture}(322,81)
\put(54,73){\footnotesize shortening: $a=P$, $b=6$, $\Gamma=\{\}$, $\Theta=\{3\}$, $\Omega=\{\}$}
\put(63,53){$\gneg P$}
\put(92,53){$\gneg Q$}
\put(127,53){$P$}
\put(154,53){$\gneg R$}
\put(188,53){$Q$}
\put(219,53){$R$}
\put(101,28){\circle{5}}
\put(101,30){\line(0,1){19}}
\put(101,30){\line(5,3){31}}
\put(162,28){\circle{5}}
\put(162,30){\line(0,1){19}}
\put(162,30){\line(-5,3){31}}
\put(223,28){\circle{5}}
\put(223,30){\line(0,1){19}}
\put(223,30){\line(-5,3){31}}
\put(162,10){\line(0,1){16}}
\put(162,10){\line(4,1){60}}
\put(162,10){\line(-4,1){60}}
\put(162,10){\line(-6,1){89}}
\put(73,25){\line(0,1){24}}
\put(162,8){\circle*{5}}
\put(117,40){\circle*{5}}
\put(160,-1){\footnotesize $1$}
\put(166,26){\footnotesize $3$}
\put(92,26){\footnotesize $2$}
\put(215,26){\footnotesize $4$}
\put(115,31){\footnotesize $5$}

\end{picture}\end{center}

\begin{center} \begin{picture}(322,91)
\put(54,83){\footnotesize shortening: $a=P$, $b=5$, $\Gamma=\{\}$, $\Theta=\{2\}$, $\Omega=\{\}$}
\put(63,63){$\gneg P$}
\put(92,63){$\gneg Q$}
\put(127,63){$P$}
\put(154,63){$\gneg R$}
\put(188,63){$Q$}
\put(219,63){$R$}
\put(101,38){\circle{5}}
\put(101,40){\line(0,1){19}}
\put(101,40){\line(5,3){31}}
\put(162,38){\circle{5}}
\put(162,40){\line(0,1){19}}
\put(162,40){\line(-5,3){31}}
\put(223,38){\circle{5}}
\put(223,40){\line(0,1){19}}
\put(223,40){\line(-5,3){31}}
\put(162,20){\line(0,1){16}}
\put(162,20){\line(4,1){60}}
\put(162,20){\line(-4,1){60}}
\put(162,20){\line(-6,1){89}}
\put(73,35){\line(0,1){24}}
\put(162,18){\circle*{5}}
\put(160,9){\footnotesize $1$}
\put(166,36){\footnotesize $3$}
\put(92,36){\footnotesize $2$}
\put(215,36){\footnotesize $4$}
\end{picture}\end{center}

This is the last time in this paper that we provide a proof in all details. Subsequent proofs will be ``lazier'', with several steps often combined together, and with justifications typically reduced to indicating the names  of the rules used, without indicating the (usually easy to guess) values of the corresponding parameters. 

The reader may want to try to see why and where the above proof fails if the target is the right rather than the left cirquent of Figure 2. That cirquent simply has no proof. The difference between one shared $P$-port and two separate $P$-ports is thus crucial here.

The left cirquent of Figure 2, unlike the right cirquent, is a circuit --- each port in it has a unique label. However, this in not at all the reason why the former is provable and the latter is not. The left cirquent of the following Figure 3 is not a circuit but its proof can be mechanically obtained from the above proof by replacing all atoms by $P$. On the other hand, the right cirquent of Figure 3, just like its predecessor from Figure 2, can be shown to have no proof:

\begin{center} \begin{picture}(271,76)
\put(0,64){$\gneg P$}
\put(19,64){$\gneg P$}
\put(42,64){$P$}
\put(54,64){$\gneg P$}
\put(76,64){$P$}
\put(92,64){$P$}

\put(28,46){\line(6,5){17}}
\put(28,46){\line(0,1){14}}
\put(28,44){\circle{5}}

\put(62,46){\line(-6,5){17}}
\put(62,46){\line(0,1){14}}
\put(62,44){\circle{5}}

\put(96,46){\line(-6,5){17}}
\put(96,46){\line(0,1){14}}
\put(96,44){\circle{5}}

\put(9,37){\line(0,1){24}}
\put(45,28){\line(-4,1){36}}
\put(45,28){\line(4,1){52}}
\put(45,28){\line(-6,5){16}}
\put(45,28){\line(6,5){16}}
\put(45,26){\circle*{5}}
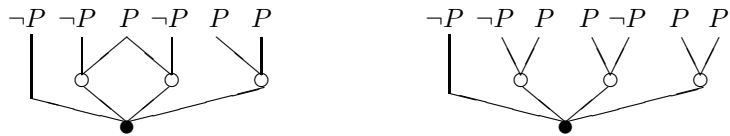
\put(10,8){{\bf Figure 3:} \lll\ proves the left but not the right cirquent}

\put(158,64){$\gneg P$}
\put(177,64){$\gneg P$}
\put(199,64){$P$}
\put(216,64){$P$}
\put(227,64){$\gneg P$}
\put(250,64){$P$}
\put(265,64){$P$}

\put(194,46){\line(-1,2){7}}
\put(194,46){\line(1,2){7}}
\put(194,44){\circle{5}}

\put(228,46){\line(-1,2){7}}
\put(228,46){\line(1,2){7}}
\put(228,44){\circle{5}}

\put(262,46){\line(-1,2){7}}
\put(262,46){\line(1,2){7}}
\put(262,44){\circle{5}}

\put(167,39){\line(0,1){22}}
\put(211,28){\line(-4,1){44}}
\put(211,28){\line(4,1){52}}
\put(211,28){\line(-6,5){16}}
\put(211,28){\line(6,5){16}}
\put(211,26){\circle*{5}}
\end{picture}
\end{center}

An (atomic-level) {\bf  instance} of a cirquent is the result of renaming (all, some or no) atoms in it. Here, of course, different occurrences (in the labels of different ports) of the same atom are required to be renamed into the same atom, but it is also possible that different atoms are renamed into the same atom. Example: the two cirquents of Figure 3 are instances of the two cirquents of Figure 2.

We noted above that the left cirquent of Figure 3 is provable because so is its more general predecessor from Figure 2. In other words, the former is provable because it is an instance of the latter which, in turn, has already been seen to have a proof. The following lemma generalizes this observation:

\begin{lemma}\label{ll1}
If a cirquent is provable, then so are all of its instances. 
\end{lemma}

\begin{proof} Consider an arbitrary cirquent $C$, and let $C'$ be an instance of it, resulting from renaming each atom $P$ of $C$ into an atom $P'$. 
Assume $\cal P$ is a proof of $C$. Note that no cirquent in $\cal P$ contains any atom that does not occur in $C$. So, let ${\cal P}'$ be the result of renaming
each atom $P$ into $P'$  in each cirquent of $\cal P$. It is not hard to see that ${\cal P}'$ is a proof of $C'$.  
\end{proof}

By a {\bf transition} we mean any binary relation $\cal T$ on cirquents. When $A{\cal T}B$, we say that $B$ {\bf follows} from $A$ by $\cal T$, and call $A$ and $B$ the {\bf premise} and the {\bf conclusion}  of the given application of the transition, respectively. Transitions are the same as rules of inference, only in a more relaxed sense than the strict sense of Section 3. Of course, every rule $\cal R$ of inference induces --- and can often be identified with --- a transition $\cal T$, such that $B$ follows from $A$ by $\cal T$ iff $B$ follows from $A$ by $\cal R$ with some (whatever) parameters. We may not always be very strict in terminologically differentiating between transitions and rules. 

A transition is said to be  {\bf strongly admissible} in a given system  if, whenever $B$ follows from $A$ by that transition, there is also a derivation of $B$ from $A$. And a transition is {\bf weakly admissible} iff, whenever $B$ follows from $A$ by that transition and $A$ is provable in the system, $B$ is also provable. 

One of the useful strongly admissible transitions is {\bf destandardization}. To obtain a premise from the conclusion $A$ of destandardization, we apply to $A$ --- in the bottom-up sense --- a series of globalizations until every non-root gate has exactly one parent; then we apply a series of deepenings until no conjunctive gate has conjunctive children and no disjunctive gate has disjunctive children; finally, we apply a series of lengthenings until there are no gates that have exactly one child.  It is easy to see that this procedure applied to $A$ yields a unique (modulo isomorphism) cirquent $B$, to which we will be referring as {\bf the standardization of $A$}. Then we say that such a $B$ follows from $A$ by destandardization. The same transition but with premise and conclusion interchanged we also call {\bf standardization}. Of course, standardization, just like destandardization,  is among the strongly admissible transitions in \lll.

Another strongly admissible transition for which we have a special name is {\bf restructuring}, which works in both top-down and bottom-up directions. We say that a cirquent $B$ follows from a cirquent $A$ by restructuring, or that ``$A$ can be restructured into $B$'', if there is a derivation of $B$ from $A$ that uses only restructuring rules. Destandardization and standardization are thus special cases of restructuring.   

One more strongly admissible transition that we are going to rely on is {\bf trade}. It is given by 
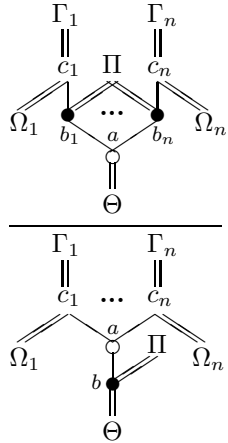
\begin{center} \begin{picture}(47,187)

\put(2,150){$c_1$}
\put(36,150){$c_n$}
\put(6,134){\circle*{5}}
\put(18,134){\bf ...}
\put(40,134){\circle*{5}}
\put(22,120){\line(3,2){18}}
\put(24,120){\line(-3,2){18}}
\put(19,150){$\Pi$}
\put(6,136){\line(0,1){10}}
\put(40,136){\line(0,1){10}}
\put(40,136){\line(-3,2){16}}
\put(37,136){\line(-3,2){16}}
\put(6,136){\line(3,2){16}}
\put(9,136){\line(3,2){16}}
\put(23,118){\circle{5}}
\put(22,106){\line(0,1){10}}
\put(24,106){\line(0,1){10}}
\put(19,97){$\Theta$}

\put(-16,92){\line(1,0){80}}

\put(2,63){$c_1$}
\put(18,63){\bf ...}
\put(36,63){$c_n$}
\put(23,48){\line(3,2){16}}
\put(23,48){\line(-3,2){16}}
\put(36,44){$\Pi$}
\put(23,46){\circle{5}}
\put(23,34){\line(0,1){10}}
\put(21,32){\line(3,2){16}}
\put(24,32){\line(3,2){16}}
\put(-13,48){\line(3,2){16}}
\put(-10,48){\line(3,2){16}}
\put(55,48){\line(-3,2){16}}
\put(58,48){\line(-3,2){16}}
\put(-13,136){\line(3,2){16}}
\put(-10,136){\line(3,2){16}}
\put(56,136){\line(-3,2){16}}
\put(59,136){\line(-3,2){16}}
\put(4,69){\line(0,1){10}}
\put(6,69){\line(0,1){10}}
\put(39,69){\line(0,1){10}}
\put(41,69){\line(0,1){10}}
\put(4,156){\line(0,1){10}}
\put(6,156){\line(0,1){10}}
\put(39,156){\line(0,1){10}}
\put(41,156){\line(0,1){10}}
\put(23,32){\circle*{5}}
\put(22,20){\line(0,1){10}}
\put(24,20){\line(0,1){10}}
\put(19,11){$\Theta$}
\put(-16,39){$\Omega_1$}
\put(53,39){$\Omega_n$}
\put(0,82){$\Gamma_1$}
\put(36,82){$\Gamma_n$}
\put(-16,128){$\Omega_1$}
\put(54,128){$\Omega_n$}
\put(0,169){$\Gamma_1$}
\put(36,169){$\Gamma_n$}
\put(21,124){\footnotesize $a$}
\put(21,51){\footnotesize $a$}
\put(15,30){\footnotesize $b$}
\put(3,124){\footnotesize $b_1$}
\put(38,124){\footnotesize $b_n$}
\end{picture}\end{center}
where $n\geq 0$, and the conventions of Section 3 continue to be in force, except that, as we see, here the number of parameters is not fixed (so that trade is not just a single rule in the strict sense of Section 3 but rather a collection of rules, one for each $n\in\{0,1,2,\ldots\}$). Below is an example of an application of trade where both the premise and the conclusion can be written as hyperformulas:

\begin{center} \begin{picture}(78,36)
\put(0,25){$(P\mld \overline{R})\mlc(Q\mld\overline{R})$}
\put(0,19){\line(1,0){78}}
\put(11,6){$(P\mlc Q)\mld R$}
\end{picture}\end{center}
Referring to the nodes of the above cirquents by the corresponding subformulas, in this application of trade  $n=2$, $\Pi=\{R\}$, the other peripheral parameters are empty,  
$c_1=P$, $c_2=Q$, $b=(P\mlc Q)\mld R$, $b_1=P\mld \overline{R}$, $b_2=Q\mld \overline{R}$, and $a$ is $P\mlc Q$ in the conclusion and $(P\mld \overline{R})\mlc(Q\mld\overline{R})$ in the premise.

The $a$ gate of the conclusion of trade will be said to be the {\bf principal gate} of a given application of this rule.  Note that when $n=0$, i.e., when the principal gate is childless, trade is simply 

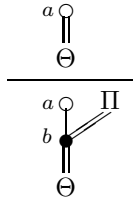
\begin{center} \begin{picture}(47,85)

\put(22,71){\circle{5}}
\put(21,59){\line(0,1){10}}
\put(23,59){\line(0,1){10}}
\put(18,50){$\Theta$}

\put(0,45){\line(1,0){47}}

\put(35,34){$\Pi$}
\put(22,36){\circle{5}}
\put(22,24){\line(0,1){10}}
\put(20,22){\line(3,2){16}}
\put(23,22){\line(3,2){16}}
\put(22,22){\circle*{5}}
\put(21,10){\line(0,1){10}}
\put(23,10){\line(0,1){10}}
\put(18,1){$\Theta$}
\put(13,69){\footnotesize $a$}
\put(13,34){\footnotesize $a$}
\put(13,21){\footnotesize $b$}
\end{picture}\end{center}
whose strong admissibility is seen from the following transformations:

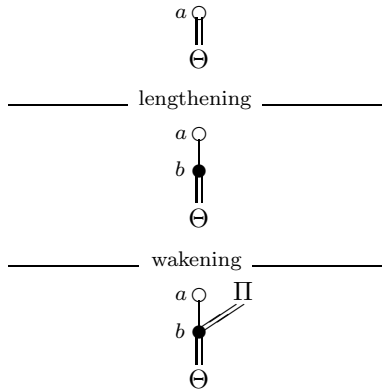
\begin{center} \begin{picture}(47,22)
\put(22,21){\circle{5}}
\put(21,9){\line(0,1){10}}
\put(23,9){\line(0,1){10}}
\put(18,0){$\Theta$}
\put(13,19){\footnotesize $a$}
\end{picture}\end{center}

\begin{center} \begin{picture}(47,50)
\put(-50,47){\line(1,0){45}}
\put(46,47){\line(1,0){45}}
\put(-1,47){\footnotesize lengthening}
\put(22,36){\circle{5}}
\put(22,24){\line(0,1){10}}
\put(22,22){\circle*{5}}
\put(21,10){\line(0,1){10}}
\put(23,10){\line(0,1){10}}
\put(18,1){$\Theta$}
\put(13,20){\footnotesize $b$}
\put(13,34){\footnotesize $a$}
\end{picture}\end{center}

\begin{center} \begin{picture}(47,50)
\put(-50,47){\line(1,0){49}}
\put(42,47){\line(1,0){49}}
\put(4,47){\footnotesize wakening}
\put(35,34){$\Pi$}
\put(22,36){\circle{5}}
\put(22,24){\line(0,1){10}}
\put(20,22){\line(3,2){16}}
\put(23,22){\line(3,2){16}}
\put(22,22){\circle*{5}}
\put(21,10){\line(0,1){10}}
\put(23,10){\line(0,1){10}}
\put(18,1){$\Theta$}
\put(13,20){\footnotesize $b$}
\put(13,34){\footnotesize $a$}
\end{picture}\end{center}
And the following transformations show the strong admissibility of trade for the case $n\geq 1$:  

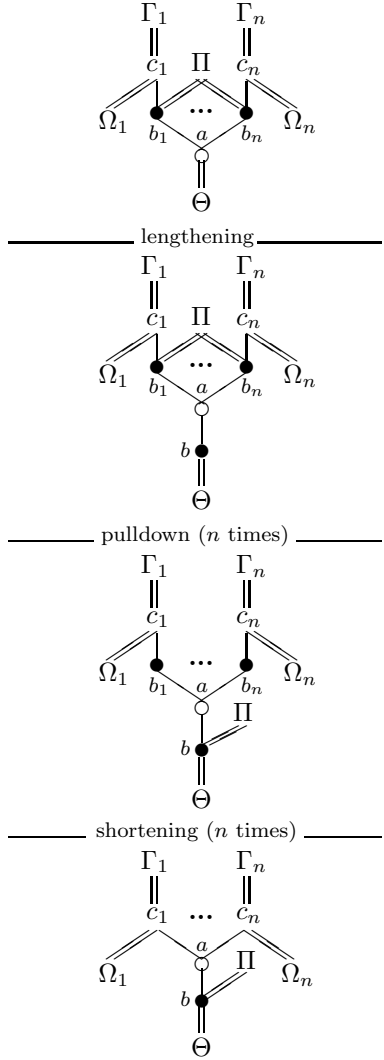
\begin{center} \begin{picture}(47,88)
\put(2,53){$c_1$}
\put(36,53){$c_n$}
\put(6,37){\circle*{5}}
\put(18,37){\bf ...}
\put(40,37){\circle*{5}}
\put(22,23){\line(3,2){18}}
\put(24,23){\line(-3,2){18}}
\put(19,53){$\Pi$}
\put(40,39){\line(0,1){10}}
\put(6,39){\line(0,1){10}}
\put(40,39){\line(-3,2){16}}
\put(37,39){\line(-3,2){16}}
\put(6,39){\line(3,2){16}}
\put(9,39){\line(3,2){16}}
\put(23,21){\circle{5}}
\put(22,9){\line(0,1){10}}
\put(24,9){\line(0,1){10}}
\put(19,0){$\Theta$}

\put(-13,39){\line(3,2){16}}
\put(-10,39){\line(3,2){16}}
\put(56,39){\line(-3,2){16}}
\put(59,39){\line(-3,2){16}}
\put(4,59){\line(0,1){10}}
\put(6,59){\line(0,1){10}}
\put(39,59){\line(0,1){10}}
\put(41,59){\line(0,1){10}}

\put(-16,31){$\Omega_1$}
\put(54,31){$\Omega_n$}
\put(0,72){$\Gamma_1$}
\put(36,72){$\Gamma_n$}
\put(21,27){\footnotesize $a$}
\put(3,27){\footnotesize $b_1$}
\put(38,27){\footnotesize $b_n$}
\end{picture}\end{center}

\begin{center} \begin{picture}(47,102)
\put(-50,101){\line(1,0){47}}
\put(0,101){\footnotesize lengthening}
\put(44,101){\line(1,0){47}}
\put(2,70){$c_1$}
\put(36,70){$c_n$}
\put(6,54){\circle*{5}}
\put(18,54){\bf ...}
\put(40,54){\circle*{5}}
\put(22,40){\line(3,2){18}}
\put(24,40){\line(-3,2){18}}
\put(19,70){$\Pi$}
\put(40,56){\line(0,1){10}}
\put(6,56){\line(0,1){10}}
\put(40,56){\line(-3,2){16}}
\put(37,56){\line(-3,2){16}}
\put(6,56){\line(3,2){16}}
\put(9,56){\line(3,2){16}}

\put(-13,56){\line(3,2){16}}
\put(-10,56){\line(3,2){16}}
\put(56,56){\line(-3,2){16}}
\put(59,56){\line(-3,2){16}}
\put(4,76){\line(0,1){10}}
\put(6,76){\line(0,1){10}}
\put(39,76){\line(0,1){10}}
\put(41,76){\line(0,1){10}}
\put(-16,48){$\Omega_1$}
\put(54,48){$\Omega_n$}
\put(0,89){$\Gamma_1$}
\put(36,89){$\Gamma_n$}
\put(21,44){\footnotesize $a$}
\put(3,44){\footnotesize $b_1$}
\put(38,44){\footnotesize $b_n$}
\put(15,20){\footnotesize $b$}

\put(23,38){\circle{5}}

\put(22,9){\line(0,1){10}}
\put(24,9){\line(0,1){10}}
\put(19,0){$\Theta$}

\put(23,22){\circle*{5}}
\put(23,25){\line(0,1){10}}
\end{picture}\end{center}

\begin{center} \begin{picture}(47,102)
\put(-50,101){\line(1,0){32}}
\put(-15,101){\footnotesize pulldown ($n$ times)}
\put(59,101){\line(1,0){32}}
\put(2,70){$c_1$}
\put(36,70){$c_n$}
\put(6,54){\circle*{5}}
\put(18,54){\bf ...}
\put(40,54){\circle*{5}}
\put(22,40){\line(3,2){18}}
\put(24,40){\line(-3,2){18}}
\put(35,33){$\Pi$}
\put(40,56){\line(0,1){10}}
\put(6,56){\line(0,1){10}}
\put(23,24){\line(2,1){14}}
\put(26,24){\line(2,1){14}}

\put(-13,56){\line(3,2){16}}
\put(-10,56){\line(3,2){16}}
\put(56,56){\line(-3,2){16}}
\put(59,56){\line(-3,2){16}}
\put(4,76){\line(0,1){10}}
\put(6,76){\line(0,1){10}}
\put(39,76){\line(0,1){10}}
\put(41,76){\line(0,1){10}}
\put(-16,48){$\Omega_1$}
\put(54,48){$\Omega_n$}
\put(0,89){$\Gamma_1$}
\put(36,89){$\Gamma_n$}
\put(21,44){\footnotesize $a$}
\put(3,44){\footnotesize $b_1$}
\put(38,44){\footnotesize $b_n$}
\put(15,20){\footnotesize $b$}
\put(23,38){\circle{5}}

\put(22,9){\line(0,1){10}}
\put(24,9){\line(0,1){10}}
\put(19,0){$\Theta$}

\put(23,22){\circle*{5}}
\put(23,25){\line(0,1){10}}
\end{picture}\end{center}

\begin{center} \begin{picture}(47,94)
\put(-50,94){\line(1,0){29}}
\put(-17,94){\footnotesize shortening ($n$ times)}
\put(62,94){\line(1,0){29}}
\put(2,63){$c_1$}
\put(18,63){\bf ...}
\put(36,63){$c_n$}
\put(23,48){\line(3,2){16}}
\put(23,48){\line(-3,2){16}}
\put(36,44){$\Pi$}
\put(23,46){\circle{5}}
\put(23,34){\line(0,1){10}}
\put(21,32){\line(3,2){16}}
\put(24,32){\line(3,2){16}}
\put(-13,48){\line(3,2){16}}
\put(-10,48){\line(3,2){16}}
\put(55,48){\line(-3,2){16}}
\put(58,48){\line(-3,2){16}}

\put(4,69){\line(0,1){10}}
\put(6,69){\line(0,1){10}}
\put(39,69){\line(0,1){10}}
\put(41,69){\line(0,1){10}}

\put(23,32){\circle*{5}}
\put(22,20){\line(0,1){10}}
\put(24,20){\line(0,1){10}}
\put(19,11){$\Theta$}
\put(-16,39){$\Omega_1$}
\put(53,39){$\Omega_n$}
\put(0,82){$\Gamma_1$}
\put(36,82){$\Gamma_n$}

\put(21,51){\footnotesize $a$}
\put(15,30){\footnotesize $b$}

\end{picture}\end{center}

\section{Semantics} In this section we define a semantics for cirquents, termed {\bf abstract resource semantics}. This is a generalization, to all cirquents,  of the same-name semantics introduced in \cite{Jap06} for the earlier mentioned special, ``shallow'', class of cirquents. 

The main purpose of a good semantics should be serving as a bridge between the real world and the otherwise meaningless formal expressions of logic. And, correspondingly, the value of a semantics should be judged by how successfully it achieves this purpose, which, in turn, depends on how naturally and adequately it formalizes certain basic intuitions connecting logic with the outside world. Such intuitions behind abstract resource semantics have been amply explained and illustrated in Section 8 of \cite{Jap06}. The reader is strongly recommended to get familiar with that piece of literature in order to appreciate the claim of abstract resource semantics that it is a ``real'' semantics of resources, formalizing the resource philosophy traditionally (and,  as argued in \cite{Jap06},  somewhat wrongly) associated with linear logic and its variations. In this paper we just provide formal definitions, only occasionally making very brief intuitive comments, and otherwise fully relying on \cite{Jap06} for extended explanations of the intuitions, motivations and philosophy underlying the semantics. 

Abstract resource semantics can be seen as a conservative generalization of the semantics of classical logic from circuits to all cirquents. The starting point of the semantics is the concept of a {\bf truth assignment} for a given cirquent $C$. This is a function that assigns one of two values --- {\em true} or {\em false} --- to each port of $C$. Any such function $f$ is a legitimate truth assignment, including the cases when $f$ assigns different truth values to ports that have identical labels. Intuitively this is perfectly meaningful in the world of resources because, say, one $25c$-port (slot of the vending machine) may receive a true coin while the other $25c$-port may receive a false coin or no coin at all.

Each truth assignment for a cirquent extends from its ports to all gates and the cirquent itself in the following, expected, way:  
  
\begin{itemize}
\item A disjunctive gate is true iff it has at least one true child. 
\item A conjunctive gate is true iff so are all of its children.
\item The cirquent is true iff so is its root.
\end{itemize}

An {\bf allocation} for a given cirquent $C$  is an unordered pair $\{a,b\}$ of ports of $C$ with opposite labels (labels $P$ and $\gneg P$ for some --- the same --- atom $P$). And an {\bf arrangement} for $C$ is any set of pairwise disjoint allocations for $C$. We call the  condition requiring  all allocations to be disjoint the {\bf monogamicity condition}. 

Let $C$ be a cirquent, $f$ a truth assignment for $C$, and $\alpha$ an arrangement for $C$. We say that $f$ is {\bf consistent with} $\alpha$ iff, for every allocation $\{a,b\}\in\alpha$,  $f(a)\not=f(b)$. That is, if ports $a$ and $b$ are allocated to each other (meaning that $\{a,b\}\in\alpha$), a truth assignment consistent with $\alpha$ should assign opposite truth values to $a$ and $b$.\footnote{In \cite{Jap06}, a weaker condition was adopted, according to which at least one (but possibly both) of the nodes $a,b$ should be assigned {\em true}. It is easy to see that either condition yields the same concept of validity, so that this difference is unimportant.} And we say that $\alpha$ is {\bf validating}\footnote{The corresponding term used in \cite{Jap06} was ``{\em trivializing}''.} (for $C$) iff $C$ is true under every truth assignment consistent with $\alpha$. To see an example, consider the following cirquent:

\begin{center} \begin{picture}(213,90)

\put(7,78){\footnotesize $1$}
\put(29,78){\footnotesize $2$}
\put(66,78){\footnotesize $3$}
\put(89,78){\footnotesize $4$}
\put(125,78){\footnotesize $5$}
\put(149,78){\footnotesize $6$}
\put(185,78){\footnotesize $7$}
\put(210,78){\footnotesize $8$}
\put(0,67){$\gneg P$}
\put(21,67){$\gneg P$}
\put(19,54){\line(-1,1){10}}
\put(19,54){\line(1,1){10}}
\put(19,52){\circle*{5}}
\put(60,67){$\gneg P$}
\put(81,67){$\gneg P$}
\put(79,54){\line(-1,1){10}}
\put(79,54){\line(1,1){10}}
\put(79,52){\circle*{5}}
\put(48,40){\circle{5}}
\put(48,42){\line(-4,1){30}}
\put(48,42){\line(4,1){30}}

\put(124,67){$P$}
\put(147,67){$P$}
\put(139,54){\line(-1,1){10}}
\put(139,54){\line(1,1){10}}
\put(139,52){\circle*{5}}
\put(184,67){$P$}
\put(207,67){$P$}
\put(199,54){\line(-1,1){10}}
\put(199,54){\line(1,1){10}}
\put(199,52){\circle*{5}}
\put(168,40){\circle{5}}
\put(168,42){\line(-4,1){30}}
\put(168,42){\line(4,1){30}}

\put(109,28){\line(-6,1){60}}
\put(109,28){\line(6,1){60}}
\put(109,26){\circle*{5}}
\put(47,8){{\bf Figure 4:} A valid cirquent}
\end{picture}\end{center}

\noindent And consider the following two arrangements for this cirquent:  
\[\alpha\ =\ \{\{1,5\},\ \{2,6\},\ \{3,7\},\ \{4,8\}\};\]
\[\beta \ =\ \{\{1,5\},\ \{2,7\},\ \{3,6\},\ \{4,8\}\}.\]
Here $\alpha$ is not a validating arrangement. Specifically, the following truth assignment $f$, while obviously consistent with $\alpha$, makes the cirquent false:
\[\mbox{$f(1)=f(2)=f(7)=f(8)=${\em false}; \ \ $f(3)=f(4)=f(5)=f(6)=${\em true}.}\]
This assignment, on the other hand, is not consistent with $\beta$. Moreover, with some thought, one can see that no truth assignment that makes the cirquent of Figure 4 false can be consistent with $\beta$. This means that $\beta$, unlike $\alpha$, {\em is} a validating arrangement for that cirquent. 

As explained and illustrated in \cite{Jap06}, our formal concept of an allocation corresponds to the intuition of allocating one resource to another: a coin ($25c$) to a coin-receiving  slot ($\gneg 25c$), a memory ($100MB$) to a memory-requesting process ($\gneg 100MB$), a power source ($100w$) to a power-consuming utensil ($\gneg 100w$), an USB-interface external device ($USB$) to an USB port of a computer ($\gneg USB$), etc. A  justification behind the monogamicity condition for arrangements is that if a resource $a$ is used by (allocated to) $b$, then it cannot be also used by (allocated to) another $c\not=b$. 
And the intuition behind a validating arrangement is that of a successful resource-management strategy/solution.

\begin{definition}\label{march19}
We say that a cirquent is a {\bf valid}\footnote{The corresponding term used in \cite{Jap06} was ``{\em trivial}''.} (in abstract resource semantics) iff there is a validating arrangement for it. 
 \end{definition}

For example, naming the ports (in the left to right order) of the cirquents of Figure 2 by the consecutive numbers $1,2,...$, the set 
\[\{\{1,3\}, \ \{2,5\}, \ \{4,6\}\}\]
is a validating arrangement for the left cirquent, which makes that cirquent valid. On the other hand, with a little thought, one can see that no possible arrangement for the right cirquent of the same figure is validating, so that that cirquent is not valid. Note that the monogamicity condition plays a crucial role in precluding the right cirquent of Figure 2 from being valid: because of monogamicity, one of the two $P$-ports of that cirquent will have to be left unallocated. 

The above arrangement is also validating for the left cirquent of Figure 3. And, again with some (this time a little more) thought, one can see that the right cirquent of the same figure has no validating arrangement, thus being non-valid.

The following cirquent is not valid, either, even though it looks so ``similar'' to the valid cirquent of \mbox{Figure 4:}

\begin{center} \begin{picture}(280,85)

\put(0,73){$\gneg P$}
\put(21,73){$\gneg P$}
\put(19,60){\line(-1,1){10}}
\put(19,60){\line(1,1){10}}
\put(19,58){\circle*{5}}
\put(50,73){$\gneg P$}
\put(71,73){$\gneg P$}
\put(69,60){\line(-1,1){10}}
\put(69,60){\line(1,1){10}}
\put(69,58){\circle*{5}}
\put(100,73){$\gneg P$}
\put(121,73){$\gneg P$}
\put(119,60){\line(-1,1){10}}
\put(119,60){\line(1,1){10}}
\put(119,58){\circle*{5}}

\put(69,43){\circle{5}}
\put(69,45){\line(-5,1){50}}
\put(69,45){\line(5,1){50}}
\put(69,45){\line(0,1){10}}

\put(156,73){$P$}
\put(178,73){$P$}
\put(169,60){\line(-1,1){10}}
\put(169,60){\line(1,1){10}}
\put(169,58){\circle*{5}}
\put(206,73){$P$}
\put(227,73){$P$}
\put(219,60){\line(-1,1){10}}
\put(219,60){\line(1,1){10}}
\put(219,58){\circle*{5}}

\put(256,73){$P$}
\put(277,73){$P$}
\put(269,60){\line(-1,1){10}}
\put(269,60){\line(1,1){10}}
\put(269,58){\circle*{5}}
\put(219,43){\circle{5}}
\put(219,45){\line(-5,1){50}}
\put(219,45){\line(5,1){50}}
\put(219,45){\line(0,1){10}}

\put(144,28){\line(-6,1){75}}
\put(144,28){\line(6,1){75}}
\put(144,26){\circle*{5}}
\put(72,8){{\bf Figure 5:} A non-valid cirquent}
\end{picture}\end{center}

Again, as illustrated in \cite{Jap06}, valid cirquents  are resource-management problems (such as the problem of getting a candy from a vending machine with a given collection of available coins) that have successful solutions. And among the potential practical values of sound and complete deductive systems such as our \lll\ or the system {\bf CL5} constructed in \cite{Jap06} is that they present tools for systematically finding such solutions.

Let $C$ be a circuit, and $\mu$ be the set of all possible allocations for $C$. This set satisfies the monogamicity condition and hence is an arrangement for $C$, because the latter, being a circuit, has at most one $P$-port and at most one $\gneg P$-port for any given atom $P$. Of course, any other arrangement for $C$ will be a subset of $\mu$, which, in turn, easily implies that $C$ is valid if and only if the arrangement $\mu$ is validating for it. In other words, 
\begin{equation}\label{march20}
\mbox{\em $C$ is valid iff it is true under every truth assignment consistent with $\mu$.} 
\end{equation}   
But notice that truth assignments consistent with $\mu$ are nothing but truth assignments in the kind old classical sense, meaning functions that assign opposite truth values to $P$ and $\gneg P$, for any atom $P$. In view of (\ref{march20}), we thus find that:
\begin{fact}\label{m20}
Validity in our sense and validity (tautologicity) in the classical sense mean the same for circuits, and hence for formulas of classical logic understood as circuits according to the stipulations of Section 2.
\end{fact} 

So, as promised, our abstract resource semantics is a conservative extension of classical semantics from circuits to all cirquents. 
 
\begin{lemma}\label{ma20}
A cirquent is valid iff it is an instance of a valid circuit.
\end{lemma}

\begin{proof} Consider an arbitrary cirquent $A$.

($\Leftarrow$:) Assume $A$ is an instance of a valid circuit $B$. Let $\alpha$ be a validating arrangement for $B$. It is not hard to see that then the same $\alpha$ is also a validating arrangement for $A$, so that $A$ is valid. 

($\Rightarrow$:) Suppose $A$ is valid. Let $\alpha$ be a validating arrangement for it. Let then $B$ be the result of renaming the occurrences of atoms within the labels of the ports of $A$ in such a way that no atom (with or without a negation) occurs in the labels of two different ports $a$ and $b$ unless $\{a,b\}\in\alpha$, in which case both occurrences of the (same) atom in the labels of $a,b$ within $A$ are renamed into the same atom. Thus, $B$ is a circuit. 
With a little thought, one can also see that the same arrangement $\alpha$ remains validating for $B$, so that $B$ is, in fact, a valid circuit. Now, it remains to notice that $A$ is an instance of $B$. 
\end{proof}

\begin{theorem}\label{th2}
A cirquent is provable in $\lll$ iff it is valid in abstract resource semantics. 
\end{theorem}

\begin{proof} Let $C$ be an arbitrary cirquent.\vspace{5pt}

{\em Soundness}: Assume $\lll\vdash C$. Let $\cal P$ be a $\lll$-proof of $C$. Let us rename the atoms (occurring in the  labels) of the cirquents of $\cal P$  in such a way that every time coupling  is used, the atom $P$ it introduces is new, in the sense that the premise does not have any ports labeled with $P$ or $\gneg P$. Let us further rename the atoms of $\cal P$ so that every time weakening introduces some new ports (ones that did not exist in the premise), the labels of such ports are new and different from each other.  Let us call the resulting sequence of cirquents ${\cal P}'$. It is not hard to see that then ${\cal P}'$ is a proof of a cirquent $C'$ such that $C$ is an instance of $C'$. The axiom $\circ$ is, of course, a circuit, and every rule of inference obviously preserves the circuit property (``{\em circuitness}'') of cirquents except coupling and weakening. But with the conditions that we imposed on those two rules when obtaining ${\cal P}'$ from $\cal P$, all of the cirquents in ${\cal P}'$ are circuits. It is also easy to see that all inference rules preserve truth and hence validity of circuits.  Thus, all cirquents in ${\cal P}'$ are valid circuits, including $C'$. And, as $C$ is an instance of $C'$, Lemma \ref{ma20} implies that $C$ is valid.\vspace{5pt}

{\em Completeness}: Assume $C$ is valid. Then, by Lemma \ref{ma20}, there is a valid circuit $C'$ such that $C$ is an instance of $C'$. Fix this $C'$. We are going to show that $C'$ is provable, which, by Lemma \ref{ll1}, immediately implies that $C$ is also provable. 

We construct, bottom-up, a proof of $C'$ as follows. First, applying (bottom-up) destandardization, we proceed from $C'$ to its standardization $J$. Let us fix  a ``sufficiently large'' integer $s$, such that $s\geq 2$ and $s$ exceeds the total number of nodes in $J$. Given a cirquent $H$, we define an {\bf active gate} of $H$ to be a disjunctive gate $a$ of $H$ that has no disjunctive ancestors.  We define the {\bf rank} of such an $a$ to be $s^{m}$, where $m$ is the number of conjunctive gates that are descendants of $a$. And we define the {\bf rank} of $H$ to be the sum of the ranks of its active gates. 

Our construction of a proof of $C'$ continues upward from $J$ as follows. We repeat the following two steps while there are non-root conjunctive gates 
in the current (topmost in the so far constructed proof) cirquent: 
  
{\em Step 1.} Pick an arbitrary conjunctive child $c$ of an arbitrary active node of the current cirquent, and apply (bottom-up) trade so that $c$ is the principal gate of the application. 

{\em Step 2.} Apply  (bottom-up) destandardization to the resulting cirquent.

With some thought, one can see that every time the above two steps are performed, the rank of the current (topmost) cirquent decreases.  Hence, the procedure will end sooner or later, and the resulting cirquent $D$ will have no non-root conjunctive gates. It is easy to see that destandardization and trade preserve both validity and circuitness (not only in the top-down but also) in the bottom-up direction. So, $D$ is a valid circuit. The pathological case when $D$ has no conjunctive gates is simple and we do not consider it here. Otherwise, $D$ is a circuit with a conjunctive root, where each child of the root is a disjunctive gate and each grandchild of the root is a port, as shown in the following example:

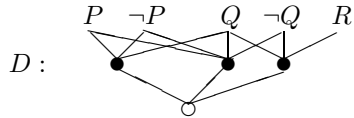
\begin{center} \begin{picture}(128,56)

\put(0,20){$D:$}

\put(96,38){$\gneg Q$}
\put(80,38){$Q$}
\put(83,25){\line(0,1){10}}
\put(83,25){\line(2,1){20}}
\put(83,25){\line(-3,1){32}}
\put(83,25){\line(-5,1){53}}
\put(83,23){\circle*{5}}
\put(104,23){\circle*{5}}
\put(104,25){\line(0,1){10}}
\put(104,25){\line(-2,1){20}}
\put(104,25){\line(2,1){20}}
\put(122,38){$R$}

\put(45,38){$\gneg P$}
\put(28,38){$P$}
\put(41,25){\line(4,1){42}}
\put(41,25){\line(1,1){10}}
\put(41,25){\line(-1,1){10}}
\put(41,23){\circle*{5}}
\put(68,8){\line(-2,1){26}}
\put(68,8){\line(1,1){17}}
\put(68,8){\line(3,1){36}}
\put(68,6){\circle{5}}

\end{picture}\end{center}

The validity of $D$ obviously implies that among the children of each disjunctive gate is a pair of ports with opposite labels. We select one such pair for each disjunctive gate, and remove all other children using weakenings.\footnote{At this point we see that weakening in \lll\ can be restricted to {\bf port weakening}, i.e., the version of weakening that permits deleting only arcs to ports (rather than any nodes). This is relevant to the claim made in Subsection 6.2.} Now, the resulting cirquent $E$ has a conjunctive gate at its root, whose every child is a disjunctive gate with exactly two children, with those two children being ports with opposite labels, as shown below:
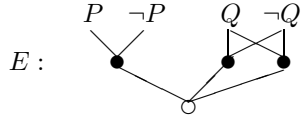
\begin{center} \begin{picture}(128,56)
\put(0,20){$E:$}
\put(96,38){$\gneg Q$}
\put(80,38){$Q$}
\put(83,25){\line(0,1){10}}
\put(83,25){\line(2,1){20}}
\put(83,23){\circle*{5}}
\put(104,23){\circle*{5}}
\put(104,25){\line(0,1){10}}
\put(104,25){\line(-2,1){20}}

\put(45,38){$\gneg P$}
\put(28,38){$P$}
\put(41,25){\line(1,1){10}}
\put(41,25){\line(-1,1){10}}
\put(41,23){\circle*{5}}
\put(68,8){\line(-2,1){26}}
\put(68,8){\line(1,1){17}}
\put(68,8){\line(3,1){36}}
\put(68,6){\circle{5}}

\end{picture}\end{center}

Furthermore, $E$, of course, inherits circuitness from $D$. And $E$'s being a circuit obviously implies that whenever two disjunctive gates share a child, they share both of their children. Applying (bottom-up) localizations to $E$, we proceed from $E$ to $F$, where $F$ is just like $E$, only without any sharing of children between different disjunctive gates:
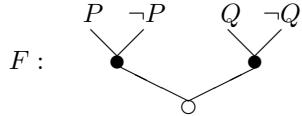
\begin{center} \begin{picture}(128,56)
\put(0,20){$F:$}
\put(96,38){$\gneg Q$}
\put(80,38){$Q$}
\put(93,25){\line(1,1){10}}
\put(93,25){\line(-1,1){10}}
\put(93,23){\circle*{5}}
\put(45,38){$\gneg P$}
\put(28,38){$P$}
\put(41,25){\line(1,1){10}}
\put(41,25){\line(-1,1){10}}
\put(41,23){\circle*{5}}
\put(68,8){\line(-2,1){26}}
\put(68,8){\line(2,1){26}}
\put(68,6){\circle{5}}
\end{picture}\end{center}

Now, applying (bottom-up) couplings to $F$, we replace in it each disjunctive gate by a childless conjunctive gate, obtaining a cirquent $G$ where all nodes are conjunctive gates:
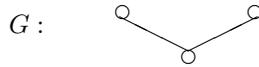
\begin{center} \begin{picture}(128,56)
\put(0,15){$G:$}

\put(93,23){\circle{5}}
\put(43,23){\circle{5}}
\put(68,8){\line(-2,1){26}}
\put(68,8){\line(2,1){26}}
\put(68,6){\circle{5}}
\end{picture}\end{center}

 Applying (bottom-up) to $G$ a series of deepenings yields the axiom cirquent $\circ$.
\end{proof}

An alternative proof of the completeness of \lll\ could rely on the forthcoming Theorem \ref{march23}. The latter, in view of the known completeness of the system {\bf G} considered there, implies that, for every tautological formula $F$ of classical logic, $\lll\vdash \underline{F}$. The cirquent $J$ constructed in our proof of Theorem  \ref{th2} can be seen to be $\underline{F}$ for some tautology $F$ and hence, in view of Theorem \ref{march23}, \lll-provable. However, such a proof, albeit shorter, would not be as direct as the one presented above. 

It should be remembered that, as noted earlier, the initial impulse to cirquent calculus was given by the needs of computability logic. Therefore, this paper would not be complete without officially establishing a connection between the latter and \lll. The original semantics of computability logic deals with formulas rather than cirquents. And, as shown in \cite{Jap06}, the class of formulas (in the sense of our Section 2) valid in computability logic coincides with the class of formulas valid in abstract resource semantics. This, in view of Theorem \ref{th2},  means that:

\begin{theorem}\label{march25}
A formula (in our present sense) is valid in computability logic iff it --- seen as a tree-like cirquent according to the stipulations of Section 2 --- is provable in \lll.
\end{theorem}

\cite{Jap06} further showed 
how to  extend the semantics of computability logic from formulas to cirquents. While ``cirquents'' there only meant special sorts of cirquents in our present, more general, sense, the generalization of the semantics of computability logic outlined in \cite{Jap06} almost automatically extends to all cirquents in our present sense as well: details can be very easily filled by anyone familiar with computability logic. And we claim  without a proof  that, with this generalized semantics of computability logic in mind, Theorem \ref{march25} can  be  strengthened by replacing ``formula'' with ``cirquent''.

Those familiar with computability logic will also remember that the language of the latter has two sorts of atoms: $P,Q,R,S,\ldots$,  called {\em general}, and 
$p,q,r,s,\ldots$, called {\em elementary}. The two sorts of atoms have two different semantic interpretations, which  result in a resource-conscious logical behavior of general atoms and classical behavior of elementary atoms. In this paper, which is notationally fully synchronized with computability logic, we have been using uppercase rather than lowercase letters for atoms. Hence, ``formula'' in Theorem \ref{march25}, as a formula of computability logic, is to be understood as one where all atoms are general. But, according to the following claim that we further make without a proof, \lll\ in fact captures a much more expressive fragment of computability logic than implied by Theorem \ref{march25}:

\begin{claim}\label{april1}
Let $F$ be a formula of the $\gneg,\mlc,\mld$-fragment of the language of computability logic, which may contain either sorts of atoms. For simplicity, here we assume  that $F$ is written in a form where $\gneg$ is only applied to atoms. Let then $\breve{F}$ be the cirquent represented --- according to the stipulations of Section 2 --- by the hyperformula obtained from $F$ through overlining all elementary (but not general) atoms and their negations, with $p,q,\ldots$, along with $P,Q,\ldots$, now treated as ordinary atoms of the language of $\lll$. Then $F$ is valid in computability logic iff $\lll\vdash \breve{F}$.   
\end{claim} 

\section{Other deep cirquent calculus systems}  

\subsection{A symmetric version of \lll} The {\bf dual} of a given inference rule is obtained by interchanging premise with conclusion and conjunctive gates with disjunctive gates. Each restructuring rule comes together with its dual, as those rules work in both directions and for either sort of gates. 

System \llls\ that we define here is a fully symmetric version of \lll, obtained by adding to the latter the duals of the main rules:
\begin{center} \begin{picture}(330,215)
\put(100,201){DUALS OF THE MAIN RULES:}

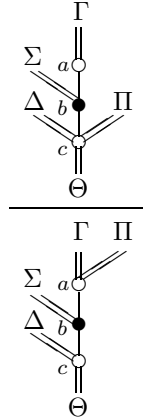
\put(318,142){\footnotesize $a$}
\put(318,59){\footnotesize $a$}
\put(318,42){\footnotesize $b$}
\put(318,125){\footnotesize $b$}
\put(318,27){\footnotesize $c$}
\put(318,110){\footnotesize $c$}

\put(156,62){\footnotesize $a$}
\put(156,112){\footnotesize $a$}

\put(6,83){\footnotesize $a$}
\put(6,112){\footnotesize $a$}
\put(2,139){\footnotesize $b$}
\put(26,139){\footnotesize $c$}

\put(-12,175){\bf Cocoupling}
\put(14,85){\circle*{5}}
\put(15,73){\line(0,1){10}}
\put(13,73){\line(0,1){10}}
\put(10,65){$\Theta$}
\put(0,90){\line(1,0){31}}
\put(18,129){$\gneg P$}
\put(0,129){$P$}
\put(14,116){\line(1,1){10}}
\put(14,116){\line(-1,1){10}}
\put(14,114){\circle{5}}
\put(15,102){\line(0,1){10}}
\put(13,102){\line(0,1){10}}
\put(10,94){$\Theta$}

\put(131,175){\bf Coweakening}
\put(161,80){$\Gamma$}
\put(160,45){$\Theta$}
\put(165,53){\line(0,1){10}}
\put(163,53){\line(0,1){10}}
\put(165,67){\line(0,1){10}}
\put(163,67){\line(0,1){10}}
\put(150,90){\line(1,0){31}}
\put(164,114){\circle{5}}
\put(165,102){\line(0,1){10}}
\put(163,102){\line(0,1){10}}
\put(160,94){$\Theta$}
\put(172,129){$\Delta$}
\put(151,129){$\Gamma$}
\put(166,116){\line(1,1){10}}
\put(163,116){\line(1,1){10}}
\put(163,116){\line(-1,1){10}}
\put(166,116){\line(-1,1){10}}
\put(164,65){\circle{5}}

\put(294,175){\bf Copulldown}

\put(324,161){$\Gamma$}
\put(339,127){$\Pi$}
\put(325,146){\line(0,1){12}}
\put(327,146){\line(0,1){12}}
\put(327,117){\line(3,2){12}}
\put(328,115){\line(3,2){15}}

\put(326,144){\circle{5}}
\put(326,131){\line(0,1){10}}
\put(324,131){\line(-3,2){16}}
\put(327,131){\line(-3,2){16}}
\put(305,127){$\Delta$}
\put(326,129){\circle*{5}}
\put(326,117){\line(0,1){10}}
\put(324,115){\line(-3,2){15}}
\put(325,117){\line(-3,2){12}}
\put(305,144){$\Sigma$}
\put(326,115){\circle{5}}
\put(325,103){\line(0,1){10}}
\put(327,103){\line(0,1){10}}
\put(322,94){$\Theta$}

\put(300,90){\line(1,0){50}}

\put(324,78){$\Gamma$}
\put(327,63){\line(0,1){10}}
\put(325,63){\line(0,1){10}}
\put(326,61){\circle{5}}
\put(305,60){$\Sigma$}
\put(326,48){\line(0,1){10}}
\put(324,46){\line(-3,2){16}}
\put(325,48){\line(-3,2){13}}
\put(305,44){$\Delta$}
\put(339,78){$\Pi$}
\put(326,46){\circle*{5}}
\put(326,34){\line(0,1){10}}
\put(324,32){\line(-3,2){16}}
\put(325,34){\line(-3,2){13}}
\put(325,63){\line(3,2){16}}
\put(328,63){\line(3,2){16}}
\put(326,32){\circle{5}}
\put(325,20){\line(0,1){10}}
\put(327,20){\line(0,1){10}}
\put(322,11){$\Theta$}
\end{picture}\end{center}

It is easy to see that each of the above three rules preserves  validity.  Therefore, in view of the already proven completeness, these rules are weakly admissible in \lll.

The {\bf negation} $\gneg C$ of a given cirquent $C$ is obtained by changing the label of each port to its opposite ($P$ to $\gneg P$ and vice versa), and changing the type (conjunctive/disjunctive) of each gate to the other type. 

The rule of cocoupling can also be called {\em cut}, specifically, {\bf port cut}. It would not be hard to show that cut remains weakly admissible in \lll\ when extended from ports $P,\gneg P$ to any subcirquents $A,\gneg A$. In fact, non-port cut is strongly admissible in \llls, for it easily (=polynomially) reduces to the port (``atomic'') version as is the case in the calculus of structures (see \cite{Gug07,Brunnen}). An interesting question to which at present we have no answer is whether cut can be eliminated without an exponential increase of proof sizes. This question is known to have a negative answer for ordinary sequent calculus.  

The top-down symmetry in the style of  the one enjoyed by \llls\ was first achieved and exploited within the framework of the calculus of structures (see, again, \cite{Gug07,Brunnen}). Such a symmetry generates a number of nice effects, some similar to those enjoyed by natural deduction systems.  Below we observe only one such effect. 

A {\bf refutation} of a given cirquent $C$ is a derivation of $\bullet$ from $C$. When such a derivation exists, $C$ is said to be {\bf refutable}. The following fact --- which, note, does not hold for \lll\ --- is obvious in view of the full symmetry of the rules of \llls:

\begin{fact}\label{f2}
In \llls, a cirquent is provable iff its negation is refutable. 
\end{fact} 

Unlike \lll, however, \llls\ is {\em non-analytic}, in any reasonable sense of this word.  Often in the literature analyticity is just understood as enjoying the {\em subformula property}, according to which everything in the premise of any given application of any of the rules of the system is a subformula of (some formula of) the conclusion. The subformula property is meaningful for sequent calculi because there the premises and the conclusion are not formulas but rather collections (sequences, multisets or sets) of formulas. But in cirquent calculus, where the premise is a single cirquent and so is the conclusion, the subformula (``subcirquent'') property hardly makes any sense.  Indeed, if it is understood literally --- as the requirement that everything in the premise be a subcirquent of ``something in the conclusion'', then simply the whole premise itself would have to be a subcirquent of the conclusion. This would fully retard any cirquent calculus system, essentially limiting its rules to the one that (in the bottom-up view) just deletes the root and jumps to one of its children. 

And it is not only cirquent calculus where the subformula property is no longer meaningful. The same holds for deep inference systems in general, such as the calculus of structures. For this reason, \cite{Gug07} uses the term ``analytic'' in a more relaxed sense, simply meaning the absence of cut, substitution, extension or rules in the style of our coweakening. The common undesirable feature of those rejected rules  is that, when moving from a conclusion to a premise, they introduce some new components, as opposed to the rules deemed in \cite{Gug07} analytic (and all rules of \lll\ would also qualify as analytic by similar standards), which merely regroup some already existing components without creating new components. What ``components'' or ``regrouping'' should exactly mean here, however, certainly does require some additional and probably nontrivial explanations. To summarize, there appears to be no well-agreed-upon concept of anaiticity in the literature. 

To avoid accusations of taking excessive terminological liberties, here we introduce the new term ``{\em interface analyticity}'', whose meaning well might be the best that one can achieve in an attempt to define a cirquent-calculus counterpart of the more traditional meaning of the word ``analyticity''. 

Following \cite{Jap06}, by the {\bf interface} of a given cirquent $C$ we mean the set of all of its ports. Intuitively, this is the visible part of the resource $C$, such as the collection of all input/output ports on the back and front panels of one's personal computer. This collection indeed presents the active ``interface''  of the resource, with the rest of it --- the gates and internal wiring, that is --- being fixed, hidden and unavailable  in the process of resource management, which, as we remember, means setting up allocations between ports (and by no means between gates). 

Imagine a circuit optimization problem. Its typical goal would be generating a better circuit that, however, computes the same Boolean function --- and hence has the same collection of inputs (same interface) --- as the original one. There are certain quite similar intuitive reasons for wanting rules of inference to preserve --- more precisely, not to expand --- the interface of the conclusion when moving to a premise.  

Having noted this, we say that a rule of inference is {\em interface-analytic}, or {\bf i-analytic} for short, iff, in any application of the rule,  the interface of  the premise is a subset of that of the conclusion (with the labels of all ports preserved).  And a system is i-analytic iff all of its rules are so. Note that \lll\ is i-analytic. On the other hand, the rules of cocoupling (cut) and coweakening of {\bf CL8S} are not i-analytic. Nor would be the rules of substitution (\cite{Finger}) or extension (\cite{CR}) if they were present in whatever form in our system. The same can be said about the rule of contraction, traditionally considered analytic. As a matter of fact, one could question the compliance of contraction with our normal, no matter how vague, intuition of analyticity. That is  because, when moving from  conclusion to premise, contraction {\em does} introduce some new material, even if only in the form of new copies of old (sub)formulas. Yet, this
non-analytic behavior of contraction is not noticeable in sequent calculus,
because, when used ``reasonably'',
contraction, while certainly introducing new material from the perspective of the whole {\em proof tree}, 
 does not really do so from the perspective of
any particular {\em branch} of that tree. Here by ``using contraction reasonably'' we mean applying it (in the bottom-up view of proofs) only before
using $\mlc$-introduction, to just make sure that each branch of the proof tree
gets its own copies of side formulas.
But in cirquent calculus or deep inference systems in general, where all branches are
combined within one cirquent or formula, contraction loses its apparent analytic innocence. In any case, unlike the formula-based deep inference approaches such as the calculus of structures, fortunately there is no need for contraction in cirquent calculus. If this rule (in whatever precise form) was adopted by \lll, it would certainly stop being i-analytic.

\subsection{Versions with the locality property}
Certain easy modifications of \lll\ or \llls\ yield versions that are {\em local}, meaning that each inference rule only affects a bounded portion of the cirquent. More precisely, a local rule modifies (deletes, creates, or changes the label in the case of nodes) only a bounded number of nodes and arcs when moving from  premise to conclusion or vice versa. Locality is a desirable property in computer implementations. The only reason why in this paper we have not chosen local axiomatizations has been striving to minimize bureaucracy. 

To see what we mean by ``easy modifications'', let us just consider weakening and pulldown as two examples. 

Weakening is not local because the number of the arcs of the conclusion that it can delete is not bounded. But nothing can be easier than to ``fix'' this problem. Specifically, we could adopt a  new --- local --- version of weakening that deletes exactly one arc. That is, the $\Delta$ parameter of weakening now would be required to be a singleton. Then, an application of the old weakening rule that deletes $n$ arcs can be simulated with $n$ applications of the new weakening rule.
Furthermore, as pointed out in a footnote when proving Theorem \ref{th2}, weakening can be further restricted by requiring the deleted arc to be pointing at a port rather than any node.  This would eliminate the possibility that deleting an arc may result in an unbounded chain of further deletions of orphaned nodes. 

Similarly, pulldown is not local as it is allowed to move around an unbounded number of arcs. We could start requiring that only a single arc be moved, that is, requiring the $\Pi$ parameter to be a singleton. Just as in the case of weakening, an application of the old rule of pulldown can then always be simulated by several applications of the new, local version of it.   

\subsection{Weakening the weakening rule} The resource philosophy associated with \lll\ and \llls\ is that one cannot use more resources than available. A more radical position is that one also has to use all available resources (nothing should be ``wasted''). Under this extreme philosophy familiar from linear logic,  the weakening rule and its dual coweakening become wrong. Removing these rules could as well be necessary when constructing systems for relevance logic. 

However, mechanically deleting weakening (and its dual, if present) from a given system may result in throwing out the baby with the bath water. So, rather than discarding the rule altogether as done in linear logic, one would apparently want to simply replace weakening by  certain weaker versions  of it ---  versions that, on one hand, are consistent with the above radical resource philosophy and, on the other hand, allow us to retain all innocent principles. Reasonable candidates for such a replacement for weakening and coweakening are the following rules:

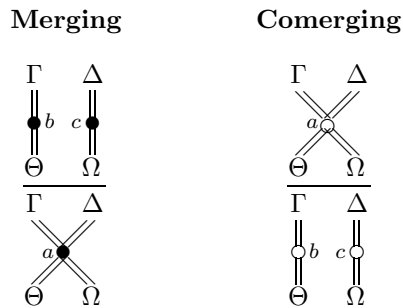
\begin{center} \begin{picture}(141,135)

\put(-5,115){\bf Merging}
\put(1,94){$\Gamma$}
\put(22,94){$\Delta$}
\put(0,10){$\Theta$}
\put(22,10){$\Omega$}
\put(3,18){\line(1,1){10}}
\put(6,18){\line(1,1){10}}
\put(24,18){\line(-1,1){10}}
\put(27,18){\line(-1,1){10}}
\put(5,81){\line(0,1){10}}
\put(3,81){\line(0,1){10}}
\put(0,55){\line(1,0){31}}
\put(4,79){\circle*{5}}
\put(5,67){\line(0,1){10}}
\put(3,67){\line(0,1){10}}
\put(0,59){$\Theta$}
\put(22,45){$\Delta$}
\put(01,45){$\Gamma$}
\put(17,32){\line(1,1){10}}
\put(14,32){\line(1,1){10}}
\put(13,32){\line(-1,1){10}}
\put(16,32){\line(-1,1){10}}
\put(15,30){\circle*{5}}
\put(27,81){\line(0,1){10}}
\put(25,81){\line(0,1){10}}
\put(26,79){\circle*{5}}
\put(27,67){\line(0,1){10}}
\put(25,67){\line(0,1){10}}
\put(22,59){$\Omega$}

\put(88,115){\bf Comerging}
\put(101,94){$\Gamma$}
\put(122,94){$\Delta$}
\put(103,32){\line(0,1){10}}
\put(105,32){\line(0,1){10}}
\put(125,32){\line(0,1){10}}
\put(127,32){\line(0,1){10}}
\put(115,78){\circle{5}}
\put(125,18){\line(0,1){10}}
\put(127,18){\line(0,1){10}}
\put(103,18){\line(0,1){10}}
\put(105,18){\line(0,1){10}}
\put(100,59){$\Theta$}
\put(122,59){$\Omega$}

\put(100,55){\line(1,0){31}}

\put(101,45){$\Gamma$}
\put(122,45){$\Delta$}
\put(114,81){\line(1,1){10}}
\put(117,81){\line(1,1){10}}
\put(113,81){\line(-1,1){10}}
\put(116,81){\line(-1,1){10}}
\put(104,30){\circle{5}}
\put(126,30){\circle{5}}
\put(103,67){\line(1,1){10}}
\put(106,67){\line(1,1){10}}
\put(124,67){\line(-1,1){10}}
\put(127,67){\line(-1,1){10}}
\put(100,10){$\Theta$}
\put(122,10){$\Omega$}

\put(8,77){\footnotesize $b$}
\put(18,77){\footnotesize $c$}
\put(7,28){\footnotesize $a$}

\put(108,28){\footnotesize $b$}
\put(118,28){\footnotesize $c$}
\put(107,77){\footnotesize $a$}
\end{picture}\end{center}

Let us look at {\em Blass's \cite{Bla92} principle}
\[\bigl((\gneg P\mld \gneg Q)\mlc(\gneg R\mld\gneg S)\bigr)\mld\bigl((P\mld R)\mlc(Q\mld S)\bigr).\]
Resources are perfectly balanced in this formula, and there are hardly any good reasons for rejecting it even from the most radical resource-philosophical point of view. It is therefore embarrassing that Blass's principle is not provable in linear logic and not even in affine logic: as shown in \cite{Jap06}, every proof of it in ordinary sequent calculus would require both contraction and weakening. This formula cannot be proven in \lll\ without weakening, either. Its provability can be however retained with the fully resource-fair rule of merging instead of weakening, as shown below: 
  
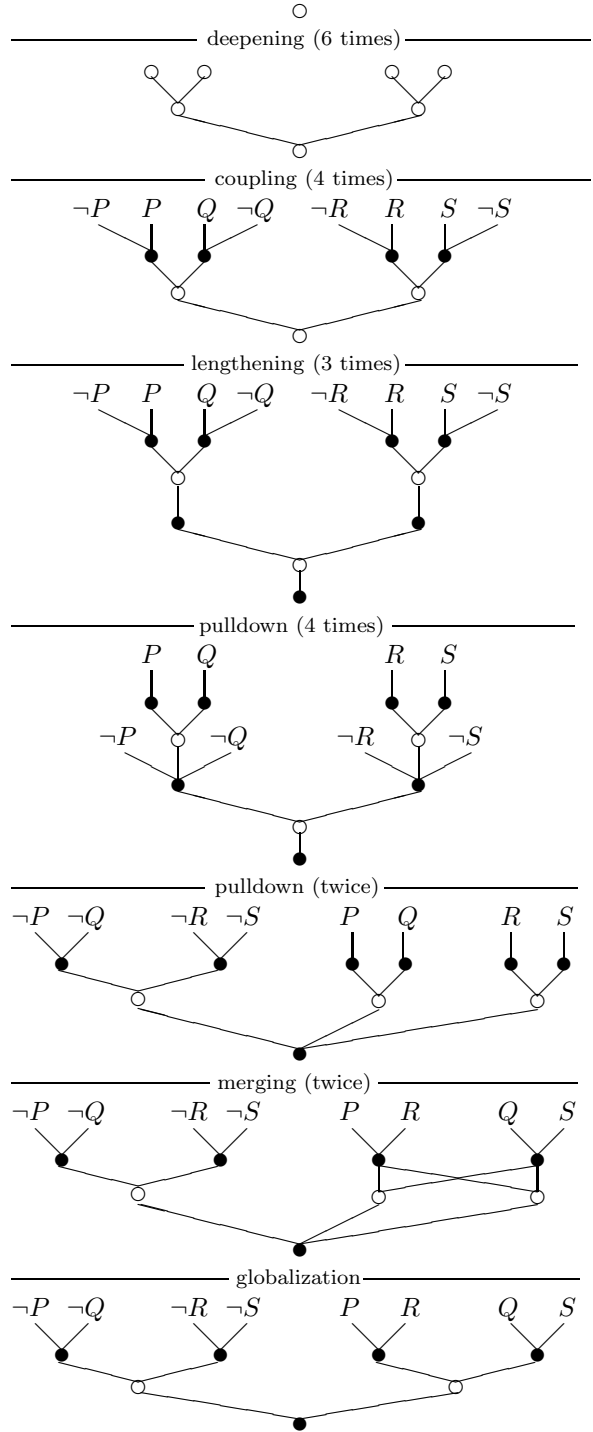
\begin{center} \begin{picture}(213,20)
\put(109,0){\circle{5}}
\end{picture}\end{center}
\begin{center} \begin{picture}(213,42)

\put(0,42){\line(1,0){70}}
\put(74,40){\footnotesize deepening (6 times)}
\put(149,42){\line(1,0){70}}
\put(109,0){\circle{5}}
\put(109,2){\line(-4,1){46}}
\put(109,2){\line(4,1){46}}
\put(53,30){\circle{5}}
\put(73,30){\circle{5}}
\put(144,30){\circle{5}}
\put(164,30){\circle{5}}
\put(154,16){\circle{5}}
\put(154,18){\line(-1,1){10}}
\put(154,18){\line(1,1){10}}
\put(63,16){\circle{5}}
\put(63,18){\line(1,1){10}}
\put(63,18){\line(-1,1){10}}

\end{picture}\end{center}

\begin{center} \begin{picture}(213,59)

\put(0,59){\line(1,0){73}}
\put(77,57){\footnotesize coupling (4 times)}
\put(146,59){\line(1,0){73}}
\put(23,45){$\gneg P$}
\put(85,45){$\gneg Q$}
\put(53,32){\line(-2,1){20}}
\put(73,32){\line(2,1){20}}
\put(113,45){$\gneg R$}
\put(176,45){$\gneg S$}
\put(144,32){\line(-2,1){20}}
\put(164,32){\line(2,1){20}}
\put(109,0){\circle{5}}
\put(109,2){\line(-4,1){46}}
\put(109,2){\line(4,1){46}}

\put(49,45){$P$}
\put(70,45){$Q$}
\put(73,32){\line(0,1){10}}
\put(53,32){\line(0,1){10}}
\put(53,30){\circle*{5}}
\put(73,30){\circle*{5}}
\put(141,45){$R$}
\put(162,45){$S$}
\put(164,32){\line(0,1){10}}
\put(144,32){\line(0,1){10}}
\put(144,30){\circle*{5}}
\put(164,30){\circle*{5}}
\put(154,16){\circle{5}}
\put(154,18){\line(-1,1){10}}
\put(154,18){\line(1,1){10}}
\put(63,16){\circle{5}}
\put(63,18){\line(1,1){10}}
\put(63,18){\line(-1,1){10}}

\end{picture}\end{center}

\begin{center} \begin{picture}(213,88)

\put(0,88){\line(1,0){65}}
\put(68,86){\footnotesize lengthening (3 times)}
\put(149,88){\line(1,0){65}}
\put(23,74){$\gneg P$}
\put(85,74){$\gneg Q$}
\put(53,61){\line(-2,1){20}}
\put(73,61){\line(2,1){20}}
\put(63,28){\circle*{5}}
\put(113,74){$\gneg R$}
\put(176,74){$\gneg S$}
\put(144,61){\line(-2,1){20}}
\put(164,61){\line(2,1){20}}
\put(154,28){\circle*{5}}
\put(109,12){\circle{5}}
\put(109,14){\line(-4,1){46}}
\put(109,14){\line(4,1){46}}

\put(49,74){$P$}
\put(70,74){$Q$}
\put(73,61){\line(0,1){10}}
\put(53,61){\line(0,1){10}}
\put(53,59){\circle*{5}}
\put(73,59){\circle*{5}}
\put(141,74){$R$}
\put(162,74){$S$}
\put(164,61){\line(0,1){10}}
\put(144,61){\line(0,1){10}}
\put(144,59){\circle*{5}}
\put(164,59){\circle*{5}}
\put(154,45){\circle{5}}
\put(154,30){\line(0,1){12}}
\put(154,47){\line(-1,1){10}}
\put(154,47){\line(1,1){10}}
\put(63,45){\circle{5}}
\put(63,30){\line(0,1){12}}
\put(63,47){\line(1,1){10}}
\put(63,47){\line(-1,1){10}}

\put(109,2){\line(0,1){8}}
\put(109,0){\circle*{5}}
\end{picture}\end{center}

\begin{center} \begin{picture}(213,88)

\put(0,88){\line(1,0){69}}
\put(71,86){\footnotesize pulldown (4 times)}
\put(144,88){\line(1,0){69}}
\put(33,43){$\gneg P$}
\put(75,43){$\gneg Q$}
\put(63,30){\line(-2,1){20}}
\put(63,30){\line(2,1){20}}
\put(63,28){\circle*{5}}
\put(123,43){$\gneg R$}
\put(165,43){$\gneg S$}
\put(154,30){\line(-2,1){20}}
\put(154,30){\line(2,1){20}}
\put(154,28){\circle*{5}}
\put(109,12){\circle{5}}
\put(109,14){\line(-4,1){46}}
\put(109,14){\line(4,1){46}}

\put(49,74){$P$}
\put(70,74){$Q$}
\put(73,61){\line(0,1){10}}
\put(53,61){\line(0,1){10}}
\put(53,59){\circle*{5}}
\put(73,59){\circle*{5}}
\put(141,74){$R$}
\put(162,74){$S$}
\put(164,61){\line(0,1){10}}
\put(144,61){\line(0,1){10}}
\put(144,59){\circle*{5}}
\put(164,59){\circle*{5}}
\put(154,45){\circle{5}}
\put(154,30){\line(0,1){12}}
\put(154,47){\line(-1,1){10}}
\put(154,47){\line(1,1){10}}
\put(63,45){\circle{5}}
\put(63,30){\line(0,1){12}}
\put(63,47){\line(1,1){10}}
\put(63,47){\line(-1,1){10}}

\put(109,2){\line(0,1){8}}
\put(109,0){\circle*{5}}
\end{picture}\end{center}

\begin{center} \begin{picture}(213,63)

\put(0,63){\line(1,0){73}}
\put(77,61){\footnotesize pulldown (twice)}
\put(141,63){\line(1,0){73}}
\put(0,49){$\gneg P$}
\put(21,49){$\gneg Q$}
\put(19,36){\line(-1,1){10}}
\put(19,36){\line(1,1){10}}
\put(19,34){\circle*{5}}
\put(60,49){$\gneg R$}
\put(81,49){$\gneg S$}
\put(79,36){\line(-1,1){10}}
\put(79,36){\line(1,1){10}}
\put(79,34){\circle*{5}}
\put(48,21){\circle{5}}
\put(48,24){\line(-4,1){30}}
\put(48,24){\line(4,1){30}}

\put(124,49){$P$}
\put(146,49){$Q$}
\put(148,36){\line(0,1){10}}
\put(129,36){\line(0,1){10}}
\put(129,34){\circle*{5}}
\put(149,34){\circle*{5}}
\put(185,49){$R$}
\put(206,49){$S$}
\put(189,36){\line(0,1){10}}
\put(209,36){\line(0,1){10}}
\put(209,34){\circle*{5}}
\put(189,34){\circle*{5}}
\put(199,20){\circle{5}}
\put(199,22){\line(-1,1){10}}
\put(199,22){\line(1,1){10}}
\put(139,20){\circle{5}}

\put(139,22){\line(1,1){10}}
\put(139,22){\line(-1,1){10}}

\put(109,2){\line(-4,1){61}}
\put(109,2){\line(6,1){90}}
\put(109,2){\line(2,1){30}}
\put(109,0){\circle*{5}}
\end{picture}\end{center}

\begin{center} \begin{picture}(213,63)

\put(0,63){\line(1,0){75}}
\put(78,61){\footnotesize merging (twice)}
\put(139,63){\line(1,0){75}}
\put(0,49){$\gneg P$}
\put(21,49){$\gneg Q$}
\put(19,36){\line(-1,1){10}}
\put(19,36){\line(1,1){10}}
\put(19,34){\circle*{5}}
\put(60,49){$\gneg R$}
\put(81,49){$\gneg S$}
\put(79,36){\line(-1,1){10}}
\put(79,36){\line(1,1){10}}
\put(79,34){\circle*{5}}
\put(48,21){\circle{5}}
\put(48,24){\line(-4,1){30}}
\put(48,24){\line(4,1){30}}

\put(124,49){$P$}
\put(147,49){$R$}
\put(139,36){\line(-1,1){10}}
\put(139,36){\line(1,1){10}}
\put(139,34){\circle*{5}}
\put(184,49){$Q$}
\put(207,49){$S$}
\put(199,36){\line(-1,1){10}}
\put(199,36){\line(1,1){10}}
\put(199,34){\circle*{5}}
\put(199,20){\circle{5}}
\put(199,22){\line(-6,1){60}}
\put(199,22){\line(0,1){10}}
\put(139,20){\circle{5}}

\put(139,22){\line(6,1){60}}
\put(139,22){\line(0,1){10}}

\put(109,2){\line(-4,1){61}}
\put(109,2){\line(6,1){90}}
\put(109,2){\line(2,1){30}}
\put(109,0){\circle*{5}}
\end{picture}\end{center}

\begin{center} \begin{picture}(213,59)

\put(0,59){\line(1,0){82}}
\put(85,57){\footnotesize globalization}
\put(133,59){\line(1,0){82}}
\put(0,45){$\gneg P$}
\put(21,45){$\gneg Q$}
\put(19,32){\line(-1,1){10}}
\put(19,32){\line(1,1){10}}
\put(19,30){\circle*{5}}
\put(60,45){$\gneg R$}
\put(81,45){$\gneg S$}
\put(79,32){\line(-1,1){10}}
\put(79,32){\line(1,1){10}}
\put(79,30){\circle*{5}}
\put(48,18){\circle{5}}
\put(48,20){\line(-4,1){30}}
\put(48,20){\line(4,1){30}}

\put(124,45){$P$}
\put(147,45){$R$}
\put(139,32){\line(-1,1){10}}
\put(139,32){\line(1,1){10}}
\put(139,30){\circle*{5}}
\put(184,45){$Q$}
\put(207,45){$S$}
\put(199,32){\line(-1,1){10}}
\put(199,32){\line(1,1){10}}
\put(199,30){\circle*{5}}
\put(168,18){\circle{5}}
\put(168,20){\line(-4,1){30}}
\put(168,20){\line(4,1){30}}

\put(109,6){\line(-6,1){60}}
\put(109,6){\line(6,1){60}}
\put(109,4){\circle*{5}}
\end{picture}\end{center}

\subsection{Cirquents with many roots} Some future treatments may call for considering cirquents that allow multiple roots (parentless nodes).  For example, the methods of cirquent calculus could be potentially used in verifying circuit equivalence, optimizing circuits, or other related problems arising in digital design.  And it should be remembered that circuits in actual computer hardware typically have not only multiple inputs (ports), but also multiple outputs (roots). Of course, there can also be many other reasons, including theoretical ones, for studying these more general sorts of cirquents.

\subsection{Cirquents with additional sorts of gates and arcs} As we already know, the introduction of cirquent calculus was originally motivated by the needs of computability logic. Cirquent calculus in the form presented in this paper captures only the modest $(\gneg,\mlc,\mld)$-fragment of  computability logic though. Extending cirquent calculus so as to accommodate incrementally more expressive fragments of computability logic would require considering cirquents with gates for choice connectives, and gates and/or arcs for recurrence connectives. Accounting for the more recently (\cite{JapIC}) introduced non-commutative sequential operators of computability logic would also require linearly ordering the outgoing edges of the corresponding gates. There is a tremendous amount of interesting and challenging work to do in this direction.

\section{\lll\ versus sequent calculus and shallow cirquent calculus systems}
This section is devoted to certain aspects of the relation between \lll\ and
 Gentzen-style sequent calculus systems, as well as  the shallow cirquent calculus systems {\bf CL5} and {\bf CCC} presented in \cite{Jap06}. 

Specifically, we first want to compare \lll\ with the classical cut-free sequent calculus system {\bf G} defined below. One difference that we already know is the greater expressiveness of \lll. But even if we are only concerned with objects that the languages of both systems can express --- Boolean functions presented in the form of classical formulas or (the corresponding) circuits, that is --- \lll\ still has distinctive advantages, related to efficiency. In Section 8 we will see  the existence of polynomial size \lll-proofs for the {\em pigeonhole principle}, the class of tautologies known to have only exponential size proofs in {\bf G} or similar systems. To appreciate this point, it would be necessary to also show that, on the other hand, no class of tautologies admits in  {\bf G} considerably shorter proofs than in \lll. In other words, we need to see that \lll\ can {\bf p-simulate} {\bf G}, meaning that there is a polynomial function $p$ such that, for any formula $F$ of classical logic, whenever $F$ has a {\bf G}-proof of size $n$, it --- more precisely, the cirquent $\underline{F}$ --- also has a \lll-proof of size $\leq p(n)$. Then and only then we can officially declare that \lll\ offers an exponential speedup (in proof efficiency) over {\bf G}. 

System {\bf G}  deals with {\bf sequents} understood as nonempty finite sets  of formulas. This version is known to be equivalent --- in the strong sense of mutual p-simulation --- to the probably more common versions of cut-free sequent calculi for classical logic where sequents are sequences or multisets (rather than sets) of formulas. An advantage of {\bf G} over such systems is the absence of structural rules. 

Below $\Gamma$ stands for any set of formulas, $P$ for any atom, and $E,F$ for any formulas. Following the standard practice, an expression such as 
``$\Gamma,E,F$'' should be understood as $\Gamma\cup\{E,F\}$. 

The axioms of {\bf G} are any sequents of the form 
\[\Gamma,\gneg P,P,\]
in addition to which the system (only) has the following two rules of inference:

\begin{center}\begin{picture}(274,68)
\put(20,53){\bf $\mld$-introduction} 
\put(40,30){$\Gamma,E,F$}
\put(35,23){\line(1,0){38}}
\put(35,10){$\Gamma,E\mld F$}

\put(177,53){\bf $\mlc$-introduction}
\put(182,30){$\Gamma,E$}
\put(226,30){$\Gamma,F$}
\put(182,23){\line(1,0){61}}
\put(195,10){$\Gamma,E\mlc F$}
\end{picture}\end{center}

The definition of {\bf provability} of a sequent $\Gamma$ in {\bf G} is standard: this means existence of a tree of sequents --- called a {\bf proof tree} for $\Gamma$ --- with $\Gamma$ at its root, in which every leaf of the tree is an axiom and every non-leaf node follows from its child or children by one of the rules of ${\mathbf G}$. A formula $F$ is considered provable in {\bf G} iff $F$, viewed as a one-element sequent, is provable.

Since we will be dealing with complexity issues, we need to agree on what the size of a formula, cirquent, sequent, derivation or proof means. We assume some reasonable encoding (computer representation) of these objects to be fixed, and agree that the {\bf size} of any such object is  the amount of bits taken by its code when written in computer memory. It is understood that all ``reasonable'' encodings are polynomially equivalent (the differences in their efficiencies are at most polynomial, that is) and, since in this paper we only care about polynomiality versus exponentiality, it is not important which particular ``reasonable'' encoding we have in mind. 

\begin{theorem}\label{march23}
\lll\ p-simulates {\bf G}.
\end{theorem}
\begin{proof}
Consider an arbitrary {\bf G}-proof tree $\cal T$ for an arbitrary formula $F$. Below we describe a procedure for converting $\cal T$ into a \lll-proof ${\cal T}^*$ of $\underline{F}$. It will be clear from our description that the size of ${\cal T}^*$ is polynomial in the size of $\cal T$. 

By abuse of terminology, in the present proof we will be often identifying a node of $\cal T$ with the corresponding sequent, even though it should be remembered that the same sequent may be ``sitting'' at more than one node.  

We construct the \lll-proof ${\cal T}^*$ of $\underline{F}$  bottom-up. The last three  cirquents of ${\cal T}^*$ are \[\mbox{$\underline{\mlc\{\mld\{F\}\}}$, \ $\underline{\mld\{F\}}$ \ and \ $\underline{F}$.}\]
 $\underline{F}$ follows from its predecessor $\underline{\mld\{F\}}$ by shortening, and so does $\underline{\mld\{F\}}$ from its predecessor $\underline{\mlc\{\mld\{F\}\}}$. 

Thus, the topmost cirquent of the bottom fragment of ${\cal T}^*$ that we have constructed so far is $\underline{\mlc\{\mld\{F\}\}}$. Let us call this cirquent $A_1$. We {\bf associate} the root of $\cal T$ with the $\underline{\mld\{F\}}$ subcirquent of $\underline{\mlc\{\mld\{F\}\}}$. 

$A_1$ is only the first cirquent of a certain series $A_1,A_2,A_3,\ldots $ of cirquents that we are going to construct one after one and include in our evolving (in the upward direction) ${\cal T}^*$. Any such $A_i$ will look like  
\[\underline{\mlc\{\mld\{E_{1}^{1},\ldots,E_{k_1}^{1}\},\ \ldots,\ \mld\{E_{1}^{n},\ldots,E_{k_n}^{n}\}\}},\]
i.e., 
\[\underline{(E_{1}^{1}\mld\ldots\mld E_{k_1}^{1})\mlc \ \ldots \ \mlc (E_{1}^{n}\mld\ldots\mld E_{k_n}^{n})},\]
where with each conjunct  $\underline{(E_{1}^{j}\mld\ldots\mld E_{k_j}^{j})}$, as in $A_1$, is associated a node of $\cal T$ such that the sequent at that node is 
\[E_{1}^{j},\ldots, E_{k_j}^{j}.\]

We describe the way of generating the $A_i$s and including them in ${\cal T}^*$ inductively. $A_1$ has already been generated. Suppose now we have already constructed the bottom portion of ${\cal T}^*$ such that $A_i$ is the top cirquent. Further suppose that there is a conjunct of $A_i$ such that the associated node of $\cal T$ is not an axiom of {\bf G} (i.e., not a leaf of $\cal T$). We may assume here that the last conjunct 
\[E_{1}^{n}\mld\ldots\mld E_{k_n}^{n}\]
of $A_i$ is such.   How we proceed from $A_i$ upward in our construction of ${\cal T}^*$ depends on whether the associated sequent $E_{1}^{n},\ldots, E_{k_n}^{n}$ is obtained by $\mld$-introduction or $\mlc$-introduction in $\cal T$. 

Suppose $E_{1}^{n},\ldots, E_{k_n}^{n}$ is obtained by $\mld$-introduction, meaning that it looks like 
\begin{equation}\label{m28a}
E_{1}^{n},\ldots, E_{k_n-1}^{n}, G\mld H\end{equation}
and the premise is 
\begin{equation}\label{m28b} E_{1}^{n},\ldots, E_{k_n-1}^{n}, G, H.\end{equation}
We then choose $A_{i+1}$ to be the cirquent 
\[\underline{(E_{1}^{1}\mld\ldots\mld E_{k_1}^{1})\mlc  \ldots \mlc (E_{1}^{n-1}\mld\ldots\mld E_{k_{n-1}}^{n-1}) \mlc  (E_{1}^{n}\mld\ldots\mld E_{k_n-1}^{n}\mld G\mld H)}.\]
The nodes of $\cal T$ associated with the conjuncts of $A_{i+1}$ remain the same as in $A_i$, with the exception of the last ($n$th) conjunct, with which we now associate the premise (\ref{m28b}) of (\ref{m28a}).  Note that $A_{i}$, which is 
\[\underline{(E_{1}^{1}\mld\ldots\mld E_{k_1}^{1})\mlc  \ldots \mlc (E_{1}^{n-1}\mld\ldots\mld E_{k_{n-1}}^{n-1}) \mlc  (E_{1}^{n}\mld\ldots\mld E_{k_n-1}^{n}\mld (G\mld H))},\]
 follows from $A_{i+1}$ by deepening. So, we include $A_{i+1}$ in front (on top) of $A_i$ in our bottom-up construction of ${\cal T}^*$, and justify the transition from $A_{i+1}$ to $A_i$ by deepening. 

Suppose now $E_{1}^{n},\ldots, E_{k_n}^{n}$ is obtained by $\mlc$-introduction, meaning that it looks like 
\begin{equation}\label{m28c} E_{1}^{n},\ldots, E_{k_n-1}^{n}, G\mlc H\end{equation}
and the two premises of it in $\cal T$ are 
\begin{equation}\label{m28d} E_{1}^{n},\ldots, E_{k_n-1}^{n}, G\end{equation}
and 
\begin{equation}\label{m28e} E_{1}^{n},\ldots, E_{k_n-1}^{n}, H.\end{equation}
In this case we choose $A_{i+1}$ to be the cirquent 
\[\underline{(E_{1}^{1}\mld\ldots\mld E_{k_1}^{1})\mlc  \ldots \mlc (E_{1}^{n-1}\mld\ldots\mld E_{k_{n-1}}^{n-1})  \mlc (E_{1}^{n}\mld\ldots\mld E_{k_n-1}^{n}\mld G)\mlc (E_{1}^{n}\mld\ldots\mld E_{k_n-1}^{n}\mld H)}.\]
The nodes of $\cal T$ associated with the first $n-1$ conjuncts of $A_{i+1}$ remain the same as in $A_i$. And   with the last two conjuncts of $A_{i+1}$ we associate the premises (\ref{m28d}) and (\ref{m28e}) of (\ref{m28c}), respectively. 
It is not hard to see that $A_i$, which is
\[\underline{(E_{1}^{1}\mld\ldots\mld E_{k_1}^{1})\mlc  \ldots \mlc (E_{1}^{n-1}\mld\ldots\mld E_{k_{n-1}}^{n-1})  \mlc (E_{1}^{n}\mld\ldots\mld E_{k_n-1}^{n}\mld (G\mlc H))},\]
 follows from $A_{i+1}$ by trade in combination with some straightforward restructuring. So, we add the corresponding (bounded number of) cirquents together with the appropriate justifications in front (on top) of $A_i$, with the new top cirquent of our bottom-up construction of ${\cal T}^*$ now being $A_{i+1}$.

We continue extending ${\cal T}^*$ upward by adding new $A_i$s in the above way until we hit the point where the topmost $A_m$ is such that 
all nodes of $\cal T$ associated with its conjuncts are leaves. It is not hard to see that this $m$ would be nothing but the total number of nodes of $\cal T$. Thus, the now topmost cirquent of the evolving ${\cal T}^*$ is 
\[A_m\ =\ \underline{(E_{1}^{1}\mld\ldots\mld E_{k_1}^{1})\mlc \ \ldots \ \mlc (E_{1}^{n}\mld\ldots\mld E_{k_n}^{n})},\]
where each 
\[E_{1}^{j},\ldots.E_{k_j}^{j}\]
is an axiom of {\bf G} and hence contains at least one pair $P,\gneg P$ of opposite literals. 

We choose one such pair of literals in each conjunct of $A_m$, and delete the arcs to all other nodes from the corresponding disjunctive gate using (bottom up) a series of weakenings. This results in a cirquent  
\[B\ =\ \underline{(P_1\mld\gneg P_1)\mlc\ldots\mlc (P_n\mld \gneg P_n)}, \]
where each $P_j$ is an atom. Not all $P_i$ and $P_j$ with $i\not=j$ may be different atoms here though. If this is indeed the case, we further apply (bottom-up) a series of localizations to $B$ and get a cirquent 
\[C\ =\ (Q_1\mld\gneg Q_1)\mlc\ldots\mlc (Q_e\mld \gneg Q_e)\]
($e< n$), where each $Q_j$ is an atom (one of the old atoms $P_1,\ldots,P_e$) different from any $Q_i$ with $i\not=j$. 

Next we apply (bottom-up) coupling to $C$ $e$ times, which results in a cirquent where all non-root nodes are childless conjunctive gates. Such gates can be eliminated by applying (bottom-up) 
a series of deepenings, and we end up with the axiom cirquent $\circ$. 
\end{proof} 

In a similar way one could show that \lll\ p-simulates the cut-free versions of the multiplicative linear and affine logics. However, as already mentioned, those are not conservative fragments of \lll. For example, the \lll-provable Blass's principle (see Section 6.3), or the cirquent of Figure 4, are both expressible in the language of linear logic, but neither linear logic nor the stronger affine logic prove them. 

Furthermore, our proof of Theorem \ref{march23} can be rather easily modified into proofs of the facts that  \lll\ also p-simulates the shallow cirquent calculus systems {\bf CL5} and {\bf CCC} of \cite{Jap06}.  At the same time, the known proofs of the nonexistence of polynomial size proofs of the pigeonhole principle in {\bf G}-style systems can be modified so as to show the nonexistence of such proofs in {\bf CCC}. And a somewhat similar argument, based on a certain resource-conscious version of the pigeonhole principle (no literal has more than one occurrence), can be used to also show an exponential speedup over {\bf CL5} offered by \lll.    Thus, \lll\ is certainly an improvement over {\bf CCC} and {\bf CL5} from the perspective of efficiency. 

But there is a much more significant difference between our present approach and the approach taken in \cite{Jap06}. While \cite{Jap06} is the official birth place of the ideas of cirquent calculus and abstract resource semantics, the particular systems elaborated in detail in \cite{Jap06} stopped only half way on the road of fully and consistently materializing those ideas. This was related to the limited syntax adopted there, which was  a somewhat unnatural mixture of circuit-style and tree-style structures.  Specifically, as mentioned earlier, the depths of cirquents were limited to two, with the root of each such cirquent required to be a conjunctive gate and its  children required to be disjunctive gates. This was a significant limitation of expressiveness and, to partially compensate for it, the ``input'' nodes (grandchildren of the root) were allowed to be any formulas rather than only literals as in our present treatment. And so, possible sharing of children between different parents was taking place only at one single (root's children) level of cirquents. Even though \cite{Jap06} proved (Theorem 20) that shallow cirquents, unlike formulas, were sufficient to represent all {\em abstract resources} (which, roughly, are the same to abstract resource semantics as Boolean functions to the semantics of classical logic), such representations were generally very inefficient, essentially requiring every abstract resource to be expressed in conjunctive normal form.\footnote{It should be noted that valid cirquents in conjunctive normal form, where children may be shared between different disjunctive nodes, are not as trivial to prove as valid conjunctive-normal-form formulas in classical logic.} From classical logic we know that conjunctive normal forms, while complete as means of expressing all Boolean functions, can generally be exponentially longer than other, more relaxed representations.  Similar reasons apply to abstract resource semantics as well, meaning that the objects of our study (abstract resources) are exponentially harder to express --- let alone prove --- in {\bf CL5} or {\bf CCC} than in \lll.  

But the most decisive improvement of the present approach over the approach of \cite{Jap06} is turning classical logic into just a special fragment of the more general logic of resource, thus eliminating conflicts between the classical and resource-conscious views, with both the semantics and the syntax of \lll\ being  single unifying and reconciling frameworks for the two diverging philosophical traditions in logic. This was impossible to achieve under the shallow cirquent calculus approach of \cite{Jap06}, for the reason of the limitations of the expressive power of shallow cirquents. And this is exactly why \cite{Jap06} had to construct two different logics: one --- {\bf CCC} --- for classical semantics and the other ---  {\bf CL5} --- for abstract resource semantics, and correspondingly prove two separate completeness theorems. The two systems had the same language but different semantics, and disagreed on many principles expressible in that common language. Specifically, {\bf CCC} was properly  stronger than {\bf CL5}, obtained from the latter by adding (a cirquent calculus version of) contraction to it, the rule that we criticized a while ago as being not ``truly analytic''.

The main purpose of the present paper is to provide a starting point and an initial impulse for what (as the author wishes to hope) may become a new line of research in proof theory and resource logics --- namely, a proof theory and a resource semantics based on circuit-style (rather than formula-style) constructs.  \cite{Jap06}, with its
limited and not fully consistent (in that it still continued to rely on formulas) materialization of this idea, had significantly lower chances to be successful in serving this purpose. 
    
\section{The pigeonhole principle}

The (propositional) pigeonhole principle is a family of classical tautologies that is known to have no polynomial size proofs in resolution systems or analytic sequent calculus systems (Haken \cite{Haken}). And existence of polynomial size proofs for this family in the cut- and substitution-free calculus of structures is an open problem, conjectured to have a negative solution (see \cite{Gug07}). While polynomial size proofs for it in Frege- and Gentzen-style systems have been found (Cook and Rechkow \cite{CR}, Buss \cite{Buss}), those proofs rely on cut and, in the case of \cite{CR}, also on an extension rule. More recently (Finger \cite{Finger}), cut-free polynomial size proofs for the pigeonhole principle were also constructed, which, however,  rely on a substitution rule.\footnote{Finger's approach also requires to switch back to the more traditional Frege- and Hilbert-style understanding of proofs (only applied to sequents rather than formulas), where proofs are seen not as trees but as sequences (which, of course, can also be viewed as DAGs) of formulas, with the possibility for each formula to serve as a premise for any number of later formulas in the sequence. The related paper \cite{Gordeev} further illustrates the efficiency advantages of proof sequences over proof trees.} All known polynomial size proofs of the pigeonhole principle thus use extension, cut, or substitution --- the ``highly non-analytic'' rules. This section presents polynomial size \lll-proofs for the pigeonhole principle. They stand out as the first known ``reasonably analytic'' --- at least in the precise sense of i-analyticity --- tractable proofs of this class of tautologies.   Our construction partly exploits certain technical ideas from \cite{CR}.  
  
Throughout this section, $n$ is an arbitrary but fixed positive integer. When we say ``polynomial'' or ``exponential'', it should be understood as polynomial or exponential in $n$. As before, out of laziness, we will only be concerned with polynomiality versus exponentiality, leaving a more accurate asymptotic analysis as an exercise for an interested reader. Such an analysis, of course, would require a more precise specification of the meaning of the concept of proof size than the one we gave in Section 7. 

The (hyper)formulas and cirquents that we consider are built from $(n+1)\times n$ atoms denoted $P_{i,j}$, one per each $i\in\{0,\ldots,n\}$ (the set of pigeons) and $j\in\{1,\ldots,n\}$ (the set of pigeonholes). The meaning associated with $P_{i,j}$ is ``pigeon $i$ is sitting in hole $j$''. 

The {\bf $n$-pigeonhole principle} is expressed by the hyperformula
\[PHP^n =\bigmld\{ \gneg P_{i,1}\mlc\ldots\mlc \gneg P_{i,n}\ \ |\ \ 0\hspace{-2pt}\leq\hspace{-2pt} i\hspace{-2pt}\leq\hspace{-2pt} n\}\ \mld\ \bigmld\{ \overline{P_{i,j}}\mlc \overline{P_{e,j}}\ \ |\ \ 0\hspace{-2pt}\leq \hspace{-2pt}i\hspace{-2pt}<\hspace{-2pt}e\hspace{-2pt}\leq \hspace{-2pt}n,\ 1\hspace{-2pt}\leq\hspace{-2pt} j\hspace{-2pt}\leq \hspace{-2pt}n\}\]
(there is no need to overline the negative occurrences of atoms because there is only one such occurrence for each atom). 
Its left disjunct asserts that there is a pigeon $i$ that is not sitting in any hole. And the right disjunct asserts that there is a hole $j$ in which some two distinct pigeons $i$ and $e$ are sitting. This is the same as to say that if every pigeon is sitting in some hole, then there is a hole with (at least) two pigeons.
 
For each $i,j$ with $0\hspace{-2pt}\leq\hspace{-2pt} i\hspace{-2pt}\leq \hspace{-2pt}n$ and $1\hspace{-2pt}\leq\hspace{-2pt} j\hspace{-2pt}\leq\hspace{-2pt} n$, we define the formulas
\[\begin{array}{l}
X^{n}_{i,j}\ =\ P_{i,j};\vspace{3pt}\\
Y^{n}_{i,j}\ = \ \gneg P_{i,j}.\end{array}\]
Next, for each $k,i,j$ with $1<\hspace{-2pt} k\hspace{-2pt}\leq\hspace{-2pt} n$, \ $0\hspace{-2pt}\leq \hspace{-2pt}i\hspace{-2pt}\leq \hspace{-2pt}k-1$ and $1\hspace{-2pt}\leq \hspace{-2pt}j\leq \hspace{-2pt}k-1$, we define the formulas 
\[\begin{array}{lll}
X^{k-1}_{i,j}& =& (X_{i,j}^{k}\mld X_{i,k}^{k})\mlc(X_{i,j}^{k}\mld X_{k,j}^{k});\vspace{3pt}\\
Y^{k-1}_{i,j}& =& Y_{i,j}^{k}\mlc(Y^{k}_{i,k}\mld Y^{k}_{k,j})\mlc(X^{k}_{i,k}\mld Y^{k}_{i,k})\mlc (X^{k}_{k,k}\mld Y^{k}_{k,k}).\end{array}\]
Finally, for each $k$ with $1\hspace{-2pt}\leq\hspace{-2pt} k\hspace{-2pt}\leq\hspace{-2pt} n$, we define the formulas
\[\begin{array}{l}
B^k\ =\ \bigmlc\{ X^{k}_{i,j}\mld Y^{k}_{i,j}\ |\ 0\hspace{-2pt}\leq\hspace{-2pt} i\hspace{-2pt}\leq\hspace{-2pt} k,\ 1\hspace{-2pt}\leq \hspace{-2pt}j\leq \hspace{-2pt}k\};\\
C^{k}\ =\  \bigmld \{  Y^{k}_{i,1}\mlc\ldots\mlc Y^{k}_{i,k}\ |\ 0\hspace{-2pt}\leq \hspace{-2pt}i\hspace{-2pt}\leq \hspace{-2pt}k\}\ \mld\ \bigmld\{ X^{k}_{i,j}\mlc X^{k}_{e,j}\ |\ 0\hspace{-2pt}\leq\hspace{-2pt} i<\hspace{-2pt}e\hspace{-2pt}\leq \hspace{-2pt}k,\ 1\hspace{-2pt}\leq\hspace{-2pt} j\leq \hspace{-2pt}k \} .
\end{array}\]
The sizes of the formulas (cirquents) $B^k$ and $C^k$ are obviously exponential. However, due to sharing, the sizes of their ``full compressions'' $\overline{B^k}$ and $\overline{C^k}$ can be seen to be only polynomial.

In what follows, we prove a number of statements claiming existence of certain polynomial size derivations. In our proofs of those statements we usually restrict ourselves to describing the derivations, without any further explicit analysis of their sizes. Such descriptions alone will be sufficient for an experienced reader to immediately see that the derivations are indeed of polynomial sizes. 

From now on, a ``derivation'' or ``proof'' means a derivation or proof in \lll. When justifying steps in derivations, we often omit explicit references to restructuring, and indicate only one of the three main rules, even though that rule needs to be combined with some restructuring steps to yield the conclusion. Mostly such a ``rule'' is going to be pulldown and, to indicate that pulldown is combined with some straightforward restructuring, we will write ``{\bf pulldown*}'' instead of just ``pulldown''. Similarly for ``{\bf weakening*}'' and ``{\bf coupling*}''.

\begin{lemma}\label{l0}
$\overline{C^n}=PHP^n$. 
\end{lemma}

\begin{proof} Immediate, as $Y^{n}_{i,j}=\gneg P_{i,j}$ and $X^{n}_{i,j}= P_{i,j}$.
\end{proof}

\begin{lemma}\label{l1}
$\overline{B^n}$ has a polynomial size proof.   
\end{lemma}

\begin{proof} 
$\overline{B^n}$, which is the same as $B^n$, has $O(n^2)$ conjuncts, each conjunct being $X^{n}_{i,j}\mld Y^{n}_{i,j}$ for some $0\hspace{-2pt}\leq\hspace{-2pt} i\hspace{-2pt}\leq \hspace{-2pt}n,\ 1\hspace{-2pt}\leq\hspace{-2pt} j\hspace{-2pt}\leq \hspace{-2pt}n$. The latter is nothing but  $P_{i,j}\mld \gneg P_{i,j}$, which can be introduced by coupling*. \end{proof}

\begin{lemma}\label{l2}
There is a polynomial size derivation of $\overline{C^{1}}$ from $\overline{B^1}$.  
\end{lemma}

\begin{proof} 
$\overline{B^1}$ is $(\overline{X^{1}_{0,1}}\mld \overline{Y^{1}_{0,1}})\mlc(\overline{X^{1}_{1,1}}\mld \overline{Y^{1}_{1,1}})$, and 
$\overline{C^{1}}$ is --- more precisely, can be restructured into --- $\overline{Y^{1}_{0,1}}\mld\overline{Y^{1}_{1,1}}  \mld (\overline{X^{1}_{0,1}}\mlc\overline{X^{1}_{1,1}})$. 
The latter follows from the former by pulldown* applied twice. 
\end{proof}

\begin{lemma}\label{l4}
For each $k$ with $1<k\leq n$, there is a polynomial size derivation of $\overline{B^{k-1}}$ from $\overline{B^k}$.  
\end{lemma}

\begin{proof}  
\(\overline{B^k}$ is  $\bigmlc\{\overline{X^{k}_{i,j}}\mld \overline{Y^{k}_{i,j}}\ \ |\ 0\hspace{-2pt}\leq \hspace{-2pt}i\hspace{-2pt}\leq \hspace{-2pt}k,\ 1\hspace{-2pt}\leq\hspace{-2pt} j\hspace{-2pt}\leq\hspace{-2pt} k\}\), which can also be written as 
\begin{equation}\label{e1.1}
\bigmlc\{ \overline{X^{k}_{i,j}}\mld \overline{Y^{k}_{i,j}}, \ \ \overline{X^{k}_{k,j}}\mld \overline{Y^{k}_{k,j}},\ \  
\overline{X^{k}_{i,k}}\mld \overline{Y^{k}_{i,k}},\ \ \overline{X^{k}_{k,k}}\mld\overline{Y^{k}_{k,k}}\ \ |\ 0\hspace{-2pt}\leq\hspace{-2pt} i\hspace{-2pt}< \hspace{-2pt}k,\ 1\hspace{-2pt}\leq\hspace{-2pt} j\hspace{-2pt}<\hspace{-2pt} k\}.\end{equation}
We introduce the abbreviation 
\[D^{k}_{i}\ =\ (X^{k}_{i,k}\mld Y^{k}_{i,k})\mlc (X^{k}_{k,k}\mld Y^{k}_{k,k})\]
and restructure (\ref{e1.1}) into the following cirquent: 
\begin{equation}\label{e1}
\Bigmlc\bigleftbrace (\overline{X^{k}_{i,j}}\mld \overline{Y^{k}_{i,j}})
\mlc  
\bigl((\overline{X^{k}_{i,k}}\mld \overline{Y^{k}_{i,k}})  
\mlc(\overline{X^{k}_{k,j}}\mld \overline{Y^{k}_{k,j}})\bigr)\mlc  
\overline{D^{k}_{i}}\ |\ 0\hspace{-2pt}\leq\hspace{-2pt} i\hspace{-2pt}< \hspace{-2pt}k,\ 1\hspace{-2pt}\leq\hspace{-2pt} j\hspace{-2pt}<\hspace{-2pt} k\vspace{2pt}\bigrightbrace.
\end{equation}

Next, for each of the $O(k^2)$ conjuncts of (\ref{e1}), in turn,  we perform the following 
 transformation, leaving the rest of the cirquent unchanged\footnote{I.e., copying and pasting those unaffected parts from one cirquent to the next one in the derivation.}  while doing so. 

\begin{center} \begin{picture}(217,15)
\put(0,0){$\bigl(\overline{X^{k}_{i,j}}\mld \overline{Y^{k}_{i,j}}\bigr)
\mlc  
\bigl((\overline{X^{k}_{i,k}}\mld \overline{Y^{k}_{i,k}})  
\mlc(\overline{X^{k}_{k,j}}\mld \overline{Y^{k}_{k,j}})\bigr)\mlc  
\overline{D^{k}_{i}}$}
\end{picture}
\end{center}
\begin{center} \begin{picture}(217,25)
\put(0,21){\line(1,0){68}}
\put(72,19){\small pulldown* (twice)}
\put(146,21){\line(1,0){68}}
\put(2,0){$
\bigl(\overline{X_{i,j}^{k}}\mld \overline{Y_{i,j}^{k}}\bigr)\mlc  
\bigl((\overline{X_{i,k}^{k}}\mlc  \overline{X_{k,j}^{k}})\mld  \overline{Y^{k}_{i,k}}\mld  \overline{Y^{k}_{k,j}}\bigr)\mlc  
\overline{D^{k}_{i}}$}
\end{picture}
\end{center}
\begin{center} \begin{picture}(334,25)
\put(0,21){\line(1,0){125}}
\put(128,19){\small weakening* (twice)}
\put(206,21){\line(1,0){125}}
\put(0,0){$\Bigl(  
\bigl((\overline{X_{i,j}^{k}}\mld \overline{X_{i,k}^{k}})\mlc (\overline{X_{i,j}^{k}}\mld \overline{X_{k,j}^{k}})\bigr)\mld  \overline{Y_{i,j}^{k}}\Bigr) \mlc 
 \Bigl((\overline{X_{i,k}^{k}}\mlc  \overline{X_{k,j}^{k}})\mld  \overline{Y^{k}_{i,k}}\mld \overline{Y^{k}_{k,j}}\Bigr)\mlc 
  \overline{D^{k}_{i}}$}
\end{picture}
\end{center}

\begin{center} \begin{picture}(410,25)
\put(0,21){\line(1,0){162}}
\put(166,19){\small weakening* (twice)}
\put(244,21){\line(1,0){162}}
\put(0,0){$ 
\Bigl(\bigl((\overline{X_{i,j}^{k}}\mld \overline{X_{i,k}^{k}})\mlc (\overline{X_{i,j}^{k}}\mld \overline{X_{k,j}^{k}})\bigr)\mld  \overline{Y_{i,j}^{k}}\Bigr) \mlc 
\Bigl( \bigl((\overline{X_{i,j}^{k}}\mld \overline{X_{i,k}^{k}})\mlc  (\overline{X_{i,j}^{k}}\mld \overline{X_{k,j}^{k}})\bigr)\mld    \overline{Y^{k}_{i,k}}\mld \overline{Y^{k}_{k,j}}\Bigr)\mlc \overline{D^{k}_{i}}$}
\end{picture}
\end{center}

\begin{center} \begin{picture}(410,25)
\put(0,21){\line(1,0){152}}
\put(158,19){\small globalization (3 times)}
\put(254,21){\line(1,0){152}}
\put(0,0){$ 
\Bigl(\overline{\bigl((X_{i,j}^{k}\mld X_{i,k}^{k})\mlc (X_{i,j}^{k}\mld X_{k,j}^{k})\bigr)}\mld  \overline{Y_{i,j}^{k}}\Bigr) \mlc 
\Bigl( \overline{\bigl((X_{i,j}^{k}\mld X_{i,k}^{k})\mlc  (X_{i,j}^{k}\mld X_{k,j}^{k})\bigr)}\mld    \overline{Y^{k}_{i,k}}\mld \overline{Y^{k}_{k,j}}\Bigr)\mlc \overline{D^{k}_{i}}$}
\end{picture}
\end{center}

\begin{center} \begin{picture}(410,25)
\put(0,21){\line(1,0){173}}
\put(179,19){\small the same as}
\put(233,21){\line(1,0){173}}
\put(112,0){$
(\overline{X^{k-1}_{i,j}}\mld  \overline{Y_{i,j}^{k}})\mlc  
(\overline{X^{k-1}_{i,j}}\mld   
\overline{Y^{k}_{i,k}}\mld \overline{Y^{k}_{k,j}})\mlc   \overline{D^{k}_{i}}$}
\end{picture}
\end{center}

\begin{center} \begin{picture}(185,25)
\put(0,21){\line(1,0){50}}
\put(54,19){\small pulldown* (twice)}
\put(130,21){\line(1,0){50}}
\put(17,0){$
\overline{X^{k-1}_{i,j}}\mld\bigl( \overline{Y_{i,j}^{k}}\mlc 
(\overline{Y^{k}_{i,k}}\mld \overline{Y^{k}_{k,j}})\mlc 
\overline{D^{k}_{i}}\bigr)$}
\end{picture}
\end{center}

\begin{center} \begin{picture}(270,25)
\put(0,21){\line(1,0){103}}
\put(109,19){\small the same as}
\put(163,21){\line(1,0){103}}
\put(0,0){$
\overline{X^{k-1}_{i,j}}\mld\Bigl( \overline{Y_{i,j}^{k}}\mlc 
(\overline{Y^{k}_{i,k}}\mld \overline{Y^{k}_{k,j}})\mlc 
\overline{\bigl((X^{k}_{i,k}\mld Y^{k}_{i,k})\mlc (X^{k}_{k,k}\mld Y^{k}_{k,k})\bigr)}\Bigr)$}
\end{picture}
\end{center}

\begin{center} \begin{picture}(270,25)
\put(0,21){\line(1,0){103}}
\put(108,19){\small restructuring}
\put(163,21){\line(1,0){103}}
\put(7,0){$
\overline{X^{k-1}_{i,j}}\mld\overline{\bigl( Y_{i,j}^{k}\mlc 
(Y^{k}_{i,k}\mld Y^{k}_{k,j})\mlc 
(X^{k}_{i,k}\mld Y^{k}_{i,k})\mlc (X^{k}_{k,k}\mld Y^{k}_{k,k})\bigr)}$}
\end{picture}
\end{center}

\begin{center} \begin{picture}(260,35)
\put(0,31){\line(1,0){99}}
\put(103,29){\small the same as}
\put(155,31){\line(1,0){99}}
\put(98,10){$
\overline{X^{k-1}_{i,j}}\mld  \overline{Y^{k-1}_{i,j}}$}
\end{picture}
\end{center}

After repeating the above transformation for all $(i,j)$, we end up with the target cirquent $\overline{B^{k-1}} \ =\   
\bigmlc\{\overline{X^{k-1}_{i,j}}\mld \overline{Y^{k-1}_{i,j}} \ |\ 0\hspace{-2pt}\leq \hspace{-2pt}i\hspace{-2pt}< \hspace{-2pt}k, \ 1\hspace{-2pt}\leq \hspace{-2pt}j\hspace{-2pt}< \hspace{-2pt}k\bigrightbrace.$
 \end{proof}

\begin{lemma}\label{l5}
For each $k$ with $1\hspace{-2pt}<\hspace{-2pt}k\hspace{-2pt}\leq\hspace{-2pt} n$, there is a polynomial size derivation of $\overline{C^{k}}$ from $\overline{C^{k-1}}$.  
\end{lemma}

\begin{proof} 
$\overline{C^{k-1}}$ can be written as  
\begin{equation}\label{i16}
\Bigmld\bigleftbrace  \bigmlc\{\overline{Y^{k-1}_{i,j}} |\ 1\hspace{-2pt}\leq\hspace{-2pt} j\hspace{-2pt}<\hspace{-2pt} k \} \ \ |\ \ 0\hspace{-2pt}\leq \hspace{-2pt}i\hspace{-2pt}< \hspace{-2pt}\hspace{-2pt}k\bigrightbrace\mld 
\Bigmld\bigleftbrace
  \overline{X^{k-1}_{i,j}}\mlc \overline{X^{k-1}_{e,j}}
\ \  |\ \  0\hspace{-2pt}\leq\hspace{-2pt} i\hspace{-2pt}<\hspace{-2pt}e\hspace{-2pt}< \hspace{-2pt}k, \ 1\hspace{-2pt}\leq\hspace{-2pt} 
j\hspace{-2pt}< \hspace{-2pt}k\bigrightbrace.
\end{equation}
This is how we  derive $\overline{C^{k}}$ from (\ref{i16}). At the beginning, for each of the $O(k^3)$ subcirquents $ \overline{X^{k-1}_{i,j}}\mlc \overline{X^{k-1}_{e,j}}$ of (\ref{i16}), in turn, we do the following transformation, leaving the rest of the cirquent unchanged:

\begin{center} \begin{picture}(296,15)
\put(118,0){$ \overline{X^{k-1}_{i,j}}\mlc \overline{X^{k-1}_{e,j}}$}
\end{picture}
\end{center}

\begin{center} \begin{picture}(276,25)
\put(0,21){\line(1,0){109}}
\put(114,19){\small the same as}
\put(167,21){\line(1,0){109}}
\put(0,0){$ \bigl((\overline{X^{k}_{i,j}}\mld \overline{X^{k}_{i,k}})\mlc   (\overline{X^{k}_{i,j}}\mld
 \overline{X^{k}_{k,j}})\bigr)
\mlc   \bigl((\overline{X^{k}_{e,j}}\mld\overline{X^{k}_{e,k}})  \mlc (\overline{X^{k}_{k,j}}\mld\overline{X^{k}_{e,j}})\bigr)$}
\end{picture}
\end{center}

\begin{center} \begin{picture}(276,25)
\put(0,21){\line(1,0){120}}
\put(126,19){\small trade}
\put(156,21){\line(1,0){120}}
\put(18,0){$\bigl( \overline{X^{k}_{i,j}}\mld (\overline{X^{k}_{i,k}}\mlc  \overline{X^{k}_{k,j}})\bigr)    
 \mlc   \bigl((\overline{X^{k}_{e,j}}\mld\overline{X^{k}_{e,k}})  \mlc (\overline{X^{k}_{k,j}}\mld\overline{X^{k}_{e,j}})\bigr)$}
\end{picture}
\end{center}

\begin{center} \begin{picture}(250,25)
\put(0,21){\line(1,0){92}}
\put(97,19){\small restructuring}
\put(152,21){\line(1,0){92}}
\put(0,0){$\Bigl( \bigl(\overline{X^{k}_{i,j}}\mld (\overline{X^{k}_{i,k}}\mlc  \overline{X^{k}_{k,j}})\bigr) \mlc  \bigl(\overline{X^{k}_{e,j}}\mld\overline{X^{k}_{e,k}}\bigr)\Bigr)\mlc \Bigl(\overline{X^{k}_{k,j}}\mld\overline{X^{k}_{e,j}}\Bigr)
$}
\end{picture}
\end{center}

\begin{center} \begin{picture}(250,25)
\put(0,21){\line(1,0){96}}
\put(102,19){\small pulldown*}
\put(148,21){\line(1,0){96}}
\put(0,0){$\Bigl(\overline{X^{k}_{i,j}}\mld  \bigl((\overline{X^{k}_{i,k}}\mlc  \overline{X^{k}_{k,j}}) \mlc  (\overline{X^{k}_{e,j}}\mld\overline{X^{k}_{e,k}})\bigr)\Bigr)\mlc  \Bigl(\overline{X^{k}_{k,j}}\mld\overline{X^{k}_{e,j}}\Bigr)
$}
\end{picture}
\end{center}

\begin{center} \begin{picture}(250,25)
\put(0,21){\line(1,0){96}}
\put(102,19){\small pulldown*}
\put(148,21){\line(1,0){96}}
\put(2,0){$   
\bigl(\overline{X^{k}_{i,j}}\mlc (\overline{X^{k}_{k,j}}\mld\overline{X^{k}_{e,j}})
\bigr)\mld \bigl((\overline{X^{k}_{i,k}}\mlc  \overline{X^{k}_{k,j}})\mlc  (\overline{X^{k}_{e,j}}\mld\overline{X^{k}_{e,k}})\bigr)$}
\end{picture}
\end{center}

\begin{center} \begin{picture}(248,25)
\put(0,21){\line(1,0){90}}
\put(96,19){\small restructuring}
\put(154,21){\line(1,0){90}}
\put(0,0){$  
\Bigl(\overline{X^{k}_{i,j}}\mlc \bigl(\overline{X^{k}_{k,j}}\mld\overline{X^{k}_{e,j}}\bigr)\Bigr)\mld \Bigl(\overline{X^{k}_{i,k}}\mlc  \bigl((\overline{X^{k}_{e,j}}\mld\overline{X^{k}_{e,k}})\mlc \overline{X^{k}_{k,j}} \bigr) \Bigr)$}
\end{picture}
\end{center}

\begin{center} \begin{picture}(248,25)
\put(0,21){\line(1,0){96}}
\put(101,19){\small pulldown*}
\put(148,21){\line(1,0){96}}
\put(0,0){$
\Bigl(\overline{X^{k}_{i,j}}\mlc  \bigl(\overline{X^{k}_{k,j}}\mld\overline{X^{k}_{e,j}}\bigr)
\Bigr)\mld \Bigl(\overline{X^{k}_{i,k}}\mlc \bigl(\overline{X^{k}_{e,k}}\mld (\overline{X^{k}_{e,j}}\mlc \overline{X^{k}_{k,j}}) \bigr) \Bigr)$}
\end{picture}
\end{center}

\begin{center} \begin{picture}(248,25)
\put(0,21){\line(1,0){96}}
\put(101,19){\small pulldown*}
\put(148,21){\line(1,0){96}}
\put(6,0){$
\bigl(\overline{X^{k}_{e,j}}\mlc \overline{X^{k}_{k,j}}\bigr)\mld
\bigl(\overline{X^{k}_{i,j}}\mlc  (\overline{X^{k}_{k,j}}\mld\overline{X^{k}_{e,j}})
\bigr)\mld \bigl(\overline{X^{k}_{i,k}}\mlc \overline{X^{k}_{e,k}}\bigr) $}
\end{picture}
\end{center}

\begin{center} \begin{picture}(280,25)
\put(0,21){\line(1,0){112}}
\put(118,19){\small pulldown*}
\put(165,21){\line(1,0){112}}
\put(0,0){$   
\Bigl(\overline{X^{k}_{e,j}}\mlc \overline{X^{k}_{k,j}}\Bigr)\mld
\Bigl(\overline{X^{k}_{i,j}}\mlc \bigl(\overline{X^{k}_{e,j}}\mld 
(\overline{X^{k}_{i,j}}\mlc \overline{X^{k}_{k,j}})\bigr)  
\Bigr)\mld
\Bigl(\overline{X^{k}_{i,k}}\mlc\overline{X^{k}_{e,k}}\Bigr) $}
\end{picture}
\end{center}

\begin{center} \begin{picture}(280,25)
\put(0,21){\line(1,0){112}}
\put(118,19){\small pulldown*}
\put(165,21){\line(1,0){112}}
\put(10,0){$  
(\overline{X^{k}_{i,j}}\mlc \overline{X^{k}_{k,j}})\mld 
(\overline{X^{k}_{e,j}}\mlc \overline{X^{k}_{k,j}})\mld
(\overline{X^{k}_{i,j}}\mlc \overline{X^{k}_{e,j}})\mld
(\overline{X^{k}_{i,k}}\mlc \overline{X^{k}_{e,k}})$}
\end{picture}
\end{center}

After repeating the above transformation for all subcirquents $ \overline{X^{k-1}_{i,j}}\mlc \overline{X^{k-1}_{e,j}}$ of (\ref{i16}), the latter turns into the following cirquent: 
\begin{equation}\label{h13}
\begin{array}{l}
\Bigmld\Bigleftbrace  \bigmlc\bigleftbrace\overline{Y^{k-1}_{i,j}} |\ 1\hspace{-2pt}\leq\hspace{-2pt} j\hspace{-2pt}<\hspace{-2pt} k \bigrightbrace \ \ |\ \ 0\hspace{-2pt}\leq \hspace{-2pt}i\hspace{-2pt}< \hspace{-2pt}\hspace{-2pt}k\Bigrightbrace\ \ \mld\vspace{3pt}\\
\Bigmld\Bigleftbrace
(\overline{X^{k}_{i,j}}\mlc \overline{X^{k}_{k,j}})\mld 
(\overline{X^{k}_{e,j}}\mlc \overline{X^{k}_{k,j}})\mld
(\overline{X^{k}_{i,j}}\mlc \overline{X^{k}_{e,j}})\mld
(\overline{X^{k}_{i,k}}\mlc \overline{X^{k}_{e,k}}) \ \   
|\  \ 0\hspace{-2pt}\leq\hspace{-2pt} i\hspace{-2pt}<\hspace{-2pt}e\hspace{-2pt}< \hspace{-2pt}k, \ 1\hspace{-2pt}\leq\hspace{-2pt} 
j\hspace{-2pt}< \hspace{-2pt}k\Bigrightbrace.
\end{array}\hspace{-6pt}
\end{equation}

Next, for each of the $k$ subcirquents $\bigmlc\bigleftbrace\overline{Y^{k-1}_{i,j}} |\ 1\hspace{-2pt}\leq\hspace{-2pt} j\hspace{-2pt}<\hspace{-2pt} k \bigrightbrace$ of (\ref{h13}), one after one, we perform the following transformation, leaving the rest of the cirquent unchanged:

\begin{center} \begin{picture}(80,15)
\put(0,0){$\bigmlc\bigleftbrace\overline{Y^{k-1}_{i,j}} |\ 1\hspace{-2pt}\leq\hspace{-2pt} j\hspace{-2pt}<\hspace{-2pt} k \bigrightbrace$}
\end{picture}
\end{center}

\begin{center} \begin{picture}(272,25)
\put(0,21){\line(1,0){108}}
\put(114,19){\small the same as}
\put(167,21){\line(1,0){108}}
\put(0,0){$\bigmlc\bigleftbrace \overline{Y^{k}_{i,j}\mlc   
(Y^{k}_{i,k}\mld Y^{k}_{k,j})\mlc (X^{k}_{i,k}\mld Y^{k}_{i,k})\mlc  (X^{k}_{k,k}\mld Y^{k}_{k,k})} \ \ |\ 1\hspace{-2pt}\leq\hspace{-2pt} j\hspace{-2pt}<\hspace{-2pt} k \bigrightbrace$}
\end{picture}
\end{center}

\begin{center} \begin{picture}(366,25)
\put(0,21){\line(1,0){153}}
\put(157,19){\small restructuring}
\put(214,21){\line(1,0){153}}
\put(0,0){$\Bigl(\Bigmlc\{\overline{Y^{k}_{i,j}}\ |\ 1\hspace{-2pt}\leq\hspace{-2pt} j\hspace{-2pt}<\hspace{-2pt} k\}\Bigr) \ \mlc \ 
\Bigl(\overline{X^{k}_{i,k}}\mld\overline{Y^{k}_{i,k}}\Bigr) \ \mlc \ 
\Bigl(\Bigmlc\bigleftbrace\overline{Y^{k}_{i,k}}\mld\overline{Y^{k}_{k,j}}\ |\ 1\hspace{-2pt}\leq\hspace{-2pt} j\hspace{-2pt}<\hspace{-2pt} k\bigrightbrace\mlc  (\overline{X^{k}_{k,k}}\mld\overline{Y^{k}_{k,k}}) \Bigr) $}
\end{picture}
\end{center}

\begin{center} \begin{picture}(376,25)
\put(0,21){\line(1,0){171}}
\put(177,19){\small trade}
\put(208,21){\line(1,0){171}}
\put(0,0){$\Bigl(\Bigmlc\{\overline{Y^{k}_{i,j}}\ |\ 1\hspace{-2pt}\leq\hspace{-2pt} j\hspace{-2pt}<\hspace{-2pt} k\}\Bigr) \ \mlc \ 
\Bigl(\overline{X^{k}_{i,k}}\mld\overline{Y^{k}_{i,k}}\Bigr) \ \mlc \ \Bigl(\bigl(\overline{Y^{k}_{i,k}}\mld\Bigmlc\{\overline{Y^{k}_{k,j}}\ |\ 1\hspace{-2pt}\leq\hspace{-2pt} j\hspace{-2pt}<\hspace{-2pt} k\}\bigr) \mlc   \bigl(\overline{X^{k}_{k,k}}\mld \overline{Y^{k}_{k,k}}\bigr)\Bigr)$}
\end{picture}
\end{center}

\begin{center} \begin{picture}(376,25)
\put(0,21){\line(1,0){163}}
\put(169,19){\small pulldown*}
\put(216,21){\line(1,0){163}}
\put(2,0){$\Bigl(\Bigmlc\{\overline{Y^{k}_{i,j}}\ |\ 1\hspace{-2pt}\leq\hspace{-2pt} j\hspace{-2pt}<\hspace{-2pt} k\}\Bigr) \ \mlc \ 
\Bigl(\overline{X^{k}_{i,k}}\mld\overline{Y^{k}_{i,k}}\Bigr) \ \mlc \ \Bigl(\bigl(\Bigmlc\{\overline{Y^{k}_{k,j}}\ |\ 1\hspace{-2pt}\leq\hspace{-2pt} j\hspace{-2pt}<\hspace{-2pt} k\}\mlc   (\overline{X^{k}_{k,k}}\mld \overline{Y^{k}_{k,k}})\bigr)\mld \overline{Y^{k}_{i,k}}\Bigr)$}
\end{picture}
\end{center}

\begin{center} \begin{picture}(376,25)
\put(0,21){\line(1,0){163}}
\put(169,19){\small pulldown*}
\put(216,21){\line(1,0){163}}
\put(5,0){$\Bigl(\Bigmlc\{\overline{Y^{k}_{i,j}}\ |\ 1\hspace{-2pt}\leq\hspace{-2pt} j\hspace{-2pt}<\hspace{-2pt} k\}\Bigr) \ \mlc \ 
\Bigl(\overline{X^{k}_{i,k}}\mld\overline{Y^{k}_{i,k}}\Bigr) \ \mlc \ \Bigl(\bigl(\Bigmlc\{\overline{Y^{k}_{k,j}}\ |\ 1\hspace{-2pt}\leq\hspace{-2pt} j\hspace{-2pt}<\hspace{-2pt} k\}\mlc   \overline{Y^{k}_{k,k}}\bigr)\mld \overline{X^{k}_{k,k}}\mld \overline{Y^{k}_{i,k}}\Bigr)$}
\end{picture}
\end{center}

\begin{center} \begin{picture}(370,25)
\put(0,21){\line(1,0){156}}
\put(161,19){\small restructuring}
\put(217,21){\line(1,0){156}}
\put(22,0){$\Bigl(\Bigmlc\{\overline{Y^{k}_{i,j}}\ |\ 1\hspace{-2pt}\leq\hspace{-2pt} j\hspace{-2pt}<\hspace{-2pt} k\}\Bigr) \ \mlc \ 
\Bigl(\overline{X^{k}_{i,k}}\mld\overline{Y^{k}_{i,k}}\Bigr) \ \mlc \ \Bigl(\Bigmlc\{\overline{Y^{k}_{k,j}}\ |\ 1\hspace{-2pt}\leq\hspace{-2pt} j\hspace{-2pt}\leq\hspace{-2pt} k\} \mld  \overline{X^{k}_{k,k}}\mld \overline{Y^{k}_{i,k}}\Bigr)$}
\end{picture}
\end{center}

\begin{center} \begin{picture}(342,25)
\put(0,21){\line(1,0){142}}
\put(146,19){\small restructuring}
\put(202,21){\line(1,0){142}}
\put(0,0){$\Bigl(\Bigmlc\{\overline{Y^{k}_{i,j}}\ |\ 1\hspace{-2pt}\leq\hspace{-2pt} j\hspace{-2pt}<\hspace{-2pt} k\}\Bigr)\ \ \mlc\ \  
 \Bigl(\bigl(\overline{X^{k}_{i,k}}\mld\overline{Y^{k}_{i,k}}\bigr)\ \mlc\  \bigl(\Bigmlc\{\overline{Y^{k}_{k,j}}\ |\ 1\hspace{-2pt}\leq\hspace{-2pt} j\hspace{-2pt}\leq\hspace{-2pt} k\} \mld \overline{X^{k}_{k,k}}\mld\overline{Y^{k}_{i,k}}\bigr)\Bigr)$}
\end{picture}
\end{center}

\begin{center} \begin{picture}(362,25)
\put(0,21){\line(1,0){157}}
\put(162,19){\small pulldown*}
\put(205,21){\line(1,0){157}}
\put(0,0){$\Bigl(\Bigmlc\{\overline{Y^{k}_{i,j}}\ |\ 1\hspace{-2pt}\leq\hspace{-2pt} j\hspace{-2pt}<\hspace{-2pt} k\}\}\Bigr)\ \ \mlc\ \   
\Bigl(\overline{Y^{k}_{i,k}}\ \mld\ 
\bigl(\Bigmlc\{\overline{Y^{k}_{k,j}}\ |\ 1\hspace{-2pt}\leq\hspace{-2pt} j\hspace{-2pt}\leq\hspace{-2pt} k\}\bigr)\ \mld\  \bigl((\overline{X^{k}_{i,k}}\mld\overline{Y^{k}_{i,k}})\mlc\overline{X^{k}_{k,k}}\bigr)\Bigr)$}
\end{picture}
\end{center}

\begin{center} \begin{picture}(362,25)
\put(0,21){\line(1,0){157}}
\put(162,19){\small pulldown*}
\put(205,21){\line(1,0){157}}
\put(16,0){$\Bigl(\Bigmlc\{\overline{Y^{k}_{i,j}}\ |\ 1\hspace{-2pt}\leq\hspace{-2pt} j\hspace{-2pt}<\hspace{-2pt} k\}\}\Bigr)\ \ \mlc\ \   
\Bigl(\overline{Y^{k}_{i,k}}\ \mld\ \bigl(
\Bigmlc\{\overline{Y^{k}_{k,j}}\ |\ 1\hspace{-2pt}\leq\hspace{-2pt} j\hspace{-2pt}\leq\hspace{-2pt} k\}\bigr)\ \mld\  \bigl(\overline{X^{k}_{i,k}}\mlc\overline{X^{k}_{k,k}}\bigr)\Bigr)$}
\end{picture}
\end{center}

\begin{center} \begin{picture}(330,35)
\put(0,31){\line(1,0){139}}
\put(145,29){\small pulldown*}
\put(189,31){\line(1,0){139}}
\put(15,10){$ \bigl(\Bigmlc\{\overline{Y^{k}_{i,j}}\ |\ 1\hspace{-2pt}\leq\hspace{-2pt} j\hspace{-2pt}<\hspace{-2pt} k\}\mlc
\overline{Y^{k}_{i,k}}\bigr)\ \mld
\ \bigl(\Bigmlc\{\overline{Y^{k}_{k,j}}\ |\ 1\hspace{-2pt}\leq\hspace{-2pt} j\hspace{-2pt}\leq\hspace{-2pt} k\}\bigr)\ \mld \ \bigl(\overline{X^{k}_{i,k}}\mlc\overline{X^{k}_{k,k}}\bigr)$}
\end{picture}
\end{center}

After repeating the above procedure for all subcirquents $\bigmlc\{\overline{Y^{k-1}_{i,j}} |\ 1\hspace{-2pt}\leq\hspace{-2pt} j\hspace{-2pt}<\hspace{-2pt} k \}$ of (\ref{h13}), the latter turns into the following cirquent:
\begin{equation}\label{g2}
\begin{array}{l}
\Bigmld\Bigleftbrace \bigl(\Bigmlc\{\overline{Y^{k}_{i,j}}\ |\ 1\hspace{-2pt}\leq\hspace{-2pt} j\hspace{-2pt}<\hspace{-2pt} k\}\mlc
\overline{Y^{k}_{i,k}}\bigr)\ \mld
\ \bigl(\Bigmlc\{\overline{Y^{k}_{k,j}}\ |\ 1\hspace{-2pt}\leq\hspace{-2pt} j\hspace{-2pt}\leq\hspace{-2pt} k\}\bigr)\ \mld \ \bigl(\overline{X^{k}_{i,k}}\mlc\overline{X^{k}_{k,k}}\bigr)\ \ |\ \ 0\hspace{-2pt}\leq \hspace{-2pt}i\hspace{-2pt}< \hspace{-2pt}\hspace{-2pt}k\Bigrightbrace\vspace{3pt}\ \ \mld\\
\Bigmld\Bigleftbrace
(\overline{X^{k}_{i,j}}\mlc \overline{X^{k}_{k,j}})\mld 
(\overline{X^{k}_{e,j}}\mlc \overline{X^{k}_{k,j}})\mld
(\overline{X^{k}_{i,j}}\mlc \overline{X^{k}_{e,j}})\mld
(\overline{X^{k}_{i,k}}\mlc \overline{X^{k}_{e,k}}) \ \   
|\  \ 0\hspace{-2pt}\leq\hspace{-2pt} i\hspace{-2pt}<\hspace{-2pt}e\hspace{-2pt}< \hspace{-2pt}k, \ 1\hspace{-2pt}\leq\hspace{-2pt} 
j\hspace{-2pt}< \hspace{-2pt}k\Bigrightbrace.\end{array}
\end{equation}

Finally, we restructure (\ref{g2}) into 
\[\Bigmld \{ \overline{Y^{k}_{i,1}}\mlc\ldots\mlc\overline{Y^{k}_{i,k}}  \ |\ 0\hspace{-2pt}\leq \hspace{-2pt}i\hspace{-2pt}\leq \hspace{-2pt}\hspace{-2pt}k\}\ \mld\   \Bigmld\{ \overline{X^{k}_{i,j}}\mlc \overline{X^{k}_{e,j}}\ |\ 0\hspace{-2pt}\leq\hspace{-2pt} i\hspace{-2pt}<\hspace{-2pt}e\hspace{-2pt}\leq \hspace{-2pt}k,\ 1\hspace{-2pt}\leq\hspace{-2pt} j\hspace{-2pt}\leq \hspace{-2pt}k\},\]
which is nothing but the desired $\overline{C^{k}}$.
\end{proof}

\begin{theorem}\label{t1}
There is a polynomial size proof of $PHP^n$. 
\end{theorem}

\begin{proof} Lemmas \ref{l1}, \ref{l4},  \ref{l2} and \ref{l5} imply that there is a polynomial size proof of $\overline{C^n}$, which,  
 by Lemma \ref{l0}, 
is the same as $PHP^n$. 
\end{proof}

\end{document}